\documentclass[pra,twocolumn,superscriptaddress,a4paper,english,longbibliography]{revtex4-2}

\usepackage{lineno}
\usepackage{times}
\usepackage{color}
\usepackage[colorlinks=true,urlcolor=blue,citecolor=blue]{hyperref}
\usepackage{url}
\usepackage{breakurl}
\usepackage{soul}
\usepackage{graphicx}
\usepackage{amsmath}
\usepackage{subfigure}
\graphicspath{ {./figures/} }
\usepackage{xcolor}
\usepackage{amssymb}
\usepackage{hyperref}
\DeclareGraphicsExtensions{.png,.eps}
\usepackage{float}

\usepackage{appendix}
\usepackage{algorithm}
\usepackage{algpseudocode}

\newcommand {\R}{\textcolor {black}}
\newcommand {\red}{\textcolor {black}}
\newcommand {\RR}{\textcolor {black}}
\newcommand {\RRed}{\textcolor {black}}
\usepackage{xcolor}
\usepackage[urlcolor=blue]{hyperref}      
\hypersetup{
    colorlinks = true,                    
    citecolor = {blue},
    linkcolor = {purple},
           }

\begin{document}


\title{Tensor network for interpretable and efficient quantum-inspired machine learning}

\author{Shi-Ju Ran} \email[Corresponding author. Email: ] {sjran@cnu.edu.cn}
\affiliation{Department of Physics, Capital Normal University, Beijing 100048, China}
\author{Gang Su} \email[Corresponding author. Email: ] {gsu@ucas.ac.cn}
\affiliation{Kavli Institute for Theoretical Sciences, and CAS Center for Excellence in Topological Quantum Computation, University of Chinese Academy of Sciences, Beijing 100190, China}
\affiliation{School of Physical Sciences, University of Chinese Academy of Sciences, P. O. Box 4588, Beijing 100049, China}
\date{\today}

\begin{abstract}
	It is a critical challenge to simultaneously gain high interpretability and efficiency with the current schemes of deep machine learning (ML). Tensor network (TN), which is a well-established mathematical tool originating from quantum mechanics, has shown its unique advantages on developing efficient ``white-box'' ML schemes. Here, we give a brief review on the inspiring progresses made in TN-based ML. On one hand, interpretability of TN ML is accommodated with the solid theoretical foundation based on quantum information and many-body physics. On the other hand, high efficiency can be rendered from the powerful TN representations and the advanced computational techniques developed in quantum many-body physics. With the fast development on quantum computers, TN is expected to conceive novel schemes runnable on quantum hardware, heading towards the ``quantum artificial intelligence'' in the forthcoming future. 
\end{abstract}

\maketitle
\section{Introduction}
	
	Deep machine learning (ML) such as \red{those based on deep} neural network (NN) has achieved tremendous successes in, e.g., computer vision and natural language process. However, the dilemma between the interpretability and efficiency, which is a long-concerned topic~\cite{BGYG+18MLinterp, ZZ18MLinterp, L18MLInterp, CPC19MLinterp}, has caused several severe challenges. \red{Generally speaking, interpretability is defined as the degree to which a human can understand the cause of a decision, which is critical to question, understand, and trust the deep ML methods~\cite{CPC19MLinterp}.}
	
	\red{Though universal characterizations of interpretability are still controversial, it is widely recognized that the powerful deep NN models are non-interpretable. Due to their high non-linearity}, rigorous understanding of the underlying mathematics of such models is mostly unlikely. \red{Interpretations in these cases are usually ``\textit{post hoc}'' (see, e.g., a recent work in Ref.~\cite{MRC23IntNN}), in contrast to the ``intrinsic'' interpretation with ``white-box'' ML models or by knowing how the models work~\cite{L18MLInterp, R19MLint}.} Consequently, the investigations are generally conducted in an inefficient trial-and-error manner that consumes significant human and computational resources. Another serious issue brought by the non-interpretability concerns the robustness. For instance, a well-trained \red{deep} NN model might be severely disturbed by noises or intentional attacks (see, e.g., Ref.~\cite{SVS19MLattack}). Currently, there exist no general theories to quantitatively characterize how far the \red{deep NN models} can be trusted or how significant different disturbances can affect the predictions. \red{From the perspective of applications, interpretability concerns several vital issues such as fairness and privacy~\cite{doshivelez2017rigorous, BGYG+18MLinterp}.}
	
	Among the potential ways of opening the black boxes in deep ML, the (classical) probabilistic theories have drawn wide attentions. From a ``revisionist'' perspective, \red{the deep} NN has been incorporated with, e.g., the mutual information, relative entropy, or the physics-inspired renormalization groups~\cite{B94MLinfo, RN17DLexpl, KR18MLRG, AYJM+19DLBI, EGJ22EntML}, to show how the models process information. \red{These interpretations are mostly \textit{post hoc}, and help to understand the learning decision-making processes to certain extent. But the results or conclusions might strongly depend on data or the specifics of models.}
	
	\red{The probabilistic ML models~\cite{G15PML}, such as Bayesian networks~\cite{B06MLbook} and Boltzmann machines~\cite{AHS85BM} (a type of Markov random fields),  are regarded to be ``white-box'' and intrinsically interpretable.} These models promise to interpret in the statistical ways that human minds can follow, where we have, e.g., the probabilistic reasoning to unveil the hidden casual relations~\cite{G03BayesML, TGK06MLBayes, LG10BayesML}. Unfortunately, the gaps between the performance of these probabilistic models and state-of-the-art deep NN's are quite huge. It seems that high efficiency and interpretability cannot be reached simultaneously with the ML models at hand~\RRed{\footnote{For the efficiency of a ML method, we mainly mean the accuracy with a proper computational complexity depending on its academic or industrial uses. For the classical ``white-box'' ML schemes, it seems that their accuracy could not beat that of the deep NN's even if we immensely increase their complexity.}}.
	
	\begin{figure}[tbp]
		\centering
		\includegraphics[angle=0,width=1\linewidth]{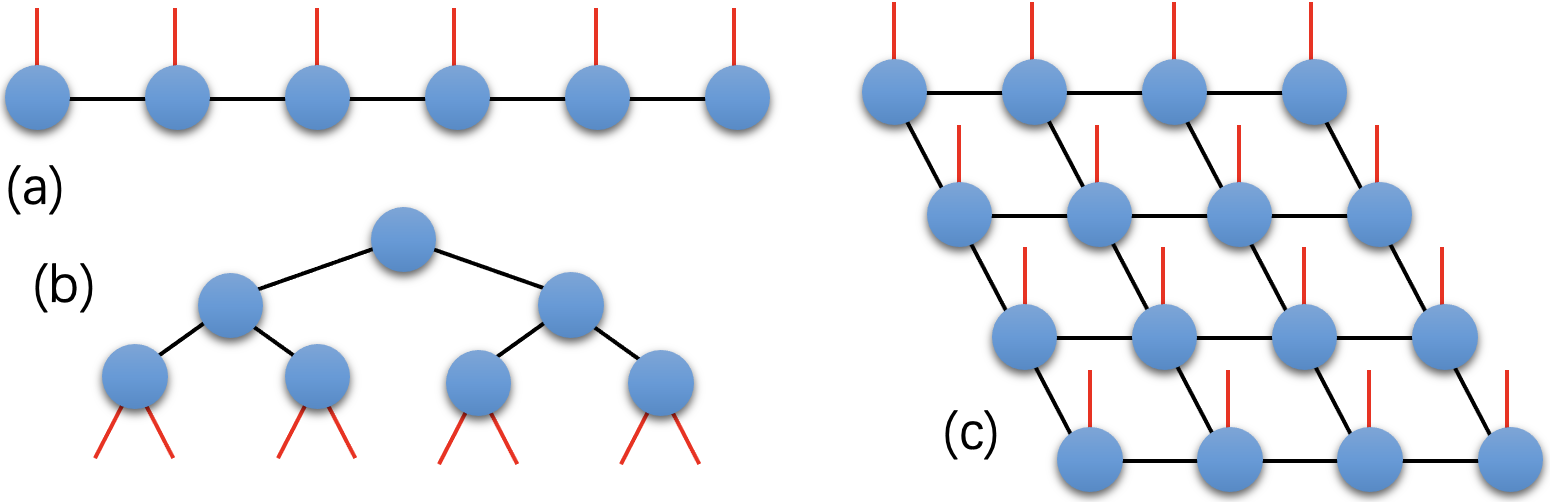}
		\caption{(Color online) Diagrammatic representations of (a) matrix product state, (b) tree TN, and (c) projected entangled pair state. \red{In (a), we mark the tensors and their indexes of an MPS consisting of 6 tensors $\{\boldsymbol{A}^{(m)}\}$ ($m=1, \ldots, 6$) according to Eq.~(\ref{eq-TN}).}}
		\label{fig-TN}
	\end{figure}
	
	\section{Tensor network: a powerful ``white-box'' mathematical tool from quantum physics}
	
	With the fast development of both the classical and quantum computations, tensor network (TN) sheds new light on getting out of the dilemma between interpretability and efficiency. A TN is defined as the contraction of multiple tensors. Its network structure determines how the tensors are contracted. Fig.~\ref{fig-TN} gives the diagrammatic representations of three kinds of TN's, namely matrix product state~\cite{PVWC07MPSRev} (MPS, which is also known as the tensor-train form~\cite{O11TTD} with the open boundary condition), tree TN~\cite{SDV06TTN}, and projected entangled pair state~\cite{VWPC06PEPSfamous}. Taking an MPS formed by $M$ tensors as an example, it results in an $M$-th order tensor $\boldsymbol{\mathcal{T}}$ after contracting the virtual indexes [the black bonds in Fig.~\ref{fig-TN} \red{(a)}] that satisfies
	\begin{equation}
		\mathcal{T}_{s_{1} s_{2} \ldots} = \sum_{\{\alpha_{m}\}} A^{(1)}_{s_{1}\alpha_{1}} A^{(2)}_{s_{2}\alpha_{1}\alpha_{2}} \ldots A^{(m)}_{s_{m}\alpha_{m-1}\alpha_{m}} \ldots A^{(M)}_{s_{M}\alpha_{M-1}}.
		\label{eq-TN}
	\end{equation}
	
	TN has achieved significant successes in quantum mechanics as an efficient representation of the states for the large-scale quantum systems~\cite{VMC08MPSPEPSRev, CV09TNSRev, RTPC+17TNrev, O19TNrev, CPDSV21TNSrev}. From the perspective of classical computation of quantum problems, the dimension of Hilbert space~\RRed{\footnote{Hilbert space is a generalization of Euclidean space. We can define vectors, operators, and the measures of their distances in a Hilbert space, while the dimensions can be infinite. Hilbert space has been used to describe the space for quantum states and operators.}} and the parameter complexity of quantum states scale exponentially with the size of the quantum system, which is known as the ``curse of dimensionality'' or ``exponential wall''. This makes large quantum systems inaccessible by the conventional methods such as exact diagonalization~\RRed{\footnote{Exact diagonalization is a widely used approach by fully diagonalizing the matrix (say quantum Hamiltonian) that contains the full information for the physics of the considered model.}}. TN reduces the parameter complexity of representing a quantum state to be just polynomial, thus efficient simulations of a large class of quantum systems became plausible. Taking MPS as an example again, the number of parameters are reduced from $\#(\boldsymbol{\mathcal{T}}) \sim O(d^{M})$ \red{(the number of elements in the $M$-th order tensor $\boldsymbol{\mathcal{T}}$)} to $\#(\text{MPS}) \sim O(Md\chi^{2})$ \red{(the total number of parameters in the tensors $\{\boldsymbol{A}^{(m)}\}$ ($m=1, \ldots, 6$)),} with $d=\text{dim}(s_{m})$ and $\chi=\text{dim}(\alpha_{m})$.
	
	\begin{figure}[tbp]
		\centering
		\includegraphics[angle=0,width=1\linewidth]{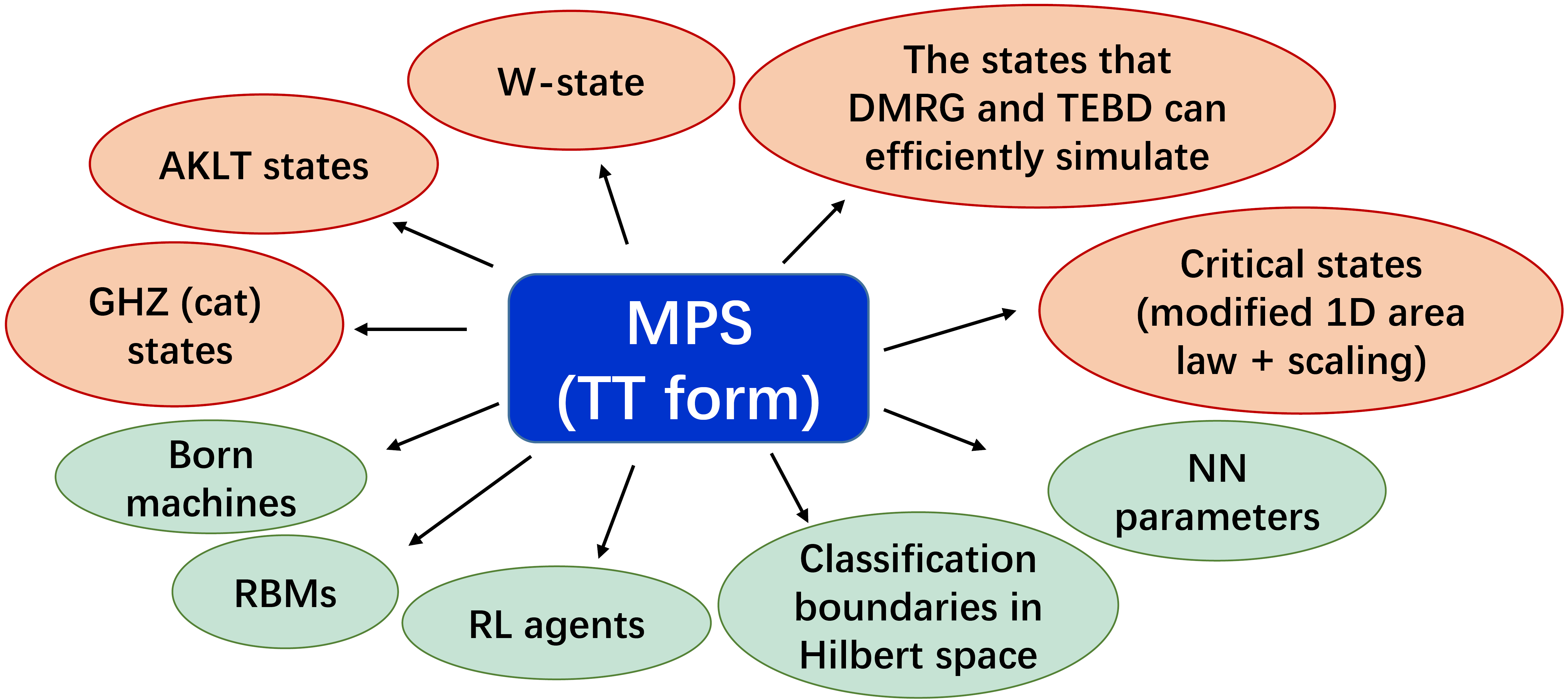}
		\caption{(Color online) \red{MPS~\cite{PVWC07MPSRev} (or TT form~\cite{O11TTD}) can efficiently represent or formulate a large group of mathematical objects. For quantum states, it can} represent the cat states~\cite{GHZ89GHZ}, AKLT state~\cite{AKLT87AKLT, AKLT88AKLT2}, W-state~\cite{W01Wstate}, and the states that can be efficiently simulated by DMRG~\cite{W92DMRG, W93DMRG} and TEBD~\cite{V03TEBD, V04TEBD}. Equipped by the scaling theories~\cite{VLRK03CritEnt, PMTM09EntScaling, TOIL08EntScaling, PMJ2010entanglement}, MPS can access the properties of the critical states obeying the 1D area law with logarithmic correction. \red{For ML, MPS can represent the restricted Boltzmann machines~\cite{AHS85BM} and the quantum Born machines~\cite{CCW18Born, CWXZ19generateTTNML}. It has been used to parameterize the classification boundaries for supervised learning~\cite{SPLRS19GTNC} and the agents for reinforcement learning (RL)~\cite{MB23TNQCML}. The high-order tensors storing the parameters of NN can also be formulated as MPS (or MPO) for model compression~\cite{TSN17TTRNN, YLMCZ2019TRDcomp, PXWY+19TRDRNN, GCHX20+NNMPO, SGLLY20NNMPO, WZLDW20TTCNN, WZCL+21nonlinearTT, qing2023compressing}.}}
		\label{fig-MPS}
	\end{figure}
	
	Among the TN theories, it has been revealed that the states satisfying the ``area laws''~\RRed{\footnote{We may consider a land consisting of many villages as an example to understand the area law. If every person in this land can only communicate with the persons in a short range nearby, the people who can communicate to a different village should live near the boarders. Therefore, the amount of exchanged information between a village and the rest ones should scale with the length of its boarders. In this case, the (two-dimensional) area law is applied. If phones are introduced, people are able to communicate with anyone in this land. The amount of exchanged information of a village should scale with its population or approximately the size of its territory. This is the volume law, where obviously the amount of exchanged information increases much faster than the area-law cases as the village expands.}} of entanglement entropy~\cite{ECP10AreaLawRev} can be efficiently approximated by the TN representations with finite bond dimensions~\cite{VWPC06PEPSfamous, TEV09TTN, PC20TNarealaw, CPDSV21TNSrev}. Entanglement entropy~\cite{NC02Qcmp} is a fundamental concept in quantum sciences~\RRed{\footnote{Entanglement can be understood as a description of quantum correlations. A strong entanglement between two quantum particles means significant affections to the state of one particle by operating another. Entanglement entropy is a measure of the strength of entanglement. For instance, zero entanglement entropy between two particles means that their state should be described by a product state. Even these two particles might be correlated, the correlations should be fully described by the classical correlations.}}. \RR{From a statistic perspective, the entanglement entropy between two subparts of a system} characterizes the amount of the gained information on one subpart by knowing the information on the rest part. Luckily, most of the states that we care about satisfy such area laws\red{, meaning the entanglement entropy scales not with the volume of the subpart but with the length of the boundary}. For instance, MPS obeys the one-dimensional area law\red{, where the boundary is represented by two zero-dimensional points and the entanglement entropy is a constant. Such an area law is satisfied by} the low-lying eigenstates of many one- and quasi-one-dimensional quantum lattice models~\cite{H07EntArea, H07MPSent, SWVC08MPSent, VC06MPSFaithfully, BOE10arealaw, PHV12exciteMPS, FWBSE15MPSMLocal}, including those with non-trivial topological properties~\cite{AKLT87AKLT, AKLT88AKLT2}. Therefore, \red{the MPS-based} algorithms including density matrix renormalization group (DMRG)~\cite{W92DMRG, W93DMRG} and time-evolving block decimation (TEBD)~\cite{V03TEBD, V04TEBD} exhibit remarkable efficiency for simulating such systems. 
	
	\red{Moreover}, a large class of artificially constructed states widely used in quantum information processing and computing can also be represented by MPS, such as Greenberger-Horne-Zeilinger (GHZ) states (also known as the cat states)~\cite{GHZ89GHZ} and W-state~\cite{W01Wstate} \red{(see the orange circles in Fig.~\ref{fig-MPS})}. The multi-scale entanglement renormalization ansatz \red{(MERA)} is designed to exhibit the logarithmic scaling of the entanglement entropy, which efficiently represent the critical states~\cite{V07EntRenor, V08MERA, EV09EntRenor}. The projected entangled pair state \red{(PEPS)} is proposed to obey the area laws in two and higher dimensions, which gained tremendous successes in studying higher-dimensional quantum systems~\cite{VWPC06PEPSfamous, JOVVC08PEPS, GLSW09StringTPS, BAV09StringTPS}. In short, the area laws of entanglement entropy provide intrinsic interpretations on the representational or computational power of TN for simulating quantum systems. Such interpretations also apply to the TN ML. Furthermore, the TN representing quantum states can be interpreted by Born's quantum-probabilistic interpretation (also known as Born rule)~\cite{Born1926}. Thus, TN is regarded as a ``white-box'' numerical tool (called Born machine~\cite{CCW18Born}), akin to the (classical) probabilistic models for ML. We will focus on this point in the next section.
	
	\section{Tensor network for quantum-inspired machine learning}
	
	Equipped with the well-established theories and efficient methods, TN illuminates a new avenue on tackling the dilemma of interpretability and efficiency in ML. To this end, two entangled lines of researches are under hot debate, which are
	\begin{enumerate}
		\item \textit{How do the quantum theories serve as the mathematical foundation for the interpretability of TNML?}
		\item \textit{How do the quantum-mechanical TN methods and quantum computing techniques conceive the TNML schemes with high efficiency?}
	\end{enumerate}
	
	\red{Focusing around these two questions, below we will introduce the recent inspiring progresses on TN for quantum-inspired ML from three aspects: feature mapping, modeling, and ML by quantum computation. These are closely related to the advantages of TN for ML on gaining both efficiency and interpretability. As the theories, models, or methods are taken from or inspired by those in quantum physics, these ML \RR{schemes} are often called ``quantum-inspired'' (see, e.g.,a recent work in Ref.~\cite{Felser2021}). But be noted that significantly more efforts are required to make towards a systematic framework of interpretability based on quantum physics.} \RR{The main methods in the TN ML mentioned below, with their relations to efficiency and interpretability, are summarized in Table~\ref{tab}.}

 \begin{table*}[tpb]
	\centering
	\caption{\red{The main methods in TN ML with their relations to efficiency and interpretability. The green shadows indicate the existence of established schemes or solid evidences on the corresponding subjects. The yellow shadows indicate that the subjects are promising but with only some preliminary results or with disputes.}}
\includegraphics[angle=0,width=1\linewidth]{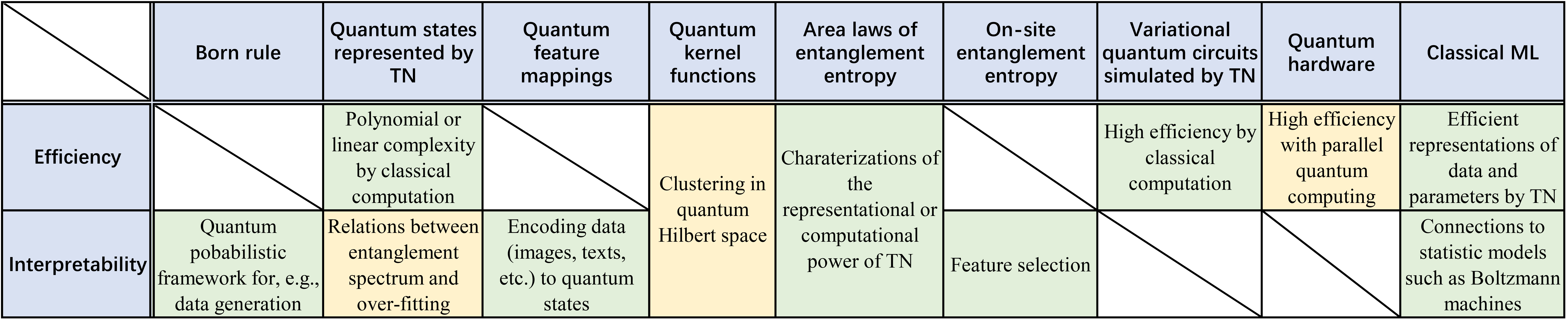}
	\label{tab}
\end{table*}
	
	\subsection{Quantum feature mappings and kernel functions}
	
	Previous works have provided us with stimulative hints on the quantum-inspired ML based on TN. A fundamental step \red{for a quantum treatment of ML} is to map the data to the Hilbert space, i.e., to encode them in quantum states~\cite{BWPR+17QML}. \red{A sample in machine learning can be an image, a sentence, a piece of time series, or \textit{etc.}, which can normally be regarded as a vector. The vector elements are called the features of the sample. The encoding of samples to quantum states is flexible~\cite{LDH11QFM, YIV16QFM, SS16TNML, DLSP22QDC, A22Qencode}.} A straightforward encoding way is to treat the features as the amplitudes of a quantum state. For instance, a sample with \red{$M$ features $\boldsymbol{x} = [x_{1}, x_{2}, \ldots, x_{M}]$ can be considered as the state of a $M$-level ``qudit'' $|\psi \rangle = \frac{1}{Z} \sum_{m} x_{m} |\phi_{m} \rangle$ with $Z=\left| \boldsymbol{x} \right| = \sqrt{\sum_{m=1}^M x_{M}^{2}}$} the normalization factor and $\{|\phi_{m} \rangle\}$ a set of complete basis states~\RRed{\footnote{With a set of quantum states that form a complete basis set in a Hilbert space, any quantum state in this space can be written as a weighted summation of the basis states.}}. The Born rule says that the probability of having the state $|\phi_{m} \rangle$ from $|\psi \rangle$ equals to the norm of the corresponding amplitudes, satisfying $P(|\phi_{m} \rangle) = |x_{m}/Z|^2$. It was also proposed to encode to the amplitudes of the state of $\lceil \log_{2} M\rceil$ qubits~\cite{A22Qencode}, to avoid the usage of high-level qudits~\RRed{\footnote{As the quantum analog of a classical bit, a qubit represents a two-level quantum system such as a quantum spin. If the number of levels is higher than two, such quantum systems are often referred as ``qudits''.}}.

\red{A different way is} to encode samples to the Hilbert space of quantum many-body states, which is dubbed as quantum many-body feature mapping (QMFM). Each feature \red{(that is a scalar)} is mapped onto a two-component vector, which can be treated as the state of one qubit~\cite{SS16TNML} as
	\begin{equation}
		x_{m} \to |x_{m}\rangle = \cos(\frac{\pi x_{m}}{2})|0\rangle + \sin(\frac{\pi x_{m}}{2})|1\rangle,
		\label{eq-FM}
	\end{equation}
where $\{| s\rangle\}$ ($s=0, 1$) \red{can be chosen as} the two eigenstates of Pauli operator $\hat{\sigma}^{z}$~\RRed{\footnote{The Pauli operators ($\hat{\sigma}^x$, $\hat{\sigma}^y$, and $\hat{\sigma}^z$) are three operators whose coefficients are given by three $(2 \times 2)$ matrices known as Pauli matrices. They are applied to describe the interactions, operations, and algebras of quantum spins. A frequently-used complete and orthogonal set of spin states is the eigenstates of one of the Pauli operators.}} and we normally assume \red{$0 \leq |x_{m}| \leq 1$. In other words, $x_{m}$ is mapped to a normalized two-component complex vector. The probability of having the state $|s\rangle$ from $|x_{m}\rangle$ satisfies $p(s) = |\langle s|x_{m}\rangle|^{2}$, with $p(0) = \cos(\frac{\pi x_{m}}{2})^{2}$ and $p(1) = \sin(\frac{\pi x_{m}}{2})^{2}$.} One may also choose an equivalent mapping $x_{m} \to |x_{m}\rangle = \hat{R}(x_{m}) |0\rangle$ with $\hat{R}$ a rotation operator and $x_{m}$ the rotational angle, for the propose of quantum computation. By means of QMFM, a sample \red{(vector)} $\boldsymbol{x}$ is mapped to a $M$-qubit product state
\begin{equation}
|\boldsymbol{x}\rangle = \prod_{\otimes m=1}^{M} |x_{m}\rangle,
		\label{eq-ProSt}
\end{equation}
with \red{$M=\dim(\boldsymbol{x})$} the number of features (or qubits in the state). \RR{One can see that a non-linear feature-mapping function will definitely introduce non-linearity in processing the sample $\boldsymbol{x}$, despite the following-up treatments on the encoded state in the Hilbert space might be linear. This non-linearity in principle would not harm the interpretability but might bring the quantum-inspired ML schemes with an accuracy competitive to the non-linear ML models if the non-linearity is key to gaining high accuracy. Developing new quantum feature mapping for high accuracy is currently an open issue.}
	
There are in general two ways to process the encoded data. One follows the ideas of the kernel-based methods such as K-means~\cite{KLLP19Qmeans}, K-nearest neighbors~\cite{WKS15QNN, DJHJZ18QKNN}, and support vector machine (SVM)~\cite{RML14QSVM}. \red{The kernel function in the quantum-inspired ML is determined by both the feature-map function} [e.g., Eq.~(\ref{eq-FM})] and the chosen measure of \red{similarity (or distance)} in the Hilbert space. \red{A frequently-used measure of similarity between two quantum states is fidelity, which is defined as the norm of the inner product of the two states~\cite{NC02Qcmp}.} With the QMFM in Eq.~(\ref{eq-FM}), the fidelity between two encoded states becomes the cosine similarity \red{between the two samples}, where the kernel function reads
\begin{equation}
	f(\boldsymbol{x}, \boldsymbol{x'}) = |\langle \boldsymbol{x}|\boldsymbol{x'}\rangle| = \prod_{m=1}^{M} \left| \cos \left[\frac{\pi}{2} (x_{m} - x'_{m}) \right] \right|.
	\label{eq-cos}
\end{equation}
The above measure of similarity decays exponentially with the number of features (or qubits) $M$, which is known as the ``catastrophe of orthogonality''. Modifications on the measure were proposed, such as the rescaled logarithmic fidelity~\cite{LR22RLF}, to avoid such a catastrophe for better stability and performance. \red{Exploring proper measures or kernel functions in the Hilbert space remains an open question in the field of quantum-inspired ML.}
	
Besides the catastrophe of orthogonality, the dimension of the Hilbert space increases exponentially \RR{with the number} of qubits. TN algorithms can be employed to efficiently represent the encoded data and evaluate the \red{similarity including fidelity~\cite{ZOV08fidTN}}. For instance, developing valid methods to distinguish quantum phases~\RRed{\footnote{In the condensed matter physics and quantum many-body physics, quantum phases are defined to distinguish the states of quantum matter that possess intrinsically different properties (such as symmetries; one may refer to Sachdev S. Physics World 1999;12:33). Quantum phases are akin to the classical phases such as solids, liquids, and gases. In the paradigm of Landau and Ginzburg (see, e.g., a review article in Hohenberg P and Krekhov A. Physics Reports 2015;572:1–42), different phases of matter are separated by the phase transitions that generally exhibit certain singular properties. A phase transition can occur by changing the physical parameters, such as temperature for the solid-liquid transition of water, and the magnetic fields for the phase transitions of quantum magnets.}} is a long-concerned issue, to which quantum-inspired ML brought new clues. Ref.~\cite{YSRS21visual} considered the unsupervised recognition of quantum phases and phase transitions. The ground states of quantum spin models with different physical parameters (such as external magnetic fields), which can be regarded as the ``quantum data'' (obtained by DMRG~\cite{W92DMRG, W93DMRG}), are represented efficiently by MPS's. The phase transitions are clearly recognized unsupervisedly by visualizing the distribution of these MPS's with non-linear dimensionality reduction~\cite{HS03tsne}, where the ground states in the same quantum phase tend to cluster towards a same sub-region of the Hilbert space. \RR{In this way, prior knowledge on the quantum phases, such as order parameters~\RRed{\footnote{In the Landau-Ginzburg paradigm, one way to identify quantum phases is to look at the corresponding order parameters. For instance, one may calculate the uniform magnetization to identify the ferromagnetic phase (where all spins are pointed in the same direction) for quantum magnets. Thus, such schemes for phase identification require the prior knowledge on which order parameter to look at. Developing the schemes with no need of such prior knowledge is an important topic to study quantum phases.}}, is not required for identifying quantum phases.}
	
Similar observations were reported for \red{the ML of} images and texts, where the encoded states (by QMFM) of the samples in \R{the} same category tend to cluster in the Hilbert space~\cite{LR22RLF}. Clustering and the exponential vastness of the Hilbert space make the classification boundary easier to locate~\cite{H01Qcluster, SPLRS19GTNC, LR22RLF, SSG22TNML}. This shares a similar spirit with the SVM's~\cite{C95SVM}. Another work showed that the classification boundary can be efficiently parametrized by the generative MPS~\cite{HWFWZ17MPSML}, which we will talk about later in this paper. These works drew forth an open question on how different quantum kernels~\cite{LDH11QFM, torabian2023compositional} would affect the ML efficiency for different kinds of data (such as images, texts, multimodal data, and quantum data) and for different ML tasks (supervised learning, unsupervised learning, reinforcement learning, etc.).
	
\subsection{Parameterized modeling and quantum probabilistic interpretation with tensor network for machine learning}
	
\red{We now turn our focus on training the quantum-inspired ML models parameterized by TN. This provides another} pathway of processing the quantum data \red{(including the ground states for quantum phase recognition and the encoded states mapped from ML samples). Some relatively early progresses were made on the data clustering} based on the quantum dynamics satisfying the Schr{\"o}dinger equation~\RRed{\footnote{Schr{\"o}dinger equation is one of the most fundamental equations in quantum physics. It is a partial differential equation of multiple variables for the time-dependent problems, or eigenvalue equation for static problems. The complexity of solving Schr{\"o}dinger equation in general increases exponentially with the size of quantum system.}} with a data-determined potential~\cite{H01Qcluster0, WH09Qcluster1}. This concerns to inversely solve the differential equations, which can be efficiently done with a small number of variables~\cite{Pronchik2003, sehanobish2020learning, HZHX+21MPNN}. For dealing with the quantum many-body states whose dimension scales exponentially, TN has been utilized to develop ML models with polynomial complexity~\cite{SS16TNML, HWFWZ17MPSML, S18MERAML, LRWP+17MLTN, CWXZ19generateTTNML, SPLRS19GTNC, CWZ21PEPSML}, thanks to its high efficiency in representing quantum many-body states and operators. 
	
By taking the advantages of \red{its connections to Born rule and quantum information theories, intrinsically interpretable ML based on TN was developed, which is referred as the quantum-inspired TN ML. The key here is to build the probabilistic ML framework from quantum states, which can be efficiently represented and simulated by TN. In this sense of probabilities, the intrinsic interpretability of such TN ML is akin to or possibly beyond the interpretability of the classical probabilistic ML.} In Ref.~\cite{HWFWZ17MPSML}, MPS was suggested to formulate the joint probabilistic distribution of features \red{provided with a dataset, and is used to implement} generative tasks. Provided with a trained MPS $|\Psi \rangle$, the probability of a given sample $\boldsymbol{x}$ is determined by the quantum probability of obtaining $|\boldsymbol{x}  \rangle$ [Eq.~(\ref{eq-ProSt})] by measuring $|\Psi \rangle$, satisfying
	\begin{equation}
		P(\boldsymbol{x}) = \left| \langle \Psi|\boldsymbol{x}  \rangle \right|^{2}.
		\label{eq-QPx}
	\end{equation}
	
	Following Born's quantum probabilistic rule, the marginal and conditional probabilities can be naturally defined. For instance, the marginal probability distribution of the $m$-th feature $x_{m}$ satisfies
	\begin{equation}
		P(x_{m}) = \langle x_{m} | \hat{\rho}^{(m)}  |x_{m} \rangle,
		\label{eq-px}
	\end{equation}
	where $|x_{m} \rangle$ is defined by Eq.~(\ref{eq-FM}) and $\hat{\rho}^{(m)} = \text{Tr}_{/x_m} | \Psi \rangle \langle \Psi|$ is the reduced density operator of the $m$-th qubit by tracing over the degrees of freedom of all other qubits except for the $m$-th. \red{The marginal probability distributions can be used for, e.g., generation~\cite{HWFWZ17MPSML} and simulating onsite entanglement~\cite{LZLR18entTNML, BTR22CPLML} (which we will introduce below).}
	
	The conditional probabilities \red{are useful for, e.g., classification~\cite{WXYRX21MPSphoto} and data fixing~\cite{HWFWZ17MPSML, RSF+20TNCS}.} The conditional probability distribution of $x_{m}$ with knowing the values of the rest features satisfies
	\begin{equation}
		P(x_{m}|x_{1}, \ldots, x_{m-1}, x_{m+1}, \ldots, x_{M}) = \left| \langle \phi^{(m)}|x_{m} \rangle \right|^{2},
		\label{eq-QPcond}
	\end{equation}
	where \R{$| \phi^{(m)} \rangle = \langle \Psi| \prod_{\otimes m' \neq m} |x_{m'} \rangle / Z$} is the quantum state by collapsing \R{all the qubits but the $m$-th to $\prod_{\otimes m' \neq m} |x_{m'} \rangle$} according to the known values.
	
	\red{The quantum probabilistic interpretation gave birth to various quantum-inspired ML schemes for, e.g.,} model optimization~\cite{SS16TNML, HWFWZ17MPSML, LRWP+17MLTN, SRG19preperation}, data generation~\cite{HWFWZ17MPSML, CWXZ19generateTTNML}, anomaly detection~\cite{WRVL20TNAD}, compressed sampling~\cite{RSF+20TNCS}, solving differential equations~\cite{HXHJR22FTN} and constrained combinatorial optimization~\cite{HHJP22TNCCO, LGCLW23TNCCO, LCP23TNCCO}. \red{Some important ML proposals based on MPS are illustrated in Fig.~\ref{fig-MPS} (see the green circles). These schemes are based on the probabilities obeying Born rule, where the probabilistic distributions can be efficiently represented and calculated with TN.}
	
	\red{The calculation of the above probability distributions involve all coefficients of the quantum state $|\Psi \rangle$, whose number increases exponentially with the number of features ($M$). Note we normally have $M \sim 10^2$ or more (say $M=784$ for the MNIST dataset of the images of hand-writing digits~\cite{MNIST_web}), meaning $2^M \simeq 10^{236}$ coefficients in the state $|\Psi\rangle$. The advantage of TN on efficiency here is the same as that for using TN for quantum simulations. Taking again MPS as an example, its parameter complexity for representing the probability distributions scale just linearly with $M$ (one may refer to Eq.~(\ref{eq-TN}) and the texts below). With the MNIST dataset, accurate generation can be made by the MPS with about $10^8$ parameters~\cite{HWFWZ17MPSML, RSF+20TNCS}. Similar advantages on efficiency can be gained with other kinds of TN, such as MERA and PEPS, using the corresponding TN contraction algorithms~\cite{RTPC+17TNrev}. We shall stress that the high efficiency of TN we refer to here is for representing or simulating the quantum states in the quantum-inspired ML. Another aspect of efficiency is from the power of the quantum computation, which we will discuss about later in Sec.~\ref{sec-3.3}.}
	
	In the construction of generative models for ML, the quantum state $| \Psi \rangle$ satisfying $P(\boldsymbol{x}^{(n)}) \simeq 1/N$ could be the equal super-position of the encoded product states \red{$\{|\boldsymbol{x}^{(n)} \rangle\}$ [Eq.~(\ref{eq-ProSt})]} as
	\begin{equation}
		| \Psi \rangle = \frac{1}{\sqrt{N}} \sum_{n=1}^{N} e^{i\theta_{n}} |\boldsymbol{x}^{(n)} \rangle,
		\label{eq-QPsi}
	\end{equation}
	\red{with $N$ the number of samples.} The equal probabilistic distribution can be deduced from above equation with the ``catastrophe'' of orthogonality $|\langle \boldsymbol{x}^{(n)}|\boldsymbol{x}^{(n')} \rangle| \simeq \delta_{nn'}$ for large $M$ [one may refer to Eq.~(\ref{eq-cos})]. Taking the phase factors $e^{i\theta_{n}} = 1$, such a state is called the lazy-learning state~\cite{SPLRS19GTNC} since it contains no variational parameters\RR{, and is directly constructed by the training samples when it is used for classification}. 
	
	The generative MPS \red{proposed in Ref.~\cite{HWFWZ17MPSML}} is essentially an approximation of the lazy-learning state with variationally-determined phase factors and bounded dimensions of the virtual indexes \red{[$\{\alpha_{m}\}$ in Eq.~(\ref{eq-TN})]. The approximation restricts the upper bound of the entanglement in the TN state} by discarding the small elements in the entanglement spectrum, which is a widely recognized method in the TN approaches for quantum physics. The reported results on supervised learning for \red{classifying images} showed that the generative MPS exhibits lower accuracy on the training set but magically higher accuracy for the testing set \red{than the lazy-learning state}~\cite{SPLRS19GTNC}. \red{In other words, the approximation made in the generative MPS suppresses over-fitting}. This implies possible connections of the quantum super-position rule and quantum entanglement to \red{generalization ability and over-fitting in ML~\cite{BPP21MLQinfo, strashko2022generalization}, which is worth exploring in the future.}
	
	The quantum probabilistic nature allows to introduce the physical concepts and \red{statistic} properties of the TN models to the investigations of ML. For instance, the entanglement entropy has been used for feature selection~\cite{LZLR18entTNML, BTR22CPLML}. The importance of a feature can be characterized by how strongly the corresponding qubit is entangled to others, using the onsite entanglement entropy~\cite{LZLR18entTNML}
	\begin{equation}
		S_{m} = - \text{Tr} (\hat{\rho}^{(m)} \ln \hat{\rho}^{(m)}),
		\label{eq-Ent1}
	\end{equation}
	with $\hat{\rho}^{(m)}$ the reduced density matrix of $|\Psi \rangle$ for the $m$-th qubit. \red{Note the onsite entanglement entropy is the von Neumann entropy of the marginal probabilities in Eq.~(\ref{eq-px}).}
	
	Other quantities \RR{and theories in quantum sciences}, such as quantum mutual information~\cite{CHLW22TNML}, quantum correlations~\cite{GAWCL22TNML}, decoherence~\cite{LCYW23TNML}, \RR{and controllability of quantum systems~\cite{PhysRevA.63.063410}} also help to enhance the \RR{ML interpretability}. These issues are still in hot debate, illuminating a promising path to characterize the ML-relevant capabilities, such as the learnability and generalization powers. The entanglement scaling was applied to unveil the properties of quantum ML models such as the quantum NN's~\cite{DLD17NNent, JWWGG20QNNent}. \RR{The controllability of quantum mechanical systems based on the dynamical Lie algebra was applied to understand and handle the over-parameterization~\cite{Larocca2023} and gradient vanishing~\cite{Larocca2022diagnosingbarren} (also known as barren plateaus~\cite{MBSBN18BarrenP}), which are critical issues for ML based on classical or quantum methods.}
	
	\red{To give a brief summary of this subsection, we introduced the probabilistic framework for the quantum-inspired TN ML. This framework allows us to borrow the theories and techniques from both the classical and quantum information sciences, so that the quantum-inspired ML can possess equal or better interpretability than the classical probabilistic ML. But many issues have not been sufficiently explored, and more efforts need to be made in the future in this newly emergent area.}
	
	\red{\subsection{Quantum-inspired tensor network machine learning with quantum computation} \label{sec-3.3}}
	
	\red{In this subsection, we will concentrate on the combination of the quantum-inspired TN ML with the quantum computational methods and techniques, mainly for the purpose of high efficiency. The key here is the ability of} TN as an efficient representation of quantum operations, which is essential for simulating the physical processes in quantum mechanics such as dynamics and thermodynamics~\cite{VGC04MPDO, ZV04MPO, BHVC09folding, RLXZS12ODTNS, CCD12FTPEPS, HM15folding, KWO17MPSopen}. For the TN ML with quantum computation, TN can serve as a mathematical representation of the quantum circuit models~\cite{HPWS18TNQML}. As the quantum counterpart of the classical logical circuits, a quantum circuit is composed of multiple quantum gates~\RRed{\footnote{A quantum gate is a specific operation on quantum state, and thus can be given by a quantum operator. When considering the quantum computation in practice, the quantum gates realizable on different quantum platforms (such as super-conducting circuits, ultra-cold atoms, and single photons) are different.}} that are usually unitary operators executable on the quantum computers. Efficient ways of deriving the quantum circuits for implementing quantum algorithms are key to quantum computation and to the ML by quantum computers.
	
	However, the quantum algorithms where the circuits can be analytically derived (e.g., Shor’s algorithm for factoring~\cite{S99shor}, Grover’s algorithm for searching~\cite{G97Grover}, and the disentangling circuits recently proposed for preparing MPSs~\cite{R20MPSencode}) are extremely rare. This puts strong limitations to quantum computation, including its applications to ML. 
	
	Variational quantum algorithms (VQAs)~\cite{CAB+21VQA} were proposed, significantly extending the scope of problems that quantum computation can handle. Among others, the quantum circuits containing variational parameters, dubbed as the variational quantum circuits (VQCs), have been used for, e.g., state preparation and tomography~\cite{LWX+20VQCtomo, ZHR21ADQC}. Variational quantum eigensolvers~\cite{PMSY+14VQE} were developed and applied to a wide range of areas such as quantum chemistry and quantum materials (one may refer to a recent review in Ref.~\cite{TCCP+22VQErev}).
	
	VQAs concern the hybrid classical-quantum optimizations~\cite{MRBA16VQA}, which also suffer from the ``curse of dimensionality''. Naturally, \red{TN has been employed as an efficient mathematical tool for classical computation, allowing to stably access equal or even much larger numbers of qubits than what the current quantum computers can handle}~\cite{MS08TNQcomp, HZNN+21TNQC, HGPC22TNQC, ZHR21ADQC, GZH21TNQC, VOA+22TNQC, PZ22TNQC, LSGVA22TNQC}. The relevant works provided valuable information on the races between the powers of classical and quantum computations in the noisy intermediate-scale quantum (NISQ) era~\cite{P18NISQ}, such as the efficiency of random-circuit sampling by the quantum computer of Google~\cite{AABB+2019GoogleQ} and by the TN simulations~\cite{GLXX+19PEPSQC, GZH21TNQC, LLLF+21TNQC, GK21QCTNC, PZ22TNQC, PCZ22TNQC}. 
	
	The previous works on TN and quantum computation have unveiled the underlying connections among quantum mechanics, quantum computing, and ML, which makes TN a uniquely suitable tool to explore \red{ML incorporated with quantum computation}~\cite{BLSF19VQCML}. As an example, let us compare the state preparation by VQC with the supervised ML by NN. The task of state preparation is to obtain the given target state ($|\psi\rangle$) on a quantum computer with high fidelity, say $|\psi\rangle \simeq \hat{U} |0\rangle$ with $\hat{U}$ a unitary transformation. The VQC is used to represent the mapping $\hat{U}$, similar to the feed-forward mapping defined by the NN. As the building blocks of VQC, the quantum gates with variational parameters are analogous to the neurons in NN. The preparation error can be evaluated by the infidelity ($f_{\text{in}} = 1-|\langle \psi| \hat{U}|0 \rangle|$, a measure of distance in quantum Hilbert space)~\cite{NC02Qcmp} between the target state and the one prepared by the VQC, analogous to the loss function (or error) in ML. The gradients and the gradient-descent process for optimizing the parameters in VQC can be implemented by the TN methods or automatic differentiation technique~\cite{R20MPSencode, ZHR21ADQC, LDGSP21TNQC, shirakawa2021automatic, dov2022approximate, rudolph2022decomposition}, analogous to the backward propagation of ML. The optimized circuit, after necessary compilation~\cite{CFM17Qcompile}, can be distributed to quantum computers for their further uses. 
	
    TN has been extensively used in the quantum-inspired ML combined with quantum computation methods including VQC~\cite{LRWP+17MLTN, chen2020hybrid, HUANG202189, 10.3389/fphy.2020.586374, PhysRevA.106.062423, WXYRX21MPSphoto, WAQ21GTNexp, LAZZARIN2022128056}. The utilizations of TN in this sense can be generally summarized into three different but relevant ways. First, the TNs with unitary constraints are used as the parameterized ML model runnable on quantum platforms~\cite{LRWP+17MLTN, HPWS18TNQML, HUANG202189, WXYRX21MPSphoto, WAQ21GTNexp}. Here, the equivalence between the unitary TNs and the quantum circuits are utilized. It is expected that the computational power (with, e.g., parallel quantum computing~\RRed{\footnote{Quantum computing is expected to possess higher parallelism over classical computing. One operation on an entangled quantum state can manage to process all the exponentially-many coefficients, even this operation might be just on one qubit or a local part of the state. In comparison, such parallelism cannot be gained if the state has no entanglement (i.e., a product state). This is one reason for regarding entanglement as the source for the superior power of quantum computing.}}) would bring high efficiency to the ML schemes running on the quantum platforms. Second, TN methods are employed to classically optimize or simulate the VQCs for machine learning~\cite{WAQ21GTNexp, PhysRevA.106.062423, MB23TNQCML}. Such classical TN simulations have already been widely applied to the races between the powers of classical and quantum computers~\cite{GLXX+19PEPSQC, GZH21TNQC, LLLF+21TNQC, GK21QCTNC, PZ22TNQC, PCZ22TNQC}, and are expected to provide useful information to guide the future investigations of the ML running on the quantum platforms. Third, TN are used for data preprocessing before entering the quantum computational procedures. For example, data compression by TN was suggested to lower the requirement of quantum hardware~\cite{chen2020hybrid}. TN is also used as an efficient representation of the quantum data (such as the states of quantum systems and the quantum states encoded from the classical data) for, e.g., classification and ML-based quantum control~\cite{DLSP22QDC, MB23TNQCML}. These researches are particularly useful for the quantum computation of ML in the NISQ era when the power of quantum hardware (number of qubits, stability, and \textit{etc.}) is for the moment limited.
	
	\section{Tensor network enhancing classical machine learning}
	
	As a fundamental mathematical tool, the wide applications of TN to ML are not limited to those obeying the quantum probabilistic interpretation. With the fact that TN can be used to efficiently represent and simulate the partition functions of classical stochastic systems, such as Ising and Potts models~(see, e.g., Refs.~\cite{N95TMRG2Dclassic, LN07TRG, EV15TNR}), the relations between TN and Boltzmann machines~\cite{AHS85BM} have been extensively studied~\cite{CCXWX18MPSbolzmann, cheng2018information, ZHRB19BolMPS, LPZZ21TNBM}. The relevant works also promoted the investigations of quantum many-body physics and ML from the perspective of, e.g., the area laws of entanglement entropy and the representation ability of TN as quantum state ansatz~\cite{DLD17NNent, GPARC18MLstringbond, LPZZ21TNBM, MVS21QSRBM}. 
	
	TN were also used to enhance NN or develop novel ML models~\cite{wang2023tensor}, ignoring any probabilistic interpretations. Tensor-train~\cite{O11TTD} and tensor-ring~\cite{ZZXZ+16tensorring} forms (which correspond to the MPS's with open and periodic boundary conditions, respectively) are applied to develop novel support vector machines~\cite{SPLRS19GTNC, CBKW19TTSVM, CBYW22TTSVM}, dimensionality reduction schemes~\cite{QZHZX22TTPCA}, and parameterized ML models~\cite{YKT17TTRNN, CBSW18TTML, WSEWA18TRNN, SBKH20TTLSTM, MZZGR23ResMPS, wu2023tensor}. Significant reductions of parameter complexity were reported, thanks to the efficiency of TN for representing higher-order tensors, which is essentially the same reason for the high efficiency of TN in describing quantum many-body systems.
	
Based on the same ground, model compression methods were proposed to decompose the variational parameters of NN's to TN's, or directly to represent the variational parameters as TN's~\cite{NPOV15TTNN, TSN17TTRNN, YLMCZ2019TRDcomp, HYSM19TNCNN, PXWY+19TRDRNN, GCHX20+NNMPO, SGLLY20NNMPO, WZLDW20TTCNN, HZ21BTNN, WZCL+21nonlinearTT, LK22TuckerNN, qing2023compressing}. Explicit decomposition procedures might not be necessary for the latter, where the parameters of NN's are not restored as tensors but directly as the tensor-train (tensor-ring) forms~\cite{NPOV15TTNN}, matrix product operators~\cite{GCHX20+NNMPO, SGLLY20NNMPO}, or deep TN's~\cite{qing2023compressing}. Non-linear activation functions were incorporated with TN to improve the performance for ML~\cite{WZCL+21nonlinearTT, MZZGR23ResMPS, qing2023compressing}, generalizing TN from a type of multi-linear models to non-linear ones.
	
\section{Discussion}
	
\begin{figure}[tbp]
	\centering
	\includegraphics[angle=0,width=1\linewidth]{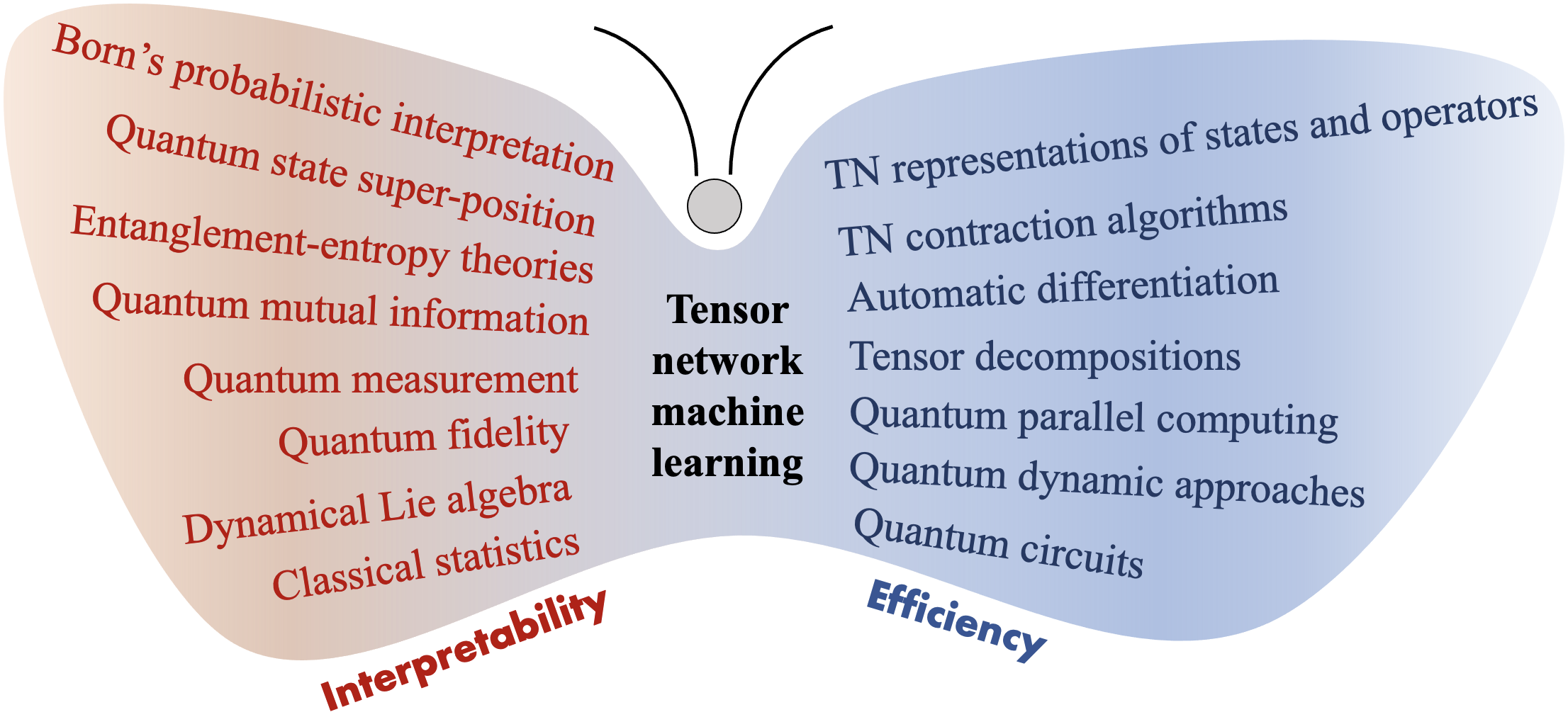}
	\caption{(Color online) The ``TN-ML butterfly'' summarizing the two unique advantages: interpretability based on quantum theories (left wing) and efficiency based on quantum methods (right wing).}
	\label{fig-butterfly}
\end{figure}

Methodologies to solve the dilemma between efficiency and interpretability in artificial intelligence \red{(AI) and particularly deep ML} have been long concerned. We here review the inspiring progresses made in TN for interpretable and efficient quantum-inspired ML. \red{The advantages of TN for ML are listed in the ``TN-ML butterfly'' in Fig.~\ref{fig-butterfly}. For the quantum-inspired ML, the advantages of TN can be summarized into two} critical points: quantum theories for interpretability, and quantum methods for efficiency. On one hand, TN enables us to apply quantum theories \red{and statistics}, such as the entanglement theories, to build a \red{probabilistic} framework for interpretability that \red{might be} beyond the description of the classical information or statistic theories. On the other hand, the powerful quantum-mechanical TN algorithms and the explosively boosted quantum computational technologies will empower the \red{quantum-inspired} TN-ML methods with high efficiency on both the classical and quantum computational platforms.
	
Particularly with the striking progresses recently made in the generative pre-trained transformer (GPT)~\cite{BMR+20GPT}, unprecedented surges in model complexity and computational power have occurred, which bring new opportunities and challenges to TN ML. The interpretability will become increasingly valuable when facing the emergent AI of GPT for not just investigations with higher efficiency but also their better use and safer control. In the current NISQ era and the forthcoming era of genuine quantum computing, TN is rapidly growing into a featured mathematical tool to explore the quantum AI from the perspective of theories, models, algorithms, software, hardware, and applications.

\section*{Acknowledgment} This work is supported in part by the NSFC (Grant No. 11834014 and No. 12004266), Beijing Natural Science Foundation (Grant No. 1232025), the National Key R\&D Program of China (Grant No. 2018FYA0305804), the Strategetic Priority Research Program of the Chinese Academy of Sciences (Grant No. XDB28000000)\R{, and Innovation Program for Quantum Science and Technology (No. 2021ZD0301800).}


\begin{thebibliography}{200}%
\makeatletter
\providecommand \@ifxundefined [1]{%
 \@ifx{#1\undefined}
}%
\providecommand \@ifnum [1]{%
 \ifnum #1\expandafter \@firstoftwo
 \else \expandafter \@secondoftwo
 \fi
}%
\providecommand \@ifx [1]{%
 \ifx #1\expandafter \@firstoftwo
 \else \expandafter \@secondoftwo
 \fi
}%
\providecommand \natexlab [1]{#1}%
\providecommand \enquote  [1]{``#1''}%
\providecommand \bibnamefont  [1]{#1}%
\providecommand \bibfnamefont [1]{#1}%
\providecommand \citenamefont [1]{#1}%
\providecommand \href@noop [0]{\@secondoftwo}%
\providecommand \href [0]{\begingroup \@sanitize@url \@href}%
\providecommand \@href[1]{\@@startlink{#1}\@@href}%
\providecommand \@@href[1]{\endgroup#1\@@endlink}%
\providecommand \@sanitize@url [0]{\catcode `\\12\catcode `\$12\catcode
  `\&12\catcode `\#12\catcode `\^12\catcode `\_12\catcode `\%12\relax}%
\providecommand \@@startlink[1]{}%
\providecommand \@@endlink[0]{}%
\providecommand \url  [0]{\begingroup\@sanitize@url \@url }%
\providecommand \@url [1]{\endgroup\@href {#1}{\urlprefix }}%
\providecommand \urlprefix  [0]{URL }%
\providecommand \Eprint [0]{\href }%
\providecommand \doibase [0]{https://doi.org/}%
\providecommand \selectlanguage [0]{\@gobble}%
\providecommand \bibinfo  [0]{\@secondoftwo}%
\providecommand \bibfield  [0]{\@secondoftwo}%
\providecommand \translation [1]{[#1]}%
\providecommand \BibitemOpen [0]{}%
\providecommand \bibitemStop [0]{}%
\providecommand \bibitemNoStop [0]{.\EOS\space}%
\providecommand \EOS [0]{\spacefactor3000\relax}%
\providecommand \BibitemShut  [1]{\csname bibitem#1\endcsname}%
\let\auto@bib@innerbib\@empty
\bibitem [{\citenamefont {Gilpin}\ \emph {et~al.}(2018)\citenamefont {Gilpin},
  \citenamefont {Bau}, \citenamefont {Yuan}, \citenamefont {Bajwa},
  \citenamefont {Specter},\ and\ \citenamefont {Kagal}}]{BGYG+18MLinterp}%
  \BibitemOpen
  \bibfield  {author} {\bibinfo {author} {\bibfnamefont {L.~H.}\ \bibnamefont
  {Gilpin}}, \bibinfo {author} {\bibfnamefont {D.}~\bibnamefont {Bau}},
  \bibinfo {author} {\bibfnamefont {B.~Z.}\ \bibnamefont {Yuan}}, \bibinfo
  {author} {\bibfnamefont {A.}~\bibnamefont {Bajwa}}, \bibinfo {author}
  {\bibfnamefont {M.}~\bibnamefont {Specter}},\ and\ \bibinfo {author}
  {\bibfnamefont {L.}~\bibnamefont {Kagal}},\ }\bibfield  {title} {\bibinfo
  {title} {Explaining explanations: An overview of interpretability of machine
  learning},\ }in\ \href@noop {} {\emph {\bibinfo {booktitle} {2018 IEEE 5th
  International Conference on Data Science and Advanced Analytics (DSAA)}}}\
  (\bibinfo {organization} {IEEE},\ \bibinfo {year} {2018})\ pp.\ \bibinfo
  {pages} {80--89}\BibitemShut {NoStop}%
\bibitem [{\citenamefont {Zhang}\ and\ \citenamefont
  {Zhu}(2018)}]{ZZ18MLinterp}%
  \BibitemOpen
  \bibfield  {author} {\bibinfo {author} {\bibfnamefont {Q.-s.}\ \bibnamefont
  {Zhang}}\ and\ \bibinfo {author} {\bibfnamefont {S.-C.}\ \bibnamefont
  {Zhu}},\ }\bibfield  {title} {\bibinfo {title} {Visual interpretability for
  deep learning: a survey},\ }\href {https://doi.org/10.1631/FITEE.1700808}
  {\bibfield  {journal} {\bibinfo  {journal} {Frontiers of Information
  Technology \& Electronic Engineering}\ }\textbf {\bibinfo {volume} {19}},\
  \bibinfo {pages} {27} (\bibinfo {year} {2018})}\BibitemShut {NoStop}%
\bibitem [{\citenamefont {Lipton}(2018)}]{L18MLInterp}%
  \BibitemOpen
  \bibfield  {author} {\bibinfo {author} {\bibfnamefont {Z.~C.}\ \bibnamefont
  {Lipton}},\ }\bibfield  {title} {\bibinfo {title} {The mythos of model
  interpretability},\ }\href {https://doi.org/10.1145/3233231} {\bibfield
  {journal} {\bibinfo  {journal} {Communications of The ACM}\ }\textbf
  {\bibinfo {volume} {61}},\ \bibinfo {pages} {36–43} (\bibinfo {year}
  {2018})}\BibitemShut {NoStop}%
\bibitem [{\citenamefont {Carvalho}\ \emph {et~al.}(2019)\citenamefont
  {Carvalho}, \citenamefont {Pereira},\ and\ \citenamefont
  {Cardoso}}]{CPC19MLinterp}%
  \BibitemOpen
  \bibfield  {author} {\bibinfo {author} {\bibfnamefont {D.~V.}\ \bibnamefont
  {Carvalho}}, \bibinfo {author} {\bibfnamefont {E.~M.}\ \bibnamefont
  {Pereira}},\ and\ \bibinfo {author} {\bibfnamefont {J.~S.}\ \bibnamefont
  {Cardoso}},\ }\bibfield  {title} {\bibinfo {title} {Machine learning
  interpretability: A survey on methods and metrics},\ }\bibfield  {journal}
  {\bibinfo  {journal} {Electronics}\ }\textbf {\bibinfo {volume} {8}},\ \href
  {https://doi.org/10.3390/electronics8080832} {10.3390/electronics8080832}
  (\bibinfo {year} {2019})\BibitemShut {NoStop}%
\bibitem [{\citenamefont {Madsen}\ \emph {et~al.}(2022)\citenamefont {Madsen},
  \citenamefont {Reddy},\ and\ \citenamefont {Chandar}}]{MRC23IntNN}%
  \BibitemOpen
  \bibfield  {author} {\bibinfo {author} {\bibfnamefont {A.}~\bibnamefont
  {Madsen}}, \bibinfo {author} {\bibfnamefont {S.}~\bibnamefont {Reddy}},\ and\
  \bibinfo {author} {\bibfnamefont {S.}~\bibnamefont {Chandar}},\ }\bibfield
  {title} {\bibinfo {title} {Post-hoc interpretability for neural nlp: A
  survey},\ }\bibfield  {journal} {\bibinfo  {journal} {ACM Computing Surveys}\
  }\textbf {\bibinfo {volume} {55}},\ \href {https://doi.org/10.1145/3546577}
  {10.1145/3546577} (\bibinfo {year} {2022})\BibitemShut {NoStop}%
\bibitem [{\citenamefont {Rudin}(2019)}]{R19MLint}%
  \BibitemOpen
  \bibfield  {author} {\bibinfo {author} {\bibfnamefont {C.}~\bibnamefont
  {Rudin}},\ }\bibfield  {title} {\bibinfo {title} {Stop explaining black box
  machine learning models for high stakes decisions and use interpretable
  models instead},\ }\href {https://doi.org/10.1038/s42256-019-0048-x}
  {\bibfield  {journal} {\bibinfo  {journal} {Nature Machine Intelligence}\
  }\textbf {\bibinfo {volume} {1}},\ \bibinfo {pages} {206} (\bibinfo {year}
  {2019})}\BibitemShut {NoStop}%
\bibitem [{\citenamefont {Su}\ \emph {et~al.}(2019)\citenamefont {Su},
  \citenamefont {Vargas},\ and\ \citenamefont {Sakurai}}]{SVS19MLattack}%
  \BibitemOpen
  \bibfield  {author} {\bibinfo {author} {\bibfnamefont {J.}~\bibnamefont
  {Su}}, \bibinfo {author} {\bibfnamefont {D.~V.}\ \bibnamefont {Vargas}},\
  and\ \bibinfo {author} {\bibfnamefont {K.}~\bibnamefont {Sakurai}},\
  }\bibfield  {title} {\bibinfo {title} {One pixel attack for fooling deep
  neural networks},\ }\href {https://doi.org/10.1109/TEVC.2019.2890858}
  {\bibfield  {journal} {\bibinfo  {journal} {IEEE Transactions on Evolutionary
  Computation}\ }\textbf {\bibinfo {volume} {23}},\ \bibinfo {pages} {828}
  (\bibinfo {year} {2019})}\BibitemShut {NoStop}%
\bibitem [{\citenamefont {Doshi-Velez}\ and\ \citenamefont
  {Kim}(2017)}]{doshivelez2017rigorous}%
  \BibitemOpen
  \bibfield  {author} {\bibinfo {author} {\bibfnamefont {F.}~\bibnamefont
  {Doshi-Velez}}\ and\ \bibinfo {author} {\bibfnamefont {B.}~\bibnamefont
  {Kim}},\ }\href@noop {} {\bibinfo {title} {Towards a rigorous science of
  interpretable machine learning}} (\bibinfo {year} {2017}),\ \Eprint
  {https://arxiv.org/abs/1702.08608} {arXiv:1702.08608 [stat.ML]} \BibitemShut
  {NoStop}%
\bibitem [{\citenamefont {Battiti}(1994)}]{B94MLinfo}%
  \BibitemOpen
  \bibfield  {author} {\bibinfo {author} {\bibfnamefont {R.}~\bibnamefont
  {Battiti}},\ }\bibfield  {title} {\bibinfo {title} {Using mutual information
  for selecting features in supervised neural net learning},\ }\href
  {https://doi.org/10.1109/72.298224} {\bibfield  {journal} {\bibinfo
  {journal} {IEEE Transactions on Neural Networks}\ }\textbf {\bibinfo {volume}
  {5}},\ \bibinfo {pages} {537} (\bibinfo {year} {1994})}\BibitemShut {NoStop}%
\bibitem [{\citenamefont {Shwartz-Ziv}\ and\ \citenamefont
  {Tishby}(2017)}]{RN17DLexpl}%
  \BibitemOpen
  \bibfield  {author} {\bibinfo {author} {\bibfnamefont {R.}~\bibnamefont
  {Shwartz-Ziv}}\ and\ \bibinfo {author} {\bibfnamefont {N.}~\bibnamefont
  {Tishby}},\ }\href@noop {} {\bibinfo {title} {Opening the black box of deep
  neural networks via information}} (\bibinfo {year} {2017}),\ \Eprint
  {https://arxiv.org/abs/1703.00810} {arXiv:1703.00810 [stat.ML]} \BibitemShut
  {NoStop}%
\bibitem [{\citenamefont {Koch-Janusz}\ and\ \citenamefont
  {Ringel}(2018)}]{KR18MLRG}%
  \BibitemOpen
  \bibfield  {author} {\bibinfo {author} {\bibfnamefont {M.}~\bibnamefont
  {Koch-Janusz}}\ and\ \bibinfo {author} {\bibfnamefont {Z.}~\bibnamefont
  {Ringel}},\ }\bibfield  {title} {\bibinfo {title} {Mutual information, neural
  networks and the renormalization group},\ }\href
  {https://doi.org/10.1038/s41567-018-0081-4} {\bibfield  {journal} {\bibinfo
  {journal} {Nature Physics}\ }\textbf {\bibinfo {volume} {14}},\ \bibinfo
  {pages} {578} (\bibinfo {year} {2018})}\BibitemShut {NoStop}%
\bibitem [{\citenamefont {Saxe}\ \emph {et~al.}(2019)\citenamefont {Saxe},
  \citenamefont {Bansal}, \citenamefont {Dapello}, \citenamefont {Advani},
  \citenamefont {Kolchinsky}, \citenamefont {Tracey},\ and\ \citenamefont
  {Cox}}]{AYJM+19DLBI}%
  \BibitemOpen
  \bibfield  {author} {\bibinfo {author} {\bibfnamefont {A.~M.}\ \bibnamefont
  {Saxe}}, \bibinfo {author} {\bibfnamefont {Y.}~\bibnamefont {Bansal}},
  \bibinfo {author} {\bibfnamefont {J.}~\bibnamefont {Dapello}}, \bibinfo
  {author} {\bibfnamefont {M.}~\bibnamefont {Advani}}, \bibinfo {author}
  {\bibfnamefont {A.}~\bibnamefont {Kolchinsky}}, \bibinfo {author}
  {\bibfnamefont {B.~D.}\ \bibnamefont {Tracey}},\ and\ \bibinfo {author}
  {\bibfnamefont {D.~D.}\ \bibnamefont {Cox}},\ }\bibfield  {title} {\bibinfo
  {title} {On the information bottleneck theory of deep learning},\ }\href
  {https://doi.org/10.1088/1742-5468/ab3985} {\bibfield  {journal} {\bibinfo
  {journal} {Journal of Statistical Mechanics: Theory and Experiment}\ }\textbf
  {\bibinfo {volume} {2019}},\ \bibinfo {pages} {124020} (\bibinfo {year}
  {2019})}\BibitemShut {NoStop}%
\bibitem [{\citenamefont {Erdmenger}\ \emph {et~al.}(2022)\citenamefont
  {Erdmenger}, \citenamefont {Grosvenor},\ and\ \citenamefont
  {Jefferson}}]{EGJ22EntML}%
  \BibitemOpen
  \bibfield  {author} {\bibinfo {author} {\bibfnamefont {J.}~\bibnamefont
  {Erdmenger}}, \bibinfo {author} {\bibfnamefont {K.~T.}\ \bibnamefont
  {Grosvenor}},\ and\ \bibinfo {author} {\bibfnamefont {R.}~\bibnamefont
  {Jefferson}},\ }\bibfield  {title} {\bibinfo {title} {{Towards quantifying
  information flows: relative entropy in deep neural networks and the
  renormalization group}},\ }\href
  {https://doi.org/10.21468/SciPostPhys.12.1.041} {\bibfield  {journal}
  {\bibinfo  {journal} {SciPost Physics}\ }\textbf {\bibinfo {volume} {12}},\
  \bibinfo {pages} {41} (\bibinfo {year} {2022})}\BibitemShut {NoStop}%
\bibitem [{\citenamefont {Ghahramani}(2015)}]{G15PML}%
  \BibitemOpen
  \bibfield  {author} {\bibinfo {author} {\bibfnamefont {Z.}~\bibnamefont
  {Ghahramani}},\ }\bibfield  {title} {\bibinfo {title} {Probabilistic machine
  learning and artificial intelligence},\ }\href
  {https://doi.org/10.1038/nature14541} {\bibfield  {journal} {\bibinfo
  {journal} {Nature}\ }\textbf {\bibinfo {volume} {521}},\ \bibinfo {pages}
  {452} (\bibinfo {year} {2015})}\BibitemShut {NoStop}%
\bibitem [{\citenamefont {Bishop}(2006)}]{B06MLbook}%
  \BibitemOpen
  \bibfield  {author} {\bibinfo {author} {\bibfnamefont {C.~M.}\ \bibnamefont
  {Bishop}},\ }\href@noop {} {\emph {\bibinfo {title} {Pattern recognition and
  machine learning}}}\ (\bibinfo  {publisher} {Springer},\ \bibinfo {year}
  {2006})\BibitemShut {NoStop}%
\bibitem [{\citenamefont {Ackley}\ \emph {et~al.}(1985)\citenamefont {Ackley},
  \citenamefont {Hinton},\ and\ \citenamefont {Sejnowski}}]{AHS85BM}%
  \BibitemOpen
  \bibfield  {author} {\bibinfo {author} {\bibfnamefont {D.~H.}\ \bibnamefont
  {Ackley}}, \bibinfo {author} {\bibfnamefont {G.~E.}\ \bibnamefont {Hinton}},\
  and\ \bibinfo {author} {\bibfnamefont {T.~J.}\ \bibnamefont {Sejnowski}},\
  }\bibfield  {title} {\bibinfo {title} {A learning algorithm for boltzmann
  machines},\ }\href
  {https://doi.org/https://doi.org/10.1016/S0364-0213(85)80012-4} {\bibfield
  {journal} {\bibinfo  {journal} {Cognitive Science}\ }\textbf {\bibinfo
  {volume} {9}},\ \bibinfo {pages} {147} (\bibinfo {year} {1985})}\BibitemShut
  {NoStop}%
\bibitem [{\citenamefont {Glymour}(2003)}]{G03BayesML}%
  \BibitemOpen
  \bibfield  {author} {\bibinfo {author} {\bibfnamefont {C.}~\bibnamefont
  {Glymour}},\ }\bibfield  {title} {\bibinfo {title} {Learning, prediction and
  causal bayes nets},\ }\href
  {https://doi.org/https://doi.org/10.1016/S1364-6613(02)00009-8} {\bibfield
  {journal} {\bibinfo  {journal} {Trends in Cognitive Sciences}\ }\textbf
  {\bibinfo {volume} {7}},\ \bibinfo {pages} {43} (\bibinfo {year}
  {2003})}\BibitemShut {NoStop}%
\bibitem [{\citenamefont {Tenenbaum}\ \emph {et~al.}(2006)\citenamefont
  {Tenenbaum}, \citenamefont {Griffiths},\ and\ \citenamefont
  {Kemp}}]{TGK06MLBayes}%
  \BibitemOpen
  \bibfield  {author} {\bibinfo {author} {\bibfnamefont {J.~B.}\ \bibnamefont
  {Tenenbaum}}, \bibinfo {author} {\bibfnamefont {T.~L.}\ \bibnamefont
  {Griffiths}},\ and\ \bibinfo {author} {\bibfnamefont {C.}~\bibnamefont
  {Kemp}},\ }\bibfield  {title} {\bibinfo {title} {Theory-based bayesian models
  of inductive learning and reasoning},\ }\href
  {https://doi.org/https://doi.org/10.1016/j.tics.2006.05.009} {\bibfield
  {journal} {\bibinfo  {journal} {Trends in Cognitive Sciences}\ }\textbf
  {\bibinfo {volume} {10}},\ \bibinfo {pages} {309} (\bibinfo {year} {2006})},\
  \bibinfo {note} {special issue: Probabilistic models of
  cognition}\BibitemShut {NoStop}%
\bibitem [{\citenamefont {Lucas}\ and\ \citenamefont
  {Griffiths}(2010)}]{LG10BayesML}%
  \BibitemOpen
  \bibfield  {author} {\bibinfo {author} {\bibfnamefont {C.~G.}\ \bibnamefont
  {Lucas}}\ and\ \bibinfo {author} {\bibfnamefont {T.~L.}\ \bibnamefont
  {Griffiths}},\ }\bibfield  {title} {\bibinfo {title} {Learning the form of
  causal relationships using hierarchical bayesian models},\ }\href
  {https://doi.org/https://doi.org/10.1111/j.1551-6709.2009.01058.x} {\bibfield
   {journal} {\bibinfo  {journal} {Cognitive Science}\ }\textbf {\bibinfo
  {volume} {34}},\ \bibinfo {pages} {113} (\bibinfo {year} {2010})}\BibitemShut
  {NoStop}%
\bibitem [{Note1()}]{Note1}%
  \BibitemOpen
  \bibinfo {note} {For the efficiency of a ML method, we mainly mean the
  accuracy with a proper computational complexity depending on its academic or
  industrial uses. For the classical ``white-box'' ML schemes, it seems that
  their accuracy could not beat that of the deep NN's even if we immensely
  increase their complexity.}\BibitemShut {Stop}%
\bibitem [{\citenamefont {P\'{e}rez-Garc\'ia}\ \emph
  {et~al.}(2007)\citenamefont {P\'{e}rez-Garc\'ia}, \citenamefont {Verstraete},
  \citenamefont {Wolf},\ and\ \citenamefont {Cirac}}]{PVWC07MPSRev}%
  \BibitemOpen
  \bibfield  {author} {\bibinfo {author} {\bibfnamefont {D.}~\bibnamefont
  {P\'{e}rez-Garc\'ia}}, \bibinfo {author} {\bibfnamefont {F.}~\bibnamefont
  {Verstraete}}, \bibinfo {author} {\bibfnamefont {M.~M.}\ \bibnamefont
  {Wolf}},\ and\ \bibinfo {author} {\bibfnamefont {J.~I.}\ \bibnamefont
  {Cirac}},\ }\bibfield  {title} {\bibinfo {title} {{Matrix product state
  representations}},\ }\href@noop {} {\bibfield  {journal} {\bibinfo  {journal}
  {Quantum Information \& Computation}\ }\textbf {\bibinfo {volume} {7}},\
  \bibinfo {pages} {401} (\bibinfo {year} {2007})}\BibitemShut {NoStop}%
\bibitem [{\citenamefont {Oseledets}(2011)}]{O11TTD}%
  \BibitemOpen
  \bibfield  {author} {\bibinfo {author} {\bibfnamefont {I.~V.}\ \bibnamefont
  {Oseledets}},\ }\bibfield  {title} {\bibinfo {title} {{Tensor-train
  decomposition}},\ }\href {https://doi.org/10.1137/090752286} {\bibfield
  {journal} {\bibinfo  {journal} {SIAM Journal on Scientific Computing}\
  }\textbf {\bibinfo {volume} {33}},\ \bibinfo {pages} {2295} (\bibinfo {year}
  {2011})}\BibitemShut {NoStop}%
\bibitem [{\citenamefont {Shi}\ \emph {et~al.}(2006)\citenamefont {Shi},
  \citenamefont {Duan},\ and\ \citenamefont {Vidal}}]{SDV06TTN}%
  \BibitemOpen
  \bibfield  {author} {\bibinfo {author} {\bibfnamefont {Y.-Y.}\ \bibnamefont
  {Shi}}, \bibinfo {author} {\bibfnamefont {L.-M.}\ \bibnamefont {Duan}},\ and\
  \bibinfo {author} {\bibfnamefont {G.}~\bibnamefont {Vidal}},\ }\bibfield
  {title} {\bibinfo {title} {Classical simulation of quantum many-body systems
  with a tree tensor network},\ }\href
  {https://doi.org/10.1103/PhysRevA.74.022320} {\bibfield  {journal} {\bibinfo
  {journal} {Physical Review A}\ }\textbf {\bibinfo {volume} {74}},\ \bibinfo
  {pages} {022320} (\bibinfo {year} {2006})}\BibitemShut {NoStop}%
\bibitem [{\citenamefont {Verstraete}\ \emph {et~al.}(2006)\citenamefont
  {Verstraete}, \citenamefont {Wolf}, \citenamefont {Perez-Garcia},\ and\
  \citenamefont {Cirac}}]{VWPC06PEPSfamous}%
  \BibitemOpen
  \bibfield  {author} {\bibinfo {author} {\bibfnamefont {F.}~\bibnamefont
  {Verstraete}}, \bibinfo {author} {\bibfnamefont {M.~M.}\ \bibnamefont
  {Wolf}}, \bibinfo {author} {\bibfnamefont {D.}~\bibnamefont {Perez-Garcia}},\
  and\ \bibinfo {author} {\bibfnamefont {J.~I.}\ \bibnamefont {Cirac}},\
  }\bibfield  {title} {\bibinfo {title} {Criticality, the area law, and the
  computational power of projected entangled pair states},\ }\href
  {https://doi.org/10.1103/PhysRevLett.96.220601} {\bibfield  {journal}
  {\bibinfo  {journal} {Physical Review Letters}\ }\textbf {\bibinfo {volume}
  {96}},\ \bibinfo {pages} {220601} (\bibinfo {year} {2006})}\BibitemShut
  {NoStop}%
\bibitem [{\citenamefont {Verstraete}\ \emph {et~al.}(2008)\citenamefont
  {Verstraete}, \citenamefont {Murg},\ and\ \citenamefont
  {Cirac}}]{VMC08MPSPEPSRev}%
  \BibitemOpen
  \bibfield  {author} {\bibinfo {author} {\bibfnamefont {F.}~\bibnamefont
  {Verstraete}}, \bibinfo {author} {\bibfnamefont {V.}~\bibnamefont {Murg}},\
  and\ \bibinfo {author} {\bibfnamefont {J.~I.}\ \bibnamefont {Cirac}},\
  }\bibfield  {title} {\bibinfo {title} {{Matrix product states, projected
  entangled pair states, and variational renormalization group methods for
  quantum spin systems}},\ }\href {https://doi.org/10.1080/14789940801912366}
  {\bibfield  {journal} {\bibinfo  {journal} {Advances in Physics}\ }\textbf
  {\bibinfo {volume} {57}},\ \bibinfo {pages} {143} (\bibinfo {year}
  {2008})}\BibitemShut {NoStop}%
\bibitem [{\citenamefont {Cirac}\ and\ \citenamefont
  {Verstraete}(2009)}]{CV09TNSRev}%
  \BibitemOpen
  \bibfield  {author} {\bibinfo {author} {\bibfnamefont {J.~I.}\ \bibnamefont
  {Cirac}}\ and\ \bibinfo {author} {\bibfnamefont {F.}~\bibnamefont
  {Verstraete}},\ }\bibfield  {title} {\bibinfo {title} {{Renormalization and
  tensor product states in spin chains and lattices}},\ }\href
  {https://doi.org/10.1088/1751-8113/42/50/504004} {\bibfield  {journal}
  {\bibinfo  {journal} {Journal of Physics A-Mathematical and Theoretical}\
  }\textbf {\bibinfo {volume} {42}},\ \bibinfo {pages} {504004} (\bibinfo
  {year} {2009})}\BibitemShut {NoStop}%
\bibitem [{\citenamefont {Ran}\ \emph {et~al.}(2020{\natexlab{a}})\citenamefont
  {Ran}, \citenamefont {Tirrito}, \citenamefont {Peng}, \citenamefont {Chen},
  \citenamefont {Tagliacozzo}, \citenamefont {Su},\ and\ \citenamefont
  {Lewenstein}}]{RTPC+17TNrev}%
  \BibitemOpen
  \bibfield  {author} {\bibinfo {author} {\bibfnamefont {S.-J.}\ \bibnamefont
  {Ran}}, \bibinfo {author} {\bibfnamefont {E.}~\bibnamefont {Tirrito}},
  \bibinfo {author} {\bibfnamefont {C.}~\bibnamefont {Peng}}, \bibinfo {author}
  {\bibfnamefont {X.}~\bibnamefont {Chen}}, \bibinfo {author} {\bibfnamefont
  {L.}~\bibnamefont {Tagliacozzo}}, \bibinfo {author} {\bibfnamefont
  {G.}~\bibnamefont {Su}},\ and\ \bibinfo {author} {\bibfnamefont
  {M.}~\bibnamefont {Lewenstein}},\ }\href@noop {} {\emph {\bibinfo {title}
  {Tensor network contractions: methods and applications to quantum many-Body
  systems}}}\ (\bibinfo  {publisher} {Springer, Cham},\ \bibinfo {year}
  {2020})\BibitemShut {NoStop}%
\bibitem [{\citenamefont {Or{\'u}s}(2019)}]{O19TNrev}%
  \BibitemOpen
  \bibfield  {author} {\bibinfo {author} {\bibfnamefont {R.}~\bibnamefont
  {Or{\'u}s}},\ }\bibfield  {title} {\bibinfo {title} {Tensor networks for
  complex quantum systems},\ }\href {https://doi.org/10.1038/s42254-019-0086-7}
  {\bibfield  {journal} {\bibinfo  {journal} {Nature Reviews Physics}\ }\textbf
  {\bibinfo {volume} {1}},\ \bibinfo {pages} {538} (\bibinfo {year}
  {2019})}\BibitemShut {NoStop}%
\bibitem [{\citenamefont {Cirac}\ \emph {et~al.}(2021)\citenamefont {Cirac},
  \citenamefont {P\'erez-Garc\'{\i}a}, \citenamefont {Schuch},\ and\
  \citenamefont {Verstraete}}]{CPDSV21TNSrev}%
  \BibitemOpen
  \bibfield  {author} {\bibinfo {author} {\bibfnamefont {J.~I.}\ \bibnamefont
  {Cirac}}, \bibinfo {author} {\bibfnamefont {D.}~\bibnamefont
  {P\'erez-Garc\'{\i}a}}, \bibinfo {author} {\bibfnamefont {N.}~\bibnamefont
  {Schuch}},\ and\ \bibinfo {author} {\bibfnamefont {F.}~\bibnamefont
  {Verstraete}},\ }\bibfield  {title} {\bibinfo {title} {Matrix product states
  and projected entangled pair states: Concepts, symmetries, theorems},\ }\href
  {https://doi.org/10.1103/RevModPhys.93.045003} {\bibfield  {journal}
  {\bibinfo  {journal} {Reviews of Modern Physics}\ }\textbf {\bibinfo {volume}
  {93}},\ \bibinfo {pages} {045003} (\bibinfo {year} {2021})}\BibitemShut
  {NoStop}%
\bibitem [{Note2()}]{Note2}%
  \BibitemOpen
  \bibinfo {note} {Hilbert space is a generalization of Euclidean space. We can
  define vectors, operators, and the measures of their distances in a Hilbert
  space, while the dimensions can be infinite. Hilbert space has been used to
  describe the space for quantum states and operators.}\BibitemShut {Stop}%
\bibitem [{Note3()}]{Note3}%
  \BibitemOpen
  \bibinfo {note} {Exact diagonalization is a widely used approach by fully
  diagonalizing the matrix (say quantum Hamiltonian) that contains the full
  information for the physics of the considered model.}\BibitemShut {Stop}%
\bibitem [{\citenamefont {Greenberger}\ \emph {et~al.}(1989)\citenamefont
  {Greenberger}, \citenamefont {Horne},\ and\ \citenamefont
  {Zeilinger}}]{GHZ89GHZ}%
  \BibitemOpen
  \bibfield  {author} {\bibinfo {author} {\bibfnamefont {D.~M.}\ \bibnamefont
  {Greenberger}}, \bibinfo {author} {\bibfnamefont {M.~A.}\ \bibnamefont
  {Horne}},\ and\ \bibinfo {author} {\bibfnamefont {A.}~\bibnamefont
  {Zeilinger}},\ }\bibinfo {title} {Going beyond bell's theorem},\ in\
  \href@noop {} {\emph {\bibinfo {booktitle} {Bell's Theorem, Quantum Theory
  and Conceptions of the Universe}}},\ \bibinfo {editor} {edited by\ \bibinfo
  {editor} {\bibfnamefont {M.}~\bibnamefont {Kafatos}}}\ (\bibinfo  {publisher}
  {Springer Netherlands},\ \bibinfo {address} {Dordrecht},\ \bibinfo {year}
  {1989})\ pp.\ \bibinfo {pages} {69--72}\BibitemShut {NoStop}%
\bibitem [{\citenamefont {Affleck}\ \emph {et~al.}(1987)\citenamefont
  {Affleck}, \citenamefont {Kennedy}, \citenamefont {Lieb},\ and\ \citenamefont
  {Tasaki}}]{AKLT87AKLT}%
  \BibitemOpen
  \bibfield  {author} {\bibinfo {author} {\bibfnamefont {I.}~\bibnamefont
  {Affleck}}, \bibinfo {author} {\bibfnamefont {T.}~\bibnamefont {Kennedy}},
  \bibinfo {author} {\bibfnamefont {E.~H.}\ \bibnamefont {Lieb}},\ and\
  \bibinfo {author} {\bibfnamefont {H.}~\bibnamefont {Tasaki}},\ }\bibfield
  {title} {\bibinfo {title} {Rigorous results on valence-bond ground states in
  antiferromagnets},\ }\href {https://doi.org/10.1103/PhysRevLett.59.799}
  {\bibfield  {journal} {\bibinfo  {journal} {Physical Review Letters}\
  }\textbf {\bibinfo {volume} {59}},\ \bibinfo {pages} {799} (\bibinfo {year}
  {1987})}\BibitemShut {NoStop}%
\bibitem [{\citenamefont {Affleck}\ \emph {et~al.}(1988)\citenamefont
  {Affleck}, \citenamefont {Kennedy}, \citenamefont {Lieb},\ and\ \citenamefont
  {Tasaki}}]{AKLT88AKLT2}%
  \BibitemOpen
  \bibfield  {author} {\bibinfo {author} {\bibfnamefont {I.}~\bibnamefont
  {Affleck}}, \bibinfo {author} {\bibfnamefont {T.}~\bibnamefont {Kennedy}},
  \bibinfo {author} {\bibfnamefont {E.~H.}\ \bibnamefont {Lieb}},\ and\
  \bibinfo {author} {\bibfnamefont {H.}~\bibnamefont {Tasaki}},\ }\bibfield
  {title} {\bibinfo {title} {Valence bond ground states in isotropic quantum
  antiferromagnets},\ }\href {https://doi.org/10.1007/BF01218021} {\bibfield
  {journal} {\bibinfo  {journal} {Communications in Mathematical Physics}\
  }\textbf {\bibinfo {volume} {115}},\ \bibinfo {pages} {477} (\bibinfo {year}
  {1988})}\BibitemShut {NoStop}%
\bibitem [{\citenamefont {D\"ur}(2001)}]{W01Wstate}%
  \BibitemOpen
  \bibfield  {author} {\bibinfo {author} {\bibfnamefont {W.}~\bibnamefont
  {D\"ur}},\ }\bibfield  {title} {\bibinfo {title} {Multipartite entanglement
  that is robust against disposal of particles},\ }\href
  {https://doi.org/10.1103/PhysRevA.63.020303} {\bibfield  {journal} {\bibinfo
  {journal} {Physical Review A}\ }\textbf {\bibinfo {volume} {63}},\ \bibinfo
  {pages} {020303} (\bibinfo {year} {2001})}\BibitemShut {NoStop}%
\bibitem [{\citenamefont {White}(1992)}]{W92DMRG}%
  \BibitemOpen
  \bibfield  {author} {\bibinfo {author} {\bibfnamefont {S.~R.}\ \bibnamefont
  {White}},\ }\bibfield  {title} {\bibinfo {title} {Density matrix formulation
  for quantum renormalization groups},\ }\href
  {https://doi.org/10.1103/PhysRevLett.69.2863} {\bibfield  {journal} {\bibinfo
   {journal} {Physical Review Letters}\ }\textbf {\bibinfo {volume} {69}},\
  \bibinfo {pages} {2863} (\bibinfo {year} {1992})}\BibitemShut {NoStop}%
\bibitem [{\citenamefont {White}(1993)}]{W93DMRG}%
  \BibitemOpen
  \bibfield  {author} {\bibinfo {author} {\bibfnamefont {S.~R.}\ \bibnamefont
  {White}},\ }\bibfield  {title} {\bibinfo {title} {{Density-matrix algorithms
  for quantum renormalization groups}},\ }\href
  {https://doi.org/10.1103/PhysRevB.48.10345} {\bibfield  {journal} {\bibinfo
  {journal} {Physical Review B}\ }\textbf {\bibinfo {volume} {48}},\ \bibinfo
  {pages} {10345} (\bibinfo {year} {1993})}\BibitemShut {NoStop}%
\bibitem [{\citenamefont {Vidal}(2003)}]{V03TEBD}%
  \BibitemOpen
  \bibfield  {author} {\bibinfo {author} {\bibfnamefont {G.}~\bibnamefont
  {Vidal}},\ }\bibfield  {title} {\bibinfo {title} {Efficient classical
  simulation of slightly entangled quantum computations},\ }\href
  {https://doi.org/10.1103/PhysRevLett.91.147902} {\bibfield  {journal}
  {\bibinfo  {journal} {Physical Review Letters}\ }\textbf {\bibinfo {volume}
  {91}},\ \bibinfo {pages} {147902} (\bibinfo {year} {2003})}\BibitemShut
  {NoStop}%
\bibitem [{\citenamefont {Vidal}(2004)}]{V04TEBD}%
  \BibitemOpen
  \bibfield  {author} {\bibinfo {author} {\bibfnamefont {G.}~\bibnamefont
  {Vidal}},\ }\bibfield  {title} {\bibinfo {title} {Efficient simulation of
  one-dimensional quantum many-body systems},\ }\href
  {https://doi.org/10.1103/PhysRevLett.93.040502} {\bibfield  {journal}
  {\bibinfo  {journal} {Physical Review Letters}\ }\textbf {\bibinfo {volume}
  {93}},\ \bibinfo {pages} {040502} (\bibinfo {year} {2004})}\BibitemShut
  {NoStop}%
\bibitem [{\citenamefont {Vidal}\ \emph {et~al.}(2003)\citenamefont {Vidal},
  \citenamefont {Latorre}, \citenamefont {Rico},\ and\ \citenamefont
  {Kitaev}}]{VLRK03CritEnt}%
  \BibitemOpen
  \bibfield  {author} {\bibinfo {author} {\bibfnamefont {G.}~\bibnamefont
  {Vidal}}, \bibinfo {author} {\bibfnamefont {J.~I.}\ \bibnamefont {Latorre}},
  \bibinfo {author} {\bibfnamefont {E.}~\bibnamefont {Rico}},\ and\ \bibinfo
  {author} {\bibfnamefont {A.}~\bibnamefont {Kitaev}},\ }\bibfield  {title}
  {\bibinfo {title} {{Entanglement in quantum critical phenomena}},\ }\href
  {https://doi.org/10.1103/PhysRevLett.90.227902} {\bibfield  {journal}
  {\bibinfo  {journal} {Physical Review Letters}\ }\textbf {\bibinfo {volume}
  {90}},\ \bibinfo {pages} {227902} (\bibinfo {year} {2003})}\BibitemShut
  {NoStop}%
\bibitem [{\citenamefont {Pollmann}\ \emph {et~al.}(2009)\citenamefont
  {Pollmann}, \citenamefont {Mukerjee}, \citenamefont {Turner},\ and\
  \citenamefont {Moore}}]{PMTM09EntScaling}%
  \BibitemOpen
  \bibfield  {author} {\bibinfo {author} {\bibfnamefont {F.}~\bibnamefont
  {Pollmann}}, \bibinfo {author} {\bibfnamefont {S.}~\bibnamefont {Mukerjee}},
  \bibinfo {author} {\bibfnamefont {A.~M.}\ \bibnamefont {Turner}},\ and\
  \bibinfo {author} {\bibfnamefont {J.~E.}\ \bibnamefont {Moore}},\ }\bibfield
  {title} {\bibinfo {title} {{Theory of finite-entanglement scaling at
  one-dimensional quantum critical points}},\ }\href
  {https://doi.org/10.1103/PhysRevLett.102.255701} {\bibfield  {journal}
  {\bibinfo  {journal} {Physical Review Letters}\ }\textbf {\bibinfo {volume}
  {102}},\ \bibinfo {pages} {255701} (\bibinfo {year} {2009})}\BibitemShut
  {NoStop}%
\bibitem [{\citenamefont {Tagliacozzo}\ \emph {et~al.}(2008)\citenamefont
  {Tagliacozzo}, \citenamefont {de~Oliveira}, \citenamefont {Iblisdir},\ and\
  \citenamefont {Latorre}}]{TOIL08EntScaling}%
  \BibitemOpen
  \bibfield  {author} {\bibinfo {author} {\bibfnamefont {L.}~\bibnamefont
  {Tagliacozzo}}, \bibinfo {author} {\bibfnamefont {T.~R.}\ \bibnamefont
  {de~Oliveira}}, \bibinfo {author} {\bibfnamefont {S.}~\bibnamefont
  {Iblisdir}},\ and\ \bibinfo {author} {\bibfnamefont {J.~I.}\ \bibnamefont
  {Latorre}},\ }\bibfield  {title} {\bibinfo {title} {{Scaling of entanglement
  support for matrix product states}},\ }\href
  {https://doi.org/10.1103/PhysRevB.78.024410} {\bibfield  {journal} {\bibinfo
  {journal} {Physical Review B}\ }\textbf {\bibinfo {volume} {78}},\ \bibinfo
  {pages} {024410} (\bibinfo {year} {2008})}\BibitemShut {NoStop}%
\bibitem [{\citenamefont {Pollmann}\ and\ \citenamefont
  {Moore}(2010)}]{PMJ2010entanglement}%
  \BibitemOpen
  \bibfield  {author} {\bibinfo {author} {\bibfnamefont {F.}~\bibnamefont
  {Pollmann}}\ and\ \bibinfo {author} {\bibfnamefont {J.~E.}\ \bibnamefont
  {Moore}},\ }\bibfield  {title} {\bibinfo {title} {{Entanglement spectra of
  critical and near-critical systems in one dimension}},\ }\href@noop {}
  {\bibfield  {journal} {\bibinfo  {journal} {New Journal of Physics}\ }\textbf
  {\bibinfo {volume} {12}},\ \bibinfo {pages} {025006} (\bibinfo {year}
  {2010})}\BibitemShut {NoStop}%
\bibitem [{\citenamefont {Cheng}\ \emph
  {et~al.}(2018{\natexlab{a}})\citenamefont {Cheng}, \citenamefont {Chen},\
  and\ \citenamefont {Wang}}]{CCW18Born}%
  \BibitemOpen
  \bibfield  {author} {\bibinfo {author} {\bibfnamefont {S.}~\bibnamefont
  {Cheng}}, \bibinfo {author} {\bibfnamefont {J.}~\bibnamefont {Chen}},\ and\
  \bibinfo {author} {\bibfnamefont {L.}~\bibnamefont {Wang}},\ }\bibfield
  {title} {\bibinfo {title} {Information perspective to probabilistic modeling:
  Boltzmann machines versus born machines},\ }\bibfield  {journal} {\bibinfo
  {journal} {Entropy}\ }\textbf {\bibinfo {volume} {20}},\ \href
  {https://doi.org/10.3390/e20080583} {10.3390/e20080583} (\bibinfo {year}
  {2018}{\natexlab{a}})\BibitemShut {NoStop}%
\bibitem [{\citenamefont {Cheng}\ \emph {et~al.}(2019)\citenamefont {Cheng},
  \citenamefont {Wang}, \citenamefont {Xiang},\ and\ \citenamefont
  {Zhang}}]{CWXZ19generateTTNML}%
  \BibitemOpen
  \bibfield  {author} {\bibinfo {author} {\bibfnamefont {S.}~\bibnamefont
  {Cheng}}, \bibinfo {author} {\bibfnamefont {L.}~\bibnamefont {Wang}},
  \bibinfo {author} {\bibfnamefont {T.}~\bibnamefont {Xiang}},\ and\ \bibinfo
  {author} {\bibfnamefont {P.}~\bibnamefont {Zhang}},\ }\bibfield  {title}
  {\bibinfo {title} {Tree tensor networks for generative modeling},\ }\href
  {https://doi.org/10.1103/PhysRevB.99.155131} {\bibfield  {journal} {\bibinfo
  {journal} {Physical Review B}\ }\textbf {\bibinfo {volume} {99}},\ \bibinfo
  {pages} {155131} (\bibinfo {year} {2019})}\BibitemShut {NoStop}%
\bibitem [{\citenamefont {Sun}\ \emph {et~al.}(2020{\natexlab{a}})\citenamefont
  {Sun}, \citenamefont {Peng}, \citenamefont {Liu}, \citenamefont {Ran},\ and\
  \citenamefont {Su}}]{SPLRS19GTNC}%
  \BibitemOpen
  \bibfield  {author} {\bibinfo {author} {\bibfnamefont {Z.-Z.}\ \bibnamefont
  {Sun}}, \bibinfo {author} {\bibfnamefont {C.}~\bibnamefont {Peng}}, \bibinfo
  {author} {\bibfnamefont {D.}~\bibnamefont {Liu}}, \bibinfo {author}
  {\bibfnamefont {S.-J.}\ \bibnamefont {Ran}},\ and\ \bibinfo {author}
  {\bibfnamefont {G.}~\bibnamefont {Su}},\ }\bibfield  {title} {\bibinfo
  {title} {Generative tensor network classification model for supervised
  machine learning},\ }\href {https://doi.org/10.1103/PhysRevB.101.075135}
  {\bibfield  {journal} {\bibinfo  {journal} {Physical Review B}\ }\textbf
  {\bibinfo {volume} {101}},\ \bibinfo {pages} {075135} (\bibinfo {year}
  {2020}{\natexlab{a}})}\BibitemShut {NoStop}%
\bibitem [{\citenamefont {Metz}\ and\ \citenamefont
  {Bukov}(2023)}]{MB23TNQCML}%
  \BibitemOpen
  \bibfield  {author} {\bibinfo {author} {\bibfnamefont {F.}~\bibnamefont
  {Metz}}\ and\ \bibinfo {author} {\bibfnamefont {M.}~\bibnamefont {Bukov}},\
  }\bibfield  {title} {\bibinfo {title} {Self-correcting quantum many-body
  control using reinforcement learning with tensor networks},\ }\href
  {https://doi.org/10.1038/s42256-023-00687-5} {\bibfield  {journal} {\bibinfo
  {journal} {Nature Machine Intelligence}\ }\textbf {\bibinfo {volume} {5}},\
  \bibinfo {pages} {780} (\bibinfo {year} {2023})}\BibitemShut {NoStop}%
\bibitem [{\citenamefont {Tjandra}\ \emph {et~al.}(2017)\citenamefont
  {Tjandra}, \citenamefont {Sakti},\ and\ \citenamefont
  {Nakamura}}]{TSN17TTRNN}%
  \BibitemOpen
  \bibfield  {author} {\bibinfo {author} {\bibfnamefont {A.}~\bibnamefont
  {Tjandra}}, \bibinfo {author} {\bibfnamefont {S.}~\bibnamefont {Sakti}},\
  and\ \bibinfo {author} {\bibfnamefont {S.}~\bibnamefont {Nakamura}},\
  }\bibfield  {title} {\bibinfo {title} {Compressing recurrent neural network
  with tensor train},\ }in\ \href@noop {} {\emph {\bibinfo {booktitle} {2017
  International Joint Conference on Neural Networks (IJCNN)}}}\ (\bibinfo
  {year} {2017})\ pp.\ \bibinfo {pages} {4451--4458}\BibitemShut {NoStop}%
\bibitem [{\citenamefont {Yuan}\ \emph {et~al.}(2019)\citenamefont {Yuan},
  \citenamefont {Li}, \citenamefont {Mandic}, \citenamefont {Cao},\ and\
  \citenamefont {Zhao}}]{YLMCZ2019TRDcomp}%
  \BibitemOpen
  \bibfield  {author} {\bibinfo {author} {\bibfnamefont {L.}~\bibnamefont
  {Yuan}}, \bibinfo {author} {\bibfnamefont {C.}~\bibnamefont {Li}}, \bibinfo
  {author} {\bibfnamefont {D.}~\bibnamefont {Mandic}}, \bibinfo {author}
  {\bibfnamefont {J.}~\bibnamefont {Cao}},\ and\ \bibinfo {author}
  {\bibfnamefont {Q.}~\bibnamefont {Zhao}},\ }\bibfield  {title} {\bibinfo
  {title} {Tensor ring decomposition with rank minimization on latent space: An
  efficient approach for tensor completion},\ }\href
  {https://doi.org/10.1609/aaai.v33i01.33019151} {\bibfield  {journal}
  {\bibinfo  {journal} {Proceedings of the AAAI Conference on Artificial
  Intelligence}\ }\textbf {\bibinfo {volume} {33}},\ \bibinfo {pages} {9151}
  (\bibinfo {year} {2019})}\BibitemShut {NoStop}%
\bibitem [{\citenamefont {Pan}\ \emph {et~al.}(2019)\citenamefont {Pan},
  \citenamefont {Xu}, \citenamefont {Wang}, \citenamefont {Ye}, \citenamefont
  {Wang}, \citenamefont {Bai},\ and\ \citenamefont {Xu}}]{PXWY+19TRDRNN}%
  \BibitemOpen
  \bibfield  {author} {\bibinfo {author} {\bibfnamefont {Y.}~\bibnamefont
  {Pan}}, \bibinfo {author} {\bibfnamefont {J.}~\bibnamefont {Xu}}, \bibinfo
  {author} {\bibfnamefont {M.}~\bibnamefont {Wang}}, \bibinfo {author}
  {\bibfnamefont {J.}~\bibnamefont {Ye}}, \bibinfo {author} {\bibfnamefont
  {F.}~\bibnamefont {Wang}}, \bibinfo {author} {\bibfnamefont {K.}~\bibnamefont
  {Bai}},\ and\ \bibinfo {author} {\bibfnamefont {Z.}~\bibnamefont {Xu}},\
  }\bibfield  {title} {\bibinfo {title} {Compressing recurrent neural networks
  with tensor ring for action recognition},\ }\href
  {https://doi.org/10.1609/aaai.v33i01.33014683} {\bibfield  {journal}
  {\bibinfo  {journal} {Proceedings of the AAAI Conference on Artificial
  Intelligence}\ }\textbf {\bibinfo {volume} {33}},\ \bibinfo {pages} {4683}
  (\bibinfo {year} {2019})}\BibitemShut {NoStop}%
\bibitem [{\citenamefont {Gao}\ \emph {et~al.}(2020)\citenamefont {Gao},
  \citenamefont {Cheng}, \citenamefont {He}, \citenamefont {Xie}, \citenamefont
  {Zhao}, \citenamefont {Lu},\ and\ \citenamefont {Xiang}}]{GCHX20+NNMPO}%
  \BibitemOpen
  \bibfield  {author} {\bibinfo {author} {\bibfnamefont {Z.-F.}\ \bibnamefont
  {Gao}}, \bibinfo {author} {\bibfnamefont {S.}~\bibnamefont {Cheng}}, \bibinfo
  {author} {\bibfnamefont {R.-Q.}\ \bibnamefont {He}}, \bibinfo {author}
  {\bibfnamefont {Z.~Y.}\ \bibnamefont {Xie}}, \bibinfo {author} {\bibfnamefont
  {H.-H.}\ \bibnamefont {Zhao}}, \bibinfo {author} {\bibfnamefont {Z.-Y.}\
  \bibnamefont {Lu}},\ and\ \bibinfo {author} {\bibfnamefont {T.}~\bibnamefont
  {Xiang}},\ }\bibfield  {title} {\bibinfo {title} {Compressing deep neural
  networks by matrix product operators},\ }\href
  {https://doi.org/10.1103/PhysRevResearch.2.023300} {\bibfield  {journal}
  {\bibinfo  {journal} {Physical Review Research}\ }\textbf {\bibinfo {volume}
  {2}},\ \bibinfo {pages} {023300} (\bibinfo {year} {2020})}\BibitemShut
  {NoStop}%
\bibitem [{\citenamefont {Sun}\ \emph {et~al.}(2020{\natexlab{b}})\citenamefont
  {Sun}, \citenamefont {Gao}, \citenamefont {Lu}, \citenamefont {Li},\ and\
  \citenamefont {Yan}}]{SGLLY20NNMPO}%
  \BibitemOpen
  \bibfield  {author} {\bibinfo {author} {\bibfnamefont {X.}~\bibnamefont
  {Sun}}, \bibinfo {author} {\bibfnamefont {Z.-F.}\ \bibnamefont {Gao}},
  \bibinfo {author} {\bibfnamefont {Z.-Y.}\ \bibnamefont {Lu}}, \bibinfo
  {author} {\bibfnamefont {J.}~\bibnamefont {Li}},\ and\ \bibinfo {author}
  {\bibfnamefont {Y.}~\bibnamefont {Yan}},\ }\bibfield  {title} {\bibinfo
  {title} {A model compression method with matrix product operators for speech
  enhancement},\ }\href {https://doi.org/10.1109/TASLP.2020.3030495} {\bibfield
   {journal} {\bibinfo  {journal} {IEEE/ACM Transactions on Audio, Speech, and
  Language Processing}\ }\textbf {\bibinfo {volume} {28}},\ \bibinfo {pages}
  {2837} (\bibinfo {year} {2020}{\natexlab{b}})}\BibitemShut {NoStop}%
\bibitem [{\citenamefont {Wang}\ \emph
  {et~al.}(2020{\natexlab{a}})\citenamefont {Wang}, \citenamefont {Zhao},
  \citenamefont {Li}, \citenamefont {Deng},\ and\ \citenamefont
  {Wu}}]{WZLDW20TTCNN}%
  \BibitemOpen
  \bibfield  {author} {\bibinfo {author} {\bibfnamefont {D.}~\bibnamefont
  {Wang}}, \bibinfo {author} {\bibfnamefont {G.}~\bibnamefont {Zhao}}, \bibinfo
  {author} {\bibfnamefont {G.}~\bibnamefont {Li}}, \bibinfo {author}
  {\bibfnamefont {L.}~\bibnamefont {Deng}},\ and\ \bibinfo {author}
  {\bibfnamefont {Y.}~\bibnamefont {Wu}},\ }\bibfield  {title} {\bibinfo
  {title} {Compressing 3dcnns based on tensor train decomposition},\ }\href
  {https://doi.org/https://doi.org/10.1016/j.neunet.2020.07.028} {\bibfield
  {journal} {\bibinfo  {journal} {Neural Networks}\ }\textbf {\bibinfo {volume}
  {131}},\ \bibinfo {pages} {215} (\bibinfo {year}
  {2020}{\natexlab{a}})}\BibitemShut {NoStop}%
\bibitem [{\citenamefont {Wang}\ \emph
  {et~al.}(2021{\natexlab{a}})\citenamefont {Wang}, \citenamefont {Zhao},
  \citenamefont {Chen}, \citenamefont {Liu}, \citenamefont {Deng},\ and\
  \citenamefont {Li}}]{WZCL+21nonlinearTT}%
  \BibitemOpen
  \bibfield  {author} {\bibinfo {author} {\bibfnamefont {D.}~\bibnamefont
  {Wang}}, \bibinfo {author} {\bibfnamefont {G.}~\bibnamefont {Zhao}}, \bibinfo
  {author} {\bibfnamefont {H.}~\bibnamefont {Chen}}, \bibinfo {author}
  {\bibfnamefont {Z.}~\bibnamefont {Liu}}, \bibinfo {author} {\bibfnamefont
  {L.}~\bibnamefont {Deng}},\ and\ \bibinfo {author} {\bibfnamefont
  {G.}~\bibnamefont {Li}},\ }\bibfield  {title} {\bibinfo {title} {Nonlinear
  tensor train format for deep neural network compression},\ }\href
  {https://doi.org/https://doi.org/10.1016/j.neunet.2021.08.028} {\bibfield
  {journal} {\bibinfo  {journal} {Neural Networks}\ }\textbf {\bibinfo {volume}
  {144}},\ \bibinfo {pages} {320} (\bibinfo {year}
  {2021}{\natexlab{a}})}\BibitemShut {NoStop}%
\bibitem [{\citenamefont {Qing}\ \emph {et~al.}(2023)\citenamefont {Qing},
  \citenamefont {Zhou}, \citenamefont {Li},\ and\ \citenamefont
  {Ran}}]{qing2023compressing}%
  \BibitemOpen
  \bibfield  {author} {\bibinfo {author} {\bibfnamefont {Y.}~\bibnamefont
  {Qing}}, \bibinfo {author} {\bibfnamefont {P.-F.}\ \bibnamefont {Zhou}},
  \bibinfo {author} {\bibfnamefont {K.}~\bibnamefont {Li}},\ and\ \bibinfo
  {author} {\bibfnamefont {S.-J.}\ \bibnamefont {Ran}},\ }\href@noop {}
  {\bibinfo {title} {Compressing neural network by tensor network with
  exponentially fewer variational parameters}} (\bibinfo {year} {2023}),\
  \Eprint {https://arxiv.org/abs/2305.06058} {arXiv:2305.06058 [cs.LG]}
  \BibitemShut {NoStop}%
\bibitem [{Note4()}]{Note4}%
  \BibitemOpen
  \bibinfo {note} {We may consider a land consisting of many villages as an
  example to understand the area law. If every person in this land can only
  communicate with the persons in a short range nearby, the people who can
  communicate to a different village should live near the boarders. Therefore,
  the amount of exchanged information between a village and the rest should
  scale with the length of its boarders. In this case, the area law is obeyed.
  Someday, phones are introduced to this land, and people are able to
  communicate with anyone in this land. The amount of exchanged information of
  a village should scale with its population or approximately the size of its
  territory. This is named as the volume law, where obviously the amount of
  exchanged information increases much faster than the area-law cases if the
  village expands its territory.}\BibitemShut {Stop}%
\bibitem [{\citenamefont {Eisert}\ \emph {et~al.}(2010)\citenamefont {Eisert},
  \citenamefont {Cramer},\ and\ \citenamefont {Plenio}}]{ECP10AreaLawRev}%
  \BibitemOpen
  \bibfield  {author} {\bibinfo {author} {\bibfnamefont {J.}~\bibnamefont
  {Eisert}}, \bibinfo {author} {\bibfnamefont {M.}~\bibnamefont {Cramer}},\
  and\ \bibinfo {author} {\bibfnamefont {M.~B.}\ \bibnamefont {Plenio}},\
  }\bibfield  {title} {\bibinfo {title} {Colloquium: Area laws for the
  entanglement entropy},\ }\href {https://doi.org/10.1103/RevModPhys.82.277}
  {\bibfield  {journal} {\bibinfo  {journal} {Reviews of Modern Physics}\
  }\textbf {\bibinfo {volume} {82}},\ \bibinfo {pages} {277} (\bibinfo {year}
  {2010})}\BibitemShut {NoStop}%
\bibitem [{\citenamefont {Tagliacozzo}\ \emph {et~al.}(2009)\citenamefont
  {Tagliacozzo}, \citenamefont {Evenbly},\ and\ \citenamefont
  {Vidal}}]{TEV09TTN}%
  \BibitemOpen
  \bibfield  {author} {\bibinfo {author} {\bibfnamefont {L.}~\bibnamefont
  {Tagliacozzo}}, \bibinfo {author} {\bibfnamefont {G.}~\bibnamefont
  {Evenbly}},\ and\ \bibinfo {author} {\bibfnamefont {G.}~\bibnamefont
  {Vidal}},\ }\bibfield  {title} {\bibinfo {title} {Simulation of
  two-dimensional quantum systems using a tree tensor network that exploits the
  entropic area law},\ }\href {https://doi.org/10.1103/PhysRevB.80.235127}
  {\bibfield  {journal} {\bibinfo  {journal} {Physical Review B}\ }\textbf
  {\bibinfo {volume} {80}},\ \bibinfo {pages} {235127} (\bibinfo {year}
  {2009})}\BibitemShut {NoStop}%
\bibitem [{\citenamefont {Piroli}\ and\ \citenamefont
  {Cirac}(2020)}]{PC20TNarealaw}%
  \BibitemOpen
  \bibfield  {author} {\bibinfo {author} {\bibfnamefont {L.}~\bibnamefont
  {Piroli}}\ and\ \bibinfo {author} {\bibfnamefont {J.~I.}\ \bibnamefont
  {Cirac}},\ }\bibfield  {title} {\bibinfo {title} {Quantum cellular automata,
  tensor networks, and area laws},\ }\href
  {https://doi.org/10.1103/PhysRevLett.125.190402} {\bibfield  {journal}
  {\bibinfo  {journal} {Physical Review Letters}\ }\textbf {\bibinfo {volume}
  {125}},\ \bibinfo {pages} {190402} (\bibinfo {year} {2020})}\BibitemShut
  {NoStop}%
\bibitem [{\citenamefont {Nielsen}\ and\ \citenamefont
  {Chuang}(2002)}]{NC02Qcmp}%
  \BibitemOpen
  \bibfield  {author} {\bibinfo {author} {\bibfnamefont {M.~A.}\ \bibnamefont
  {Nielsen}}\ and\ \bibinfo {author} {\bibfnamefont {I.}~\bibnamefont
  {Chuang}},\ }\href@noop {} {\emph {\bibinfo {title} {Quantum computation and
  quantum information}}}\ (\bibinfo  {publisher} {American Association of
  Physics Teachers},\ \bibinfo {year} {2002})\BibitemShut {NoStop}%
\bibitem [{Note5()}]{Note5}%
  \BibitemOpen
  \bibinfo {note} {Entanglement can be understood as a description of quantum
  correlations. A strong entanglement between two quantum particles means
  significant affections to the state of one particle by operating another.
  Entanglement entropy is a measure of the strength of entanglement. For
  instance, zero entanglement entropy between two particles means that their
  state should be described by a product state. Even these two particles might
  be correlated, the correlations should be fully described by the classical
  correlations.}\BibitemShut {Stop}%
\bibitem [{\citenamefont {Hastings}(2007{\natexlab{a}})}]{H07EntArea}%
  \BibitemOpen
  \bibfield  {author} {\bibinfo {author} {\bibfnamefont {M.~B.}\ \bibnamefont
  {Hastings}},\ }\bibfield  {title} {\bibinfo {title} {An area law for
  one-dimensional quantum systems},\ }\href
  {https://doi.org/10.1088/1742-5468/2007/08/p08024} {\bibfield  {journal}
  {\bibinfo  {journal} {Journal of Statistical Mechanics: Theory and
  Experiment}\ }\textbf {\bibinfo {volume} {2007}},\ \bibinfo {pages} {P08024}
  (\bibinfo {year} {2007}{\natexlab{a}})}\BibitemShut {NoStop}%
\bibitem [{\citenamefont {Hastings}(2007{\natexlab{b}})}]{H07MPSent}%
  \BibitemOpen
  \bibfield  {author} {\bibinfo {author} {\bibfnamefont {M.~B.}\ \bibnamefont
  {Hastings}},\ }\bibfield  {title} {\bibinfo {title} {Entropy and entanglement
  in quantum ground states},\ }\href
  {https://doi.org/10.1103/PhysRevB.76.035114} {\bibfield  {journal} {\bibinfo
  {journal} {Physical Review B}\ }\textbf {\bibinfo {volume} {76}},\ \bibinfo
  {pages} {035114} (\bibinfo {year} {2007}{\natexlab{b}})}\BibitemShut
  {NoStop}%
\bibitem [{\citenamefont {Schuch}\ \emph {et~al.}(2008)\citenamefont {Schuch},
  \citenamefont {Wolf}, \citenamefont {Verstraete},\ and\ \citenamefont
  {Cirac}}]{SWVC08MPSent}%
  \BibitemOpen
  \bibfield  {author} {\bibinfo {author} {\bibfnamefont {N.}~\bibnamefont
  {Schuch}}, \bibinfo {author} {\bibfnamefont {M.~M.}\ \bibnamefont {Wolf}},
  \bibinfo {author} {\bibfnamefont {F.}~\bibnamefont {Verstraete}},\ and\
  \bibinfo {author} {\bibfnamefont {J.~I.}\ \bibnamefont {Cirac}},\ }\bibfield
  {title} {\bibinfo {title} {{Entropy scaling and simulability by matrix
  product states}},\ }\href {https://doi.org/10.1103/PhysRevLett.100.030504}
  {\bibfield  {journal} {\bibinfo  {journal} {Physical Review Letters}\
  }\textbf {\bibinfo {volume} {100}},\ \bibinfo {pages} {030504} (\bibinfo
  {year} {2008})}\BibitemShut {NoStop}%
\bibitem [{\citenamefont {Verstraete}\ and\ \citenamefont
  {Cirac}(2006)}]{VC06MPSFaithfully}%
  \BibitemOpen
  \bibfield  {author} {\bibinfo {author} {\bibfnamefont {F.}~\bibnamefont
  {Verstraete}}\ and\ \bibinfo {author} {\bibfnamefont {J.~I.}\ \bibnamefont
  {Cirac}},\ }\bibfield  {title} {\bibinfo {title} {Matrix product states
  represent ground states faithfully},\ }\href
  {https://doi.org/10.1103/PhysRevB.73.094423} {\bibfield  {journal} {\bibinfo
  {journal} {Physical Review B}\ }\textbf {\bibinfo {volume} {73}},\ \bibinfo
  {pages} {094423} (\bibinfo {year} {2006})}\BibitemShut {NoStop}%
\bibitem [{\citenamefont {de~Beaudrap}\ \emph {et~al.}(2010)\citenamefont
  {de~Beaudrap}, \citenamefont {Osborne},\ and\ \citenamefont
  {Eisert}}]{BOE10arealaw}%
  \BibitemOpen
  \bibfield  {author} {\bibinfo {author} {\bibfnamefont {N.}~\bibnamefont
  {de~Beaudrap}}, \bibinfo {author} {\bibfnamefont {T.~J.}\ \bibnamefont
  {Osborne}},\ and\ \bibinfo {author} {\bibfnamefont {J.}~\bibnamefont
  {Eisert}},\ }\bibfield  {title} {\bibinfo {title} {Ground states of
  unfrustrated spin hamiltonians satisfy an area law},\ }\href
  {https://doi.org/10.1088/1367-2630/12/9/095007} {\bibfield  {journal}
  {\bibinfo  {journal} {New Journal of Physics}\ }\textbf {\bibinfo {volume}
  {12}},\ \bibinfo {pages} {095007} (\bibinfo {year} {2010})}\BibitemShut
  {NoStop}%
\bibitem [{\citenamefont {Pirvu}\ \emph {et~al.}(2012)\citenamefont {Pirvu},
  \citenamefont {Haegeman},\ and\ \citenamefont {Verstraete}}]{PHV12exciteMPS}%
  \BibitemOpen
  \bibfield  {author} {\bibinfo {author} {\bibfnamefont {B.}~\bibnamefont
  {Pirvu}}, \bibinfo {author} {\bibfnamefont {J.}~\bibnamefont {Haegeman}},\
  and\ \bibinfo {author} {\bibfnamefont {F.}~\bibnamefont {Verstraete}},\
  }\bibfield  {title} {\bibinfo {title} {Matrix product state based algorithm
  for determining dispersion relations of quantum spin chains with periodic
  boundary conditions},\ }\href {https://doi.org/10.1103/PhysRevB.85.035130}
  {\bibfield  {journal} {\bibinfo  {journal} {Physical Review B}\ }\textbf
  {\bibinfo {volume} {85}},\ \bibinfo {pages} {035130} (\bibinfo {year}
  {2012})}\BibitemShut {NoStop}%
\bibitem [{\citenamefont {Friesdorf}\ \emph {et~al.}(2015)\citenamefont
  {Friesdorf}, \citenamefont {Werner}, \citenamefont {Brown}, \citenamefont
  {Scholz},\ and\ \citenamefont {Eisert}}]{FWBSE15MPSMLocal}%
  \BibitemOpen
  \bibfield  {author} {\bibinfo {author} {\bibfnamefont {M.}~\bibnamefont
  {Friesdorf}}, \bibinfo {author} {\bibfnamefont {A.~H.}\ \bibnamefont
  {Werner}}, \bibinfo {author} {\bibfnamefont {W.}~\bibnamefont {Brown}},
  \bibinfo {author} {\bibfnamefont {V.~B.}\ \bibnamefont {Scholz}},\ and\
  \bibinfo {author} {\bibfnamefont {J.}~\bibnamefont {Eisert}},\ }\bibfield
  {title} {\bibinfo {title} {Many-body localization implies that eigenvectors
  are matrix-product states},\ }\href
  {https://doi.org/10.1103/PhysRevLett.114.170505} {\bibfield  {journal}
  {\bibinfo  {journal} {Physical Review Letters}\ }\textbf {\bibinfo {volume}
  {114}},\ \bibinfo {pages} {170505} (\bibinfo {year} {2015})}\BibitemShut
  {NoStop}%
\bibitem [{\citenamefont {Vidal}(2007)}]{V07EntRenor}%
  \BibitemOpen
  \bibfield  {author} {\bibinfo {author} {\bibfnamefont {G.}~\bibnamefont
  {Vidal}},\ }\bibfield  {title} {\bibinfo {title} {Entanglement
  renormalization},\ }\href {https://doi.org/10.1103/PhysRevLett.99.220405}
  {\bibfield  {journal} {\bibinfo  {journal} {Physical Review Letters}\
  }\textbf {\bibinfo {volume} {99}},\ \bibinfo {pages} {220405} (\bibinfo
  {year} {2007})}\BibitemShut {NoStop}%
\bibitem [{\citenamefont {Vidal}(2008)}]{V08MERA}%
  \BibitemOpen
  \bibfield  {author} {\bibinfo {author} {\bibfnamefont {G.}~\bibnamefont
  {Vidal}},\ }\bibfield  {title} {\bibinfo {title} {Class of quantum many-body
  states that can be efficiently simulated},\ }\href
  {https://doi.org/10.1103/PhysRevLett.101.110501} {\bibfield  {journal}
  {\bibinfo  {journal} {Physical Review Letters}\ }\textbf {\bibinfo {volume}
  {101}},\ \bibinfo {pages} {110501} (\bibinfo {year} {2008})}\BibitemShut
  {NoStop}%
\bibitem [{\citenamefont {Evenbly}\ and\ \citenamefont
  {Vidal}(2009)}]{EV09EntRenor}%
  \BibitemOpen
  \bibfield  {author} {\bibinfo {author} {\bibfnamefont {G.}~\bibnamefont
  {Evenbly}}\ and\ \bibinfo {author} {\bibfnamefont {G.}~\bibnamefont
  {Vidal}},\ }\bibfield  {title} {\bibinfo {title} {Entanglement
  renormalization in two spatial dimensions},\ }\href
  {https://doi.org/10.1103/PhysRevLett.102.180406} {\bibfield  {journal}
  {\bibinfo  {journal} {Physical Review Letters}\ }\textbf {\bibinfo {volume}
  {102}},\ \bibinfo {pages} {180406} (\bibinfo {year} {2009})}\BibitemShut
  {NoStop}%
\bibitem [{\citenamefont {Jordan}\ \emph {et~al.}(2008)\citenamefont {Jordan},
  \citenamefont {Or\'us}, \citenamefont {Vidal}, \citenamefont {Verstraete},\
  and\ \citenamefont {Cirac}}]{JOVVC08PEPS}%
  \BibitemOpen
  \bibfield  {author} {\bibinfo {author} {\bibfnamefont {J.}~\bibnamefont
  {Jordan}}, \bibinfo {author} {\bibfnamefont {R.}~\bibnamefont {Or\'us}},
  \bibinfo {author} {\bibfnamefont {G.}~\bibnamefont {Vidal}}, \bibinfo
  {author} {\bibfnamefont {F.}~\bibnamefont {Verstraete}},\ and\ \bibinfo
  {author} {\bibfnamefont {J.~I.}\ \bibnamefont {Cirac}},\ }\bibfield  {title}
  {\bibinfo {title} {Classical simulation of infinite-size quantum lattice
  systems in two spatial dimensions},\ }\href
  {https://doi.org/10.1103/PhysRevLett.101.250602} {\bibfield  {journal}
  {\bibinfo  {journal} {Physical Review Letters}\ }\textbf {\bibinfo {volume}
  {101}},\ \bibinfo {pages} {250602} (\bibinfo {year} {2008})}\BibitemShut
  {NoStop}%
\bibitem [{\citenamefont {Gu}\ \emph {et~al.}(2009)\citenamefont {Gu},
  \citenamefont {Levin}, \citenamefont {Swingle},\ and\ \citenamefont
  {Wen}}]{GLSW09StringTPS}%
  \BibitemOpen
  \bibfield  {author} {\bibinfo {author} {\bibfnamefont {Z.-C.}\ \bibnamefont
  {Gu}}, \bibinfo {author} {\bibfnamefont {M.}~\bibnamefont {Levin}}, \bibinfo
  {author} {\bibfnamefont {B.}~\bibnamefont {Swingle}},\ and\ \bibinfo {author}
  {\bibfnamefont {X.-G.}\ \bibnamefont {Wen}},\ }\bibfield  {title} {\bibinfo
  {title} {Tensor-product representations for string-net condensed states},\
  }\href {https://doi.org/10.1103/PhysRevB.79.085118} {\bibfield  {journal}
  {\bibinfo  {journal} {Physical Review B}\ }\textbf {\bibinfo {volume} {79}},\
  \bibinfo {pages} {085118} (\bibinfo {year} {2009})}\BibitemShut {NoStop}%
\bibitem [{\citenamefont {Buerschaper}\ \emph {et~al.}(2009)\citenamefont
  {Buerschaper}, \citenamefont {Aguado},\ and\ \citenamefont
  {Vidal}}]{BAV09StringTPS}%
  \BibitemOpen
  \bibfield  {author} {\bibinfo {author} {\bibfnamefont {O.}~\bibnamefont
  {Buerschaper}}, \bibinfo {author} {\bibfnamefont {M.}~\bibnamefont
  {Aguado}},\ and\ \bibinfo {author} {\bibfnamefont {G.}~\bibnamefont
  {Vidal}},\ }\bibfield  {title} {\bibinfo {title} {Explicit tensor network
  representation for the ground states of string-net models},\ }\href
  {https://doi.org/10.1103/PhysRevB.79.085119} {\bibfield  {journal} {\bibinfo
  {journal} {Physical Review B}\ }\textbf {\bibinfo {volume} {79}},\ \bibinfo
  {pages} {085119} (\bibinfo {year} {2009})}\BibitemShut {NoStop}%
\bibitem [{\citenamefont {Born}(1926)}]{Born1926}%
  \BibitemOpen
  \bibfield  {author} {\bibinfo {author} {\bibfnamefont {M.}~\bibnamefont
  {Born}},\ }\bibfield  {title} {\bibinfo {title} {Zur quantenmechanik der
  sto{\ss}vorg{\"a}nge},\ }\href
  {https://api.semanticscholar.org/CorpusID:119896026} {\bibfield  {journal}
  {\bibinfo  {journal} {Zeitschrift f{\"u}r Physik}\ }\textbf {\bibinfo
  {volume} {37}},\ \bibinfo {pages} {863} (\bibinfo {year} {1926})}\BibitemShut
  {NoStop}%
\bibitem [{\citenamefont {Felser}\ \emph {et~al.}(2021)\citenamefont {Felser},
  \citenamefont {Trenti}, \citenamefont {Sestini}, \citenamefont {Gianelle},
  \citenamefont {Zuliani}, \citenamefont {Lucchesi},\ and\ \citenamefont
  {Montangero}}]{Felser2021}%
  \BibitemOpen
  \bibfield  {author} {\bibinfo {author} {\bibfnamefont {T.}~\bibnamefont
  {Felser}}, \bibinfo {author} {\bibfnamefont {M.}~\bibnamefont {Trenti}},
  \bibinfo {author} {\bibfnamefont {L.}~\bibnamefont {Sestini}}, \bibinfo
  {author} {\bibfnamefont {A.}~\bibnamefont {Gianelle}}, \bibinfo {author}
  {\bibfnamefont {D.}~\bibnamefont {Zuliani}}, \bibinfo {author} {\bibfnamefont
  {D.}~\bibnamefont {Lucchesi}},\ and\ \bibinfo {author} {\bibfnamefont
  {S.}~\bibnamefont {Montangero}},\ }\bibfield  {title} {\bibinfo {title}
  {Quantum-inspired machine learning on high-energy physics data},\ }\href
  {https://doi.org/10.1038/s41534-021-00443-w} {\bibfield  {journal} {\bibinfo
  {journal} {npj Quantum Information}\ }\textbf {\bibinfo {volume} {7}},\
  \bibinfo {pages} {111} (\bibinfo {year} {2021})}\BibitemShut {NoStop}%
\bibitem [{\citenamefont {Biamonte}\ \emph {et~al.}(2017)\citenamefont
  {Biamonte}, \citenamefont {Wittek}, \citenamefont {Pancotti}, \citenamefont
  {Rebentrost}, \citenamefont {Wiebe},\ and\ \citenamefont
  {Lloyd}}]{BWPR+17QML}%
  \BibitemOpen
  \bibfield  {author} {\bibinfo {author} {\bibfnamefont {J.}~\bibnamefont
  {Biamonte}}, \bibinfo {author} {\bibfnamefont {P.}~\bibnamefont {Wittek}},
  \bibinfo {author} {\bibfnamefont {N.}~\bibnamefont {Pancotti}}, \bibinfo
  {author} {\bibfnamefont {P.}~\bibnamefont {Rebentrost}}, \bibinfo {author}
  {\bibfnamefont {N.}~\bibnamefont {Wiebe}},\ and\ \bibinfo {author}
  {\bibfnamefont {S.}~\bibnamefont {Lloyd}},\ }\bibfield  {title} {\bibinfo
  {title} {Quantum machine learning},\ }\href@noop {} {\bibfield  {journal}
  {\bibinfo  {journal} {Nature}\ }\textbf {\bibinfo {volume} {549}},\ \bibinfo
  {pages} {195} (\bibinfo {year} {2017})}\BibitemShut {NoStop}%
\bibitem [{\citenamefont {Le}\ \emph {et~al.}(2011)\citenamefont {Le},
  \citenamefont {Dong},\ and\ \citenamefont {Hirota}}]{LDH11QFM}%
  \BibitemOpen
  \bibfield  {author} {\bibinfo {author} {\bibfnamefont {P.~Q.}\ \bibnamefont
  {Le}}, \bibinfo {author} {\bibfnamefont {F.}~\bibnamefont {Dong}},\ and\
  \bibinfo {author} {\bibfnamefont {K.}~\bibnamefont {Hirota}},\ }\bibfield
  {title} {\bibinfo {title} {A flexible representation of quantum images for
  polynomial preparation, image compression, and processing operations},\
  }\href {https://doi.org/10.1007/s11128-010-0177-y} {\bibfield  {journal}
  {\bibinfo  {journal} {Quantum Information Processing}\ }\textbf {\bibinfo
  {volume} {10}},\ \bibinfo {pages} {63} (\bibinfo {year} {2011})}\BibitemShut
  {NoStop}%
\bibitem [{\citenamefont {Yan}\ \emph {et~al.}(2016)\citenamefont {Yan},
  \citenamefont {Iliyasu},\ and\ \citenamefont {Venegas-Andraca}}]{YIV16QFM}%
  \BibitemOpen
  \bibfield  {author} {\bibinfo {author} {\bibfnamefont {F.}~\bibnamefont
  {Yan}}, \bibinfo {author} {\bibfnamefont {A.~M.}\ \bibnamefont {Iliyasu}},\
  and\ \bibinfo {author} {\bibfnamefont {S.~E.}\ \bibnamefont
  {Venegas-Andraca}},\ }\bibfield  {title} {\bibinfo {title} {A survey of
  quantum image representations},\ }\href
  {https://doi.org/10.1007/s11128-015-1195-6} {\bibfield  {journal} {\bibinfo
  {journal} {Quantum Information Processing}\ }\textbf {\bibinfo {volume}
  {15}},\ \bibinfo {pages} {1} (\bibinfo {year} {2016})}\BibitemShut {NoStop}%
\bibitem [{\citenamefont {Stoudenmire}\ and\ \citenamefont
  {Schwab}(2016)}]{SS16TNML}%
  \BibitemOpen
  \bibfield  {author} {\bibinfo {author} {\bibfnamefont {E.}~\bibnamefont
  {Stoudenmire}}\ and\ \bibinfo {author} {\bibfnamefont {D.~J.}\ \bibnamefont
  {Schwab}},\ }\bibfield  {title} {\bibinfo {title} {Supervised learning with
  tensor networks},\ }in\ \href@noop {} {\emph {\bibinfo {booktitle} {Advances
  in Neural Information Processing Systems 29}}},\ \bibinfo {editor} {edited
  by\ \bibinfo {editor} {\bibfnamefont {D.~D.}\ \bibnamefont {Lee}}, \bibinfo
  {editor} {\bibfnamefont {M.}~\bibnamefont {Sugiyama}}, \bibinfo {editor}
  {\bibfnamefont {U.~V.}\ \bibnamefont {Luxburg}}, \bibinfo {editor}
  {\bibfnamefont {I.}~\bibnamefont {Guyon}},\ and\ \bibinfo {editor}
  {\bibfnamefont {R.}~\bibnamefont {Garnett}}}\ (\bibinfo  {publisher} {Curran
  Associates, Inc.},\ \bibinfo {year} {2016})\ pp.\ \bibinfo {pages}
  {4799--4807}\BibitemShut {NoStop}%
\bibitem [{\citenamefont {Dilip}\ \emph {et~al.}(2022)\citenamefont {Dilip},
  \citenamefont {Liu}, \citenamefont {Smith},\ and\ \citenamefont
  {Pollmann}}]{DLSP22QDC}%
  \BibitemOpen
  \bibfield  {author} {\bibinfo {author} {\bibfnamefont {R.}~\bibnamefont
  {Dilip}}, \bibinfo {author} {\bibfnamefont {Y.-J.}\ \bibnamefont {Liu}},
  \bibinfo {author} {\bibfnamefont {A.}~\bibnamefont {Smith}},\ and\ \bibinfo
  {author} {\bibfnamefont {F.}~\bibnamefont {Pollmann}},\ }\bibfield  {title}
  {\bibinfo {title} {Data compression for quantum machine learning},\ }\href
  {https://doi.org/10.1103/PhysRevResearch.4.043007} {\bibfield  {journal}
  {\bibinfo  {journal} {Physical Review Research}\ }\textbf {\bibinfo {volume}
  {4}},\ \bibinfo {pages} {043007} (\bibinfo {year} {2022})}\BibitemShut
  {NoStop}%
\bibitem [{\citenamefont {Ashhab}(2022)}]{A22Qencode}%
  \BibitemOpen
  \bibfield  {author} {\bibinfo {author} {\bibfnamefont {S.}~\bibnamefont
  {Ashhab}},\ }\bibfield  {title} {\bibinfo {title} {Quantum state preparation
  protocol for encoding classical data into the amplitudes of a quantum
  information processing register's wave function},\ }\href
  {https://doi.org/10.1103/PhysRevResearch.4.013091} {\bibfield  {journal}
  {\bibinfo  {journal} {Physical Review Research}\ }\textbf {\bibinfo {volume}
  {4}},\ \bibinfo {pages} {013091} (\bibinfo {year} {2022})}\BibitemShut
  {NoStop}%
\bibitem [{Note6()}]{Note6}%
  \BibitemOpen
  \bibinfo {note} {With a set of quantum states that form a complete basis set
  in a Hilbert space, any quantum state in this space can be written as a
  weighted summation of the basis states.}\BibitemShut {Stop}%
\bibitem [{Note7()}]{Note7}%
  \BibitemOpen
  \bibinfo {note} {As the quantum analog of a classical bit, a qubit represents
  a two-level quantum system such as a quantum spin. If the number of levels is
  higher than two, such quantum systems are often referred as
  ``qudits''.}\BibitemShut {Stop}%
\bibitem [{Note8()}]{Note8}%
  \BibitemOpen
  \bibinfo {note} {The Pauli operators ($\protect \hat {\sigma }^x$, $\protect
  \hat {\sigma }^y$, and $\protect \hat {\sigma }^z$) are three operators whose
  coefficients are given by three $(2 \times 2)$ matrices known as Pauli
  matrices. They are applied to describe the interactions, operations, and
  algebras of quantum spins. A frequently-used complete and orthogonal set of
  spin states is the eigenstates of one of the Pauli operators.}\BibitemShut
  {Stop}%
\bibitem [{\citenamefont {Kerenidis}\ \emph {et~al.}(2019)\citenamefont
  {Kerenidis}, \citenamefont {Landman}, \citenamefont {Luongo},\ and\
  \citenamefont {Prakash}}]{KLLP19Qmeans}%
  \BibitemOpen
  \bibfield  {author} {\bibinfo {author} {\bibfnamefont {I.}~\bibnamefont
  {Kerenidis}}, \bibinfo {author} {\bibfnamefont {J.}~\bibnamefont {Landman}},
  \bibinfo {author} {\bibfnamefont {A.}~\bibnamefont {Luongo}},\ and\ \bibinfo
  {author} {\bibfnamefont {A.}~\bibnamefont {Prakash}},\ }\bibfield  {title}
  {\bibinfo {title} {q-means: A quantum algorithm for unsupervised machine
  learning},\ }in\ \href@noop {} {\emph {\bibinfo {booktitle} {Advances in
  Neural Information Processing Systems}}},\ Vol.~\bibinfo {volume} {32},\
  \bibinfo {editor} {edited by\ \bibinfo {editor} {\bibfnamefont
  {H.}~\bibnamefont {Wallach}}, \bibinfo {editor} {\bibfnamefont
  {H.}~\bibnamefont {Larochelle}}, \bibinfo {editor} {\bibfnamefont
  {A.}~\bibnamefont {Beygelzimer}}, \bibinfo {editor} {\bibfnamefont
  {F.}~\bibnamefont {d\textquotesingle Alch\'{e}-Buc}}, \bibinfo {editor}
  {\bibfnamefont {E.}~\bibnamefont {Fox}},\ and\ \bibinfo {editor}
  {\bibfnamefont {R.}~\bibnamefont {Garnett}}}\ (\bibinfo  {publisher} {Curran
  Associates, Inc.},\ \bibinfo {year} {2019})\BibitemShut {NoStop}%
\bibitem [{\citenamefont {Wiebe}\ \emph {et~al.}(2015)\citenamefont {Wiebe},
  \citenamefont {Kapoor},\ and\ \citenamefont {Svore}}]{WKS15QNN}%
  \BibitemOpen
  \bibfield  {author} {\bibinfo {author} {\bibfnamefont {N.}~\bibnamefont
  {Wiebe}}, \bibinfo {author} {\bibfnamefont {A.}~\bibnamefont {Kapoor}},\ and\
  \bibinfo {author} {\bibfnamefont {K.~M.}\ \bibnamefont {Svore}},\ }\bibfield
  {title} {\bibinfo {title} {Quantum nearest-neighbor algorithms for machine
  learning},\ }\href
  {https://www.microsoft.com/en-us/research/publication/quantum-nearest-neighbor-algorithms-for-machine-learning/}
  {\bibfield  {journal} {\bibinfo  {journal} {Quantum Information and
  Computation}\ }\textbf {\bibinfo {volume} {15}},\ \bibinfo {pages} {318}
  (\bibinfo {year} {2015})}\BibitemShut {NoStop}%
\bibitem [{\citenamefont {Dang}\ \emph {et~al.}(2018)\citenamefont {Dang},
  \citenamefont {Jiang}, \citenamefont {Hu}, \citenamefont {Ji},\ and\
  \citenamefont {Zhang}}]{DJHJZ18QKNN}%
  \BibitemOpen
  \bibfield  {author} {\bibinfo {author} {\bibfnamefont {Y.}~\bibnamefont
  {Dang}}, \bibinfo {author} {\bibfnamefont {N.}~\bibnamefont {Jiang}},
  \bibinfo {author} {\bibfnamefont {H.}~\bibnamefont {Hu}}, \bibinfo {author}
  {\bibfnamefont {Z.}~\bibnamefont {Ji}},\ and\ \bibinfo {author}
  {\bibfnamefont {W.}~\bibnamefont {Zhang}},\ }\bibfield  {title} {\bibinfo
  {title} {Image classification based on quantum k-nearest-neighbor
  algorithm},\ }\href {https://doi.org/10.1007/s11128-018-2004-9} {\bibfield
  {journal} {\bibinfo  {journal} {Quantum Information Processing}\ }\textbf
  {\bibinfo {volume} {17}},\ \bibinfo {pages} {239} (\bibinfo {year}
  {2018})}\BibitemShut {NoStop}%
\bibitem [{\citenamefont {Rebentrost}\ \emph {et~al.}(2014)\citenamefont
  {Rebentrost}, \citenamefont {Mohseni},\ and\ \citenamefont
  {Lloyd}}]{RML14QSVM}%
  \BibitemOpen
  \bibfield  {author} {\bibinfo {author} {\bibfnamefont {P.}~\bibnamefont
  {Rebentrost}}, \bibinfo {author} {\bibfnamefont {M.}~\bibnamefont
  {Mohseni}},\ and\ \bibinfo {author} {\bibfnamefont {S.}~\bibnamefont
  {Lloyd}},\ }\bibfield  {title} {\bibinfo {title} {Quantum support vector
  machine for big data classification},\ }\href
  {https://doi.org/10.1103/PhysRevLett.113.130503} {\bibfield  {journal}
  {\bibinfo  {journal} {Physical Review Letters}\ }\textbf {\bibinfo {volume}
  {113}},\ \bibinfo {pages} {130503} (\bibinfo {year} {2014})}\BibitemShut
  {NoStop}%
\bibitem [{\citenamefont {Li}\ and\ \citenamefont {Ran}(2022)}]{LR22RLF}%
  \BibitemOpen
  \bibfield  {author} {\bibinfo {author} {\bibfnamefont {W.-M.}\ \bibnamefont
  {Li}}\ and\ \bibinfo {author} {\bibfnamefont {S.-J.}\ \bibnamefont {Ran}},\
  }\bibfield  {title} {\bibinfo {title} {Non-parametric semi-supervised
  learning in many-body hilbert space with rescaled logarithmic fidelity},\
  }\bibfield  {journal} {\bibinfo  {journal} {Mathematics}\ }\textbf {\bibinfo
  {volume} {10}},\ \href {https://doi.org/10.3390/math10060940}
  {10.3390/math10060940} (\bibinfo {year} {2022})\BibitemShut {NoStop}%
\bibitem [{\citenamefont {Zhou}\ \emph {et~al.}(2008)\citenamefont {Zhou},
  \citenamefont {Or\'{u}s},\ and\ \citenamefont {Vidal}}]{ZOV08fidTN}%
  \BibitemOpen
  \bibfield  {author} {\bibinfo {author} {\bibfnamefont {H.-Q.}\ \bibnamefont
  {Zhou}}, \bibinfo {author} {\bibfnamefont {R.}~\bibnamefont {Or\'{u}s}},\
  and\ \bibinfo {author} {\bibfnamefont {G.}~\bibnamefont {Vidal}},\ }\bibfield
   {title} {\bibinfo {title} {{Ground state fidelity from tensor network
  representations}},\ }\href {https://doi.org/10.1103/PhysRevLett.100.080601}
  {\bibfield  {journal} {\bibinfo  {journal} {Physical Review Letters}\
  }\textbf {\bibinfo {volume} {100}},\ \bibinfo {pages} {080601} (\bibinfo
  {year} {2008})}\BibitemShut {NoStop}%
\bibitem [{Note9()}]{Note9}%
  \BibitemOpen
  \bibinfo {note} {In the condensed matter physics and quantum many-body
  physics, quantum phases are defined to distinguish the states of quantum
  matter that possess intrinsically different properties (such as symmetries;
  one may refer to Sachdev S. Physics World 1999;12:33). Quantum phases are
  akin to the classical phases such as solids, liquids, and gases. In the
  paradigm of Landau and Ginzburg (see, e.g., a review article in Hohenberg P
  and Krekhov A. Physics Reports 2015;572:1–42), different phases of matter
  are separated by the phase transitions that generally exhibit certain
  singular properties. A phase transition can occur by changing the physical
  parameters, such as temperature for the solid-liquid transition of water, and
  the magnetic fields for the phase transitions of quantum
  magnets.}\BibitemShut {Stop}%
\bibitem [{\citenamefont {Yang}\ \emph {et~al.}(2021)\citenamefont {Yang},
  \citenamefont {Sun}, \citenamefont {Ran},\ and\ \citenamefont
  {Su}}]{YSRS21visual}%
  \BibitemOpen
  \bibfield  {author} {\bibinfo {author} {\bibfnamefont {Y.}~\bibnamefont
  {Yang}}, \bibinfo {author} {\bibfnamefont {Z.-Z.}\ \bibnamefont {Sun}},
  \bibinfo {author} {\bibfnamefont {S.-J.}\ \bibnamefont {Ran}},\ and\ \bibinfo
  {author} {\bibfnamefont {G.}~\bibnamefont {Su}},\ }\bibfield  {title}
  {\bibinfo {title} {Visualizing quantum phases and identifying quantum phase
  transitions by nonlinear dimensional reduction},\ }\href
  {https://doi.org/10.1103/PhysRevB.103.075106} {\bibfield  {journal} {\bibinfo
   {journal} {Physical Review B}\ }\textbf {\bibinfo {volume} {103}},\ \bibinfo
  {pages} {075106} (\bibinfo {year} {2021})}\BibitemShut {NoStop}%
\bibitem [{\citenamefont {Hinton}\ and\ \citenamefont
  {Roweis}(2003)}]{HS03tsne}%
  \BibitemOpen
  \bibfield  {author} {\bibinfo {author} {\bibfnamefont {G.~E.}\ \bibnamefont
  {Hinton}}\ and\ \bibinfo {author} {\bibfnamefont {S.~T.}\ \bibnamefont
  {Roweis}},\ }\bibfield  {title} {\bibinfo {title} {Stochastic neighbor
  embedding},\ }in\ \href@noop {} {\emph {\bibinfo {booktitle} {Advances in
  Neural Information Processing Systems 15}}},\ \bibinfo {editor} {edited by\
  \bibinfo {editor} {\bibfnamefont {S.}~\bibnamefont {Becker}}, \bibinfo
  {editor} {\bibfnamefont {S.}~\bibnamefont {Thrun}},\ and\ \bibinfo {editor}
  {\bibfnamefont {K.}~\bibnamefont {Obermayer}}}\ (\bibinfo  {publisher} {MIT
  Press},\ \bibinfo {year} {2003})\ pp.\ \bibinfo {pages}
  {857--864}\BibitemShut {NoStop}%
\bibitem [{Note10()}]{Note10}%
  \BibitemOpen
  \bibinfo {note} {In the Landau-Ginzburg paradigm, one way to identify quantum
  phases is to look at the corresponding order parameters. For instance, one
  may calculate the uniform magnetization to identify the ferromagnetic phase
  (where all spins are pointed in the same direction) for quantum magnets.
  Thus, such schemes for phase identification require the prior knowledge on
  which order parameter to look at. Developing the schemes with no need of such
  prior knowledge is an important topic to study quantum phases.}\BibitemShut
  {Stop}%
\bibitem [{\citenamefont {Horn}(2001{\natexlab{a}})}]{H01Qcluster}%
  \BibitemOpen
  \bibfield  {author} {\bibinfo {author} {\bibfnamefont {D.}~\bibnamefont
  {Horn}},\ }\bibfield  {title} {\bibinfo {title} {Clustering via hilbert
  space},\ }\href
  {https://doi.org/https://doi.org/10.1016/S0378-4371(01)00442-3} {\bibfield
  {journal} {\bibinfo  {journal} {Physica A: Statistical Mechanics and its
  Applications}\ }\textbf {\bibinfo {volume} {302}},\ \bibinfo {pages} {70}
  (\bibinfo {year} {2001}{\natexlab{a}})},\ \bibinfo {note} {proc. Int.
  Workshop on Frontiers in the Physics of Complex Systems}\BibitemShut
  {NoStop}%
\bibitem [{\citenamefont {Shi}\ \emph {et~al.}(2022)\citenamefont {Shi},
  \citenamefont {Shang},\ and\ \citenamefont {Guo}}]{SSG22TNML}%
  \BibitemOpen
  \bibfield  {author} {\bibinfo {author} {\bibfnamefont {X.}~\bibnamefont
  {Shi}}, \bibinfo {author} {\bibfnamefont {Y.}~\bibnamefont {Shang}},\ and\
  \bibinfo {author} {\bibfnamefont {C.}~\bibnamefont {Guo}},\ }\bibfield
  {title} {\bibinfo {title} {Clustering using matrix product states},\ }\href
  {https://doi.org/10.1103/PhysRevA.105.052424} {\bibfield  {journal} {\bibinfo
   {journal} {Physical Review A}\ }\textbf {\bibinfo {volume} {105}},\ \bibinfo
  {pages} {052424} (\bibinfo {year} {2022})}\BibitemShut {NoStop}%
\bibitem [{\citenamefont {Cortes}\ and\ \citenamefont {Vapnik}(1995)}]{C95SVM}%
  \BibitemOpen
  \bibfield  {author} {\bibinfo {author} {\bibfnamefont {C.}~\bibnamefont
  {Cortes}}\ and\ \bibinfo {author} {\bibfnamefont {V.}~\bibnamefont
  {Vapnik}},\ }\bibfield  {title} {\bibinfo {title} {Support-vector networks},\
  }\href {https://doi.org/10.1007/BF00994018} {\bibfield  {journal} {\bibinfo
  {journal} {Machine Learning}\ }\textbf {\bibinfo {volume} {20}},\ \bibinfo
  {pages} {273} (\bibinfo {year} {1995})}\BibitemShut {NoStop}%
\bibitem [{\citenamefont {Han}\ \emph {et~al.}(2018)\citenamefont {Han},
  \citenamefont {Wang}, \citenamefont {Fan}, \citenamefont {Wang},\ and\
  \citenamefont {Zhang}}]{HWFWZ17MPSML}%
  \BibitemOpen
  \bibfield  {author} {\bibinfo {author} {\bibfnamefont {Z.-Y.}\ \bibnamefont
  {Han}}, \bibinfo {author} {\bibfnamefont {J.}~\bibnamefont {Wang}}, \bibinfo
  {author} {\bibfnamefont {H.}~\bibnamefont {Fan}}, \bibinfo {author}
  {\bibfnamefont {L.}~\bibnamefont {Wang}},\ and\ \bibinfo {author}
  {\bibfnamefont {P.}~\bibnamefont {Zhang}},\ }\bibfield  {title} {\bibinfo
  {title} {Unsupervised generative modeling using matrix product states},\
  }\href {https://doi.org/10.1103/PhysRevX.8.031012} {\bibfield  {journal}
  {\bibinfo  {journal} {Physical Review X}\ }\textbf {\bibinfo {volume} {8}},\
  \bibinfo {pages} {031012} (\bibinfo {year} {2018})}\BibitemShut {NoStop}%
\bibitem [{\citenamefont {Torabian}\ and\ \citenamefont
  {Krems}(2023)}]{torabian2023compositional}%
  \BibitemOpen
  \bibfield  {author} {\bibinfo {author} {\bibfnamefont {E.}~\bibnamefont
  {Torabian}}\ and\ \bibinfo {author} {\bibfnamefont {R.~V.}\ \bibnamefont
  {Krems}},\ }\href@noop {} {\bibinfo {title} {Compositional optimization of
  quantum circuits for quantum kernels of support vector machines}} (\bibinfo
  {year} {2023}),\ \Eprint {https://arxiv.org/abs/2203.13848} {arXiv:2203.13848
  [quant-ph]} \BibitemShut {NoStop}%
\bibitem [{Note11()}]{Note11}%
  \BibitemOpen
  \bibinfo {note} {Schr{\"o}dinger equation is one of the most fundamental
  equations in quantum physics. It is a partial differential equation of
  multiple variables for the time-dependent problems, or eigenvalue equation
  for static problems. The complexity of solving Schr{\"o}dinger equation in
  general increases exponentially with the size of quantum system.}\BibitemShut
  {Stop}%
\bibitem [{\citenamefont {Horn}(2001{\natexlab{b}})}]{H01Qcluster0}%
  \BibitemOpen
  \bibfield  {author} {\bibinfo {author} {\bibfnamefont {D.}~\bibnamefont
  {Horn}},\ }\bibfield  {title} {\bibinfo {title} {Clustering via hilbert
  space},\ }\href
  {https://doi.org/https://doi.org/10.1016/S0378-4371(01)00442-3} {\bibfield
  {journal} {\bibinfo  {journal} {Physica A: Statistical Mechanics and its
  Applications}\ }\textbf {\bibinfo {volume} {302}},\ \bibinfo {pages} {70}
  (\bibinfo {year} {2001}{\natexlab{b}})},\ \bibinfo {note} {proc. Int.
  Workshop on Frontiers in the Physics of Complex Systems}\BibitemShut
  {NoStop}%
\bibitem [{\citenamefont {Weinstein}\ and\ \citenamefont
  {Horn}(2009)}]{WH09Qcluster1}%
  \BibitemOpen
  \bibfield  {author} {\bibinfo {author} {\bibfnamefont {M.}~\bibnamefont
  {Weinstein}}\ and\ \bibinfo {author} {\bibfnamefont {D.}~\bibnamefont
  {Horn}},\ }\bibfield  {title} {\bibinfo {title} {Dynamic quantum clustering:
  A method for visual exploration of structures in data},\ }\href
  {https://doi.org/10.1103/PhysRevE.80.066117} {\bibfield  {journal} {\bibinfo
  {journal} {Physical Review E}\ }\textbf {\bibinfo {volume} {80}},\ \bibinfo
  {pages} {066117} (\bibinfo {year} {2009})}\BibitemShut {NoStop}%
\bibitem [{\citenamefont {Pronchik}\ and\ \citenamefont
  {Williams}(2003)}]{Pronchik2003}%
  \BibitemOpen
  \bibfield  {author} {\bibinfo {author} {\bibfnamefont {J.~N.}\ \bibnamefont
  {Pronchik}}\ and\ \bibinfo {author} {\bibfnamefont {B.~W.}\ \bibnamefont
  {Williams}},\ }\bibfield  {title} {\bibinfo {title} {Exactly solvable quantum
  mechanical potentials: An alternative approach},\ }\href
  {https://doi.org/10.1021/ed080p918} {\bibfield  {journal} {\bibinfo
  {journal} {Journal of Chemical Education}\ }\textbf {\bibinfo {volume}
  {80}},\ \bibinfo {pages} {918} (\bibinfo {year} {2003})}\BibitemShut
  {NoStop}%
\bibitem [{\citenamefont {Sehanobish}\ \emph {et~al.}(2020)\citenamefont
  {Sehanobish}, \citenamefont {Corzo}, \citenamefont {Kara},\ and\
  \citenamefont {van Dijk}}]{sehanobish2020learning}%
  \BibitemOpen
  \bibfield  {author} {\bibinfo {author} {\bibfnamefont {A.}~\bibnamefont
  {Sehanobish}}, \bibinfo {author} {\bibfnamefont {H.~H.}\ \bibnamefont
  {Corzo}}, \bibinfo {author} {\bibfnamefont {O.}~\bibnamefont {Kara}},\ and\
  \bibinfo {author} {\bibfnamefont {D.}~\bibnamefont {van Dijk}},\ }\href@noop
  {} {\bibinfo {title} {Learning potentials of quantum systems using deep
  neural networks}} (\bibinfo {year} {2020}),\ \Eprint
  {https://arxiv.org/abs/2006.13297} {arXiv:2006.13297 [cs.LG]} \BibitemShut
  {NoStop}%
\bibitem [{\citenamefont {Hong}\ \emph {et~al.}(2021)\citenamefont {Hong},
  \citenamefont {Zhou}, \citenamefont {Xi}, \citenamefont {Hu}, \citenamefont
  {Ji},\ and\ \citenamefont {Ran}}]{HZHX+21MPNN}%
  \BibitemOpen
  \bibfield  {author} {\bibinfo {author} {\bibfnamefont {R.}~\bibnamefont
  {Hong}}, \bibinfo {author} {\bibfnamefont {P.-F.}\ \bibnamefont {Zhou}},
  \bibinfo {author} {\bibfnamefont {B.}~\bibnamefont {Xi}}, \bibinfo {author}
  {\bibfnamefont {J.}~\bibnamefont {Hu}}, \bibinfo {author} {\bibfnamefont
  {A.-C.}\ \bibnamefont {Ji}},\ and\ \bibinfo {author} {\bibfnamefont {S.-J.}\
  \bibnamefont {Ran}},\ }\bibfield  {title} {\bibinfo {title} {{Predicting
  quantum potentials by deep neural network and metropolis sampling}},\ }\href
  {https://doi.org/10.21468/SciPostPhysCore.4.3.022} {\bibfield  {journal}
  {\bibinfo  {journal} {SciPost Physics Core}\ }\textbf {\bibinfo {volume}
  {4}},\ \bibinfo {pages} {022} (\bibinfo {year} {2021})}\BibitemShut {NoStop}%
\bibitem [{\citenamefont {Stoudenmire}(2018)}]{S18MERAML}%
  \BibitemOpen
  \bibfield  {author} {\bibinfo {author} {\bibfnamefont {E.~M.}\ \bibnamefont
  {Stoudenmire}},\ }\bibfield  {title} {\bibinfo {title} {Learning relevant
  features of data with multi-scale tensor networks},\ }\href
  {https://doi.org/10.1088/2058-9565/aaba1a} {\bibfield  {journal} {\bibinfo
  {journal} {Quantum Science and Technology}\ }\textbf {\bibinfo {volume}
  {3}},\ \bibinfo {pages} {034003} (\bibinfo {year} {2018})}\BibitemShut
  {NoStop}%
\bibitem [{\citenamefont {Liu}\ \emph {et~al.}(2019)\citenamefont {Liu},
  \citenamefont {Ran}, \citenamefont {Wittek}, \citenamefont {Peng},
  \citenamefont {Garc{\'{\i}}a}, \citenamefont {Su},\ and\ \citenamefont
  {Lewenstein}}]{LRWP+17MLTN}%
  \BibitemOpen
  \bibfield  {author} {\bibinfo {author} {\bibfnamefont {D.}~\bibnamefont
  {Liu}}, \bibinfo {author} {\bibfnamefont {S.-J.}\ \bibnamefont {Ran}},
  \bibinfo {author} {\bibfnamefont {P.}~\bibnamefont {Wittek}}, \bibinfo
  {author} {\bibfnamefont {C.}~\bibnamefont {Peng}}, \bibinfo {author}
  {\bibfnamefont {R.~B.}\ \bibnamefont {Garc{\'{\i}}a}}, \bibinfo {author}
  {\bibfnamefont {G.}~\bibnamefont {Su}},\ and\ \bibinfo {author}
  {\bibfnamefont {M.}~\bibnamefont {Lewenstein}},\ }\bibfield  {title}
  {\bibinfo {title} {Machine learning by unitary tensor network of hierarchical
  tree structure},\ }\href {https://doi.org/10.1088/1367-2630/ab31ef}
  {\bibfield  {journal} {\bibinfo  {journal} {New Journal of Physics}\ }\textbf
  {\bibinfo {volume} {21}},\ \bibinfo {pages} {073059} (\bibinfo {year}
  {2019})}\BibitemShut {NoStop}%
\bibitem [{\citenamefont {Cheng}\ \emph {et~al.}(2021)\citenamefont {Cheng},
  \citenamefont {Wang},\ and\ \citenamefont {Zhang}}]{CWZ21PEPSML}%
  \BibitemOpen
  \bibfield  {author} {\bibinfo {author} {\bibfnamefont {S.}~\bibnamefont
  {Cheng}}, \bibinfo {author} {\bibfnamefont {L.}~\bibnamefont {Wang}},\ and\
  \bibinfo {author} {\bibfnamefont {P.}~\bibnamefont {Zhang}},\ }\bibfield
  {title} {\bibinfo {title} {Supervised learning with projected entangled pair
  states},\ }\href {https://doi.org/10.1103/PhysRevB.103.125117} {\bibfield
  {journal} {\bibinfo  {journal} {Physical Review B}\ }\textbf {\bibinfo
  {volume} {103}},\ \bibinfo {pages} {125117} (\bibinfo {year}
  {2021})}\BibitemShut {NoStop}%
\bibitem [{\citenamefont {Liu}\ \emph {et~al.}(2021{\natexlab{a}})\citenamefont
  {Liu}, \citenamefont {Li}, \citenamefont {Zhang}, \citenamefont {Lewenstein},
  \citenamefont {Su},\ and\ \citenamefont {Ran}}]{LZLR18entTNML}%
  \BibitemOpen
  \bibfield  {author} {\bibinfo {author} {\bibfnamefont {Y.}~\bibnamefont
  {Liu}}, \bibinfo {author} {\bibfnamefont {W.-J.}\ \bibnamefont {Li}},
  \bibinfo {author} {\bibfnamefont {X.}~\bibnamefont {Zhang}}, \bibinfo
  {author} {\bibfnamefont {M.}~\bibnamefont {Lewenstein}}, \bibinfo {author}
  {\bibfnamefont {G.}~\bibnamefont {Su}},\ and\ \bibinfo {author}
  {\bibfnamefont {S.-J.}\ \bibnamefont {Ran}},\ }\bibfield  {title} {\bibinfo
  {title} {Entanglement-based feature extraction by tensor network machine
  learning},\ }\bibfield  {journal} {\bibinfo  {journal} {Frontiers in Applied
  Mathematics and Statistics}\ }\textbf {\bibinfo {volume} {7}},\ \href
  {https://doi.org/10.3389/fams.2021.716044} {10.3389/fams.2021.716044}
  (\bibinfo {year} {2021}{\natexlab{a}})\BibitemShut {NoStop}%
\bibitem [{\citenamefont {Bai}\ \emph {et~al.}(2022)\citenamefont {Bai},
  \citenamefont {Tang},\ and\ \citenamefont {Ran}}]{BTR22CPLML}%
  \BibitemOpen
  \bibfield  {author} {\bibinfo {author} {\bibfnamefont {S.-C.}\ \bibnamefont
  {Bai}}, \bibinfo {author} {\bibfnamefont {Y.-C.}\ \bibnamefont {Tang}},\ and\
  \bibinfo {author} {\bibfnamefont {S.-J.}\ \bibnamefont {Ran}},\ }\bibfield
  {title} {\bibinfo {title} {Unsupervised recognition of informative features
  via tensor network machine learning and quantum entanglement variations},\
  }\href {https://doi.org/10.1088/0256-307X/39/10/100701} {\bibfield  {journal}
  {\bibinfo  {journal} {Chinese Physics Letters}\ }\textbf {\bibinfo {volume}
  {39}},\ \bibinfo {eid} {100701} (\bibinfo {year} {2022})}\BibitemShut
  {NoStop}%
\bibitem [{\citenamefont {Wang}\ \emph
  {et~al.}(2021{\natexlab{b}})\citenamefont {Wang}, \citenamefont {Xiao},
  \citenamefont {Yi}, \citenamefont {Ran},\ and\ \citenamefont
  {Xue}}]{WXYRX21MPSphoto}%
  \BibitemOpen
  \bibfield  {author} {\bibinfo {author} {\bibfnamefont {K.}~\bibnamefont
  {Wang}}, \bibinfo {author} {\bibfnamefont {L.}~\bibnamefont {Xiao}}, \bibinfo
  {author} {\bibfnamefont {W.}~\bibnamefont {Yi}}, \bibinfo {author}
  {\bibfnamefont {S.-J.}\ \bibnamefont {Ran}},\ and\ \bibinfo {author}
  {\bibfnamefont {P.}~\bibnamefont {Xue}},\ }\bibfield  {title} {\bibinfo
  {title} {Experimental realization of a quantum image classifier via
  tensor-network-based machine learning},\ }\href
  {https://doi.org/10.1364/PRJ.434217} {\bibfield  {journal} {\bibinfo
  {journal} {Photonics Research}\ }\textbf {\bibinfo {volume} {9}},\ \bibinfo
  {pages} {2332} (\bibinfo {year} {2021}{\natexlab{b}})}\BibitemShut {NoStop}%
\bibitem [{\citenamefont {Ran}\ \emph {et~al.}(2020{\natexlab{b}})\citenamefont
  {Ran}, \citenamefont {Sun}, \citenamefont {Fei}, \citenamefont {Su},\ and\
  \citenamefont {Lewenstein}}]{RSF+20TNCS}%
  \BibitemOpen
  \bibfield  {author} {\bibinfo {author} {\bibfnamefont {S.-J.}\ \bibnamefont
  {Ran}}, \bibinfo {author} {\bibfnamefont {Z.-Z.}\ \bibnamefont {Sun}},
  \bibinfo {author} {\bibfnamefont {S.-M.}\ \bibnamefont {Fei}}, \bibinfo
  {author} {\bibfnamefont {G.}~\bibnamefont {Su}},\ and\ \bibinfo {author}
  {\bibfnamefont {M.}~\bibnamefont {Lewenstein}},\ }\bibfield  {title}
  {\bibinfo {title} {Tensor network compressed sensing with unsupervised
  machine learning},\ }\href {https://doi.org/10.1103/PhysRevResearch.2.033293}
  {\bibfield  {journal} {\bibinfo  {journal} {Physical Review Research}\
  }\textbf {\bibinfo {volume} {2}},\ \bibinfo {pages} {033293} (\bibinfo {year}
  {2020}{\natexlab{b}})}\BibitemShut {NoStop}%
\bibitem [{\citenamefont {Sun}\ \emph {et~al.}(2020{\natexlab{c}})\citenamefont
  {Sun}, \citenamefont {Ran},\ and\ \citenamefont {Su}}]{SRG19preperation}%
  \BibitemOpen
  \bibfield  {author} {\bibinfo {author} {\bibfnamefont {Z.-Z.}\ \bibnamefont
  {Sun}}, \bibinfo {author} {\bibfnamefont {S.-J.}\ \bibnamefont {Ran}},\ and\
  \bibinfo {author} {\bibfnamefont {G.}~\bibnamefont {Su}},\ }\bibfield
  {title} {\bibinfo {title} {Tangent-space gradient optimization of tensor
  network for machine learning},\ }\href
  {https://doi.org/10.1103/PhysRevE.102.012152} {\bibfield  {journal} {\bibinfo
   {journal} {Physical Review E}\ }\textbf {\bibinfo {volume} {102}},\ \bibinfo
  {pages} {012152} (\bibinfo {year} {2020}{\natexlab{c}})}\BibitemShut
  {NoStop}%
\bibitem [{\citenamefont {Wang}\ \emph
  {et~al.}(2020{\natexlab{b}})\citenamefont {Wang}, \citenamefont {Roberts},
  \citenamefont {Vidal},\ and\ \citenamefont {Leichenauer}}]{WRVL20TNAD}%
  \BibitemOpen
  \bibfield  {author} {\bibinfo {author} {\bibfnamefont {J.}~\bibnamefont
  {Wang}}, \bibinfo {author} {\bibfnamefont {C.}~\bibnamefont {Roberts}},
  \bibinfo {author} {\bibfnamefont {G.}~\bibnamefont {Vidal}},\ and\ \bibinfo
  {author} {\bibfnamefont {S.}~\bibnamefont {Leichenauer}},\ }\href@noop {}
  {\bibinfo {title} {Anomaly detection with tensor networks}} (\bibinfo {year}
  {2020}{\natexlab{b}}),\ \Eprint {https://arxiv.org/abs/2006.02516}
  {arXiv:2006.02516 [cs.LG]} \BibitemShut {NoStop}%
\bibitem [{\citenamefont {Hong}\ \emph {et~al.}(2022)\citenamefont {Hong},
  \citenamefont {Xiao}, \citenamefont {Hu}, \citenamefont {Ji},\ and\
  \citenamefont {Ran}}]{HXHJR22FTN}%
  \BibitemOpen
  \bibfield  {author} {\bibinfo {author} {\bibfnamefont {R.}~\bibnamefont
  {Hong}}, \bibinfo {author} {\bibfnamefont {Y.-X.}\ \bibnamefont {Xiao}},
  \bibinfo {author} {\bibfnamefont {J.}~\bibnamefont {Hu}}, \bibinfo {author}
  {\bibfnamefont {A.-C.}\ \bibnamefont {Ji}},\ and\ \bibinfo {author}
  {\bibfnamefont {S.-J.}\ \bibnamefont {Ran}},\ }\bibfield  {title} {\bibinfo
  {title} {Functional tensor network solving many-body schr\"odinger
  equation},\ }\href {https://doi.org/10.1103/PhysRevB.105.165116} {\bibfield
  {journal} {\bibinfo  {journal} {Physical Review B}\ }\textbf {\bibinfo
  {volume} {105}},\ \bibinfo {pages} {165116} (\bibinfo {year}
  {2022})}\BibitemShut {NoStop}%
\bibitem [{\citenamefont {Hao}\ \emph {et~al.}(2022)\citenamefont {Hao},
  \citenamefont {Huang}, \citenamefont {Jia},\ and\ \citenamefont
  {Peng}}]{HHJP22TNCCO}%
  \BibitemOpen
  \bibfield  {author} {\bibinfo {author} {\bibfnamefont {T.}~\bibnamefont
  {Hao}}, \bibinfo {author} {\bibfnamefont {X.}~\bibnamefont {Huang}}, \bibinfo
  {author} {\bibfnamefont {C.}~\bibnamefont {Jia}},\ and\ \bibinfo {author}
  {\bibfnamefont {C.}~\bibnamefont {Peng}},\ }\bibfield  {title} {\bibinfo
  {title} {A quantum-inspired tensor network algorithm for constrained
  combinatorial optimization problems},\ }\bibfield  {journal} {\bibinfo
  {journal} {Frontiers in Physics}\ }\textbf {\bibinfo {volume} {10}},\ \href
  {https://doi.org/10.3389/fphy.2022.906590} {10.3389/fphy.2022.906590}
  (\bibinfo {year} {2022})\BibitemShut {NoStop}%
\bibitem [{\citenamefont {Liu}\ \emph {et~al.}(2023)\citenamefont {Liu},
  \citenamefont {Gao}, \citenamefont {Cain}, \citenamefont {Lukin},\ and\
  \citenamefont {Wang}}]{LGCLW23TNCCO}%
  \BibitemOpen
  \bibfield  {author} {\bibinfo {author} {\bibfnamefont {J.-G.}\ \bibnamefont
  {Liu}}, \bibinfo {author} {\bibfnamefont {X.}~\bibnamefont {Gao}}, \bibinfo
  {author} {\bibfnamefont {M.}~\bibnamefont {Cain}}, \bibinfo {author}
  {\bibfnamefont {M.~D.}\ \bibnamefont {Lukin}},\ and\ \bibinfo {author}
  {\bibfnamefont {S.-T.}\ \bibnamefont {Wang}},\ }\bibfield  {title} {\bibinfo
  {title} {Computing solution space properties of combinatorial optimization
  problems via generic tensor networks},\ }\href
  {https://doi.org/10.1137/22M1501787} {\bibfield  {journal} {\bibinfo
  {journal} {SIAM Journal on Scientific Computing}\ }\textbf {\bibinfo {volume}
  {45}},\ \bibinfo {pages} {A1239} (\bibinfo {year} {2023})}\BibitemShut
  {NoStop}%
\bibitem [{\citenamefont {Lopez-Piqueres}\ \emph {et~al.}(2023)\citenamefont
  {Lopez-Piqueres}, \citenamefont {Chen},\ and\ \citenamefont
  {Perdomo-Ortiz}}]{LCP23TNCCO}%
  \BibitemOpen
  \bibfield  {author} {\bibinfo {author} {\bibfnamefont {J.}~\bibnamefont
  {Lopez-Piqueres}}, \bibinfo {author} {\bibfnamefont {J.}~\bibnamefont
  {Chen}},\ and\ \bibinfo {author} {\bibfnamefont {A.}~\bibnamefont
  {Perdomo-Ortiz}},\ }\href@noop {} {\bibinfo {title} {Symmetric tensor
  networks for generative modeling and constrained combinatorial optimization}}
  (\bibinfo {year} {2023}),\ \Eprint {https://arxiv.org/abs/2211.09121}
  {arXiv:2211.09121 [quant-ph]} \BibitemShut {NoStop}%
\bibitem [{MNI()}]{MNIST_web}%
  \BibitemOpen
  \href@noop {} {}\bibinfo {note}
  {\url{http://yann.lecun.com/exdb/mnist}}\BibitemShut {NoStop}%
\bibitem [{\citenamefont {Banchi}\ \emph {et~al.}(2021)\citenamefont {Banchi},
  \citenamefont {Pereira},\ and\ \citenamefont {Pirandola}}]{BPP21MLQinfo}%
  \BibitemOpen
  \bibfield  {author} {\bibinfo {author} {\bibfnamefont {L.}~\bibnamefont
  {Banchi}}, \bibinfo {author} {\bibfnamefont {J.}~\bibnamefont {Pereira}},\
  and\ \bibinfo {author} {\bibfnamefont {S.}~\bibnamefont {Pirandola}},\
  }\bibfield  {title} {\bibinfo {title} {Generalization in quantum machine
  learning: A quantum information standpoint},\ }\href
  {https://doi.org/10.1103/PRXQuantum.2.040321} {\bibfield  {journal} {\bibinfo
   {journal} {PRX Quantum}\ }\textbf {\bibinfo {volume} {2}},\ \bibinfo {pages}
  {040321} (\bibinfo {year} {2021})}\BibitemShut {NoStop}%
\bibitem [{\citenamefont {Strashko}\ and\ \citenamefont
  {Stoudenmire}(2022)}]{strashko2022generalization}%
  \BibitemOpen
  \bibfield  {author} {\bibinfo {author} {\bibfnamefont {A.}~\bibnamefont
  {Strashko}}\ and\ \bibinfo {author} {\bibfnamefont {E.~M.}\ \bibnamefont
  {Stoudenmire}},\ }\href@noop {} {\bibinfo {title} {Generalization and
  overfitting in matrix product state machine learning architectures}}
  (\bibinfo {year} {2022}),\ \Eprint {https://arxiv.org/abs/2208.04372}
  {arXiv:2208.04372 [cs.LG]} \BibitemShut {NoStop}%
\bibitem [{\citenamefont {Convy}\ \emph {et~al.}(2022)\citenamefont {Convy},
  \citenamefont {Huggins}, \citenamefont {Liao},\ and\ \citenamefont
  {Whaley}}]{CHLW22TNML}%
  \BibitemOpen
  \bibfield  {author} {\bibinfo {author} {\bibfnamefont {I.}~\bibnamefont
  {Convy}}, \bibinfo {author} {\bibfnamefont {W.}~\bibnamefont {Huggins}},
  \bibinfo {author} {\bibfnamefont {H.}~\bibnamefont {Liao}},\ and\ \bibinfo
  {author} {\bibfnamefont {K.~B.}\ \bibnamefont {Whaley}},\ }\bibfield  {title}
  {\bibinfo {title} {Mutual information scaling for tensor network machine
  learning},\ }\href {https://doi.org/10.1088/2632-2153/ac44a9} {\bibfield
  {journal} {\bibinfo  {journal} {Machine Learning: Science and Technology}\
  }\textbf {\bibinfo {volume} {3}},\ \bibinfo {pages} {015017} (\bibinfo {year}
  {2022})}\BibitemShut {NoStop}%
\bibitem [{\citenamefont {Gao}\ \emph {et~al.}(2022)\citenamefont {Gao},
  \citenamefont {Anschuetz}, \citenamefont {Wang}, \citenamefont {Cirac},\ and\
  \citenamefont {Lukin}}]{GAWCL22TNML}%
  \BibitemOpen
  \bibfield  {author} {\bibinfo {author} {\bibfnamefont {X.}~\bibnamefont
  {Gao}}, \bibinfo {author} {\bibfnamefont {E.~R.}\ \bibnamefont {Anschuetz}},
  \bibinfo {author} {\bibfnamefont {S.-T.}\ \bibnamefont {Wang}}, \bibinfo
  {author} {\bibfnamefont {J.~I.}\ \bibnamefont {Cirac}},\ and\ \bibinfo
  {author} {\bibfnamefont {M.~D.}\ \bibnamefont {Lukin}},\ }\bibfield  {title}
  {\bibinfo {title} {Enhancing generative models via quantum correlations},\
  }\href {https://doi.org/10.1103/PhysRevX.12.021037} {\bibfield  {journal}
  {\bibinfo  {journal} {Physical Review X}\ }\textbf {\bibinfo {volume} {12}},\
  \bibinfo {pages} {021037} (\bibinfo {year} {2022})}\BibitemShut {NoStop}%
\bibitem [{\citenamefont {Liao}\ \emph {et~al.}(2023)\citenamefont {Liao},
  \citenamefont {Convy}, \citenamefont {Yang},\ and\ \citenamefont
  {Whaley}}]{LCYW23TNML}%
  \BibitemOpen
  \bibfield  {author} {\bibinfo {author} {\bibfnamefont {H.}~\bibnamefont
  {Liao}}, \bibinfo {author} {\bibfnamefont {I.}~\bibnamefont {Convy}},
  \bibinfo {author} {\bibfnamefont {Z.}~\bibnamefont {Yang}},\ and\ \bibinfo
  {author} {\bibfnamefont {K.~B.}\ \bibnamefont {Whaley}},\ }\bibfield  {title}
  {\bibinfo {title} {Decohering tensor network quantum machine learning
  models},\ }\href {https://doi.org/10.1007/s42484-022-00095-9} {\bibfield
  {journal} {\bibinfo  {journal} {Quantum Machine Intelligence}\ }\textbf
  {\bibinfo {volume} {5}},\ \bibinfo {pages} {7} (\bibinfo {year}
  {2023})}\BibitemShut {NoStop}%
\bibitem [{\citenamefont {Schirmer}\ \emph {et~al.}(2001)\citenamefont
  {Schirmer}, \citenamefont {Fu},\ and\ \citenamefont
  {Solomon}}]{PhysRevA.63.063410}%
  \BibitemOpen
  \bibfield  {author} {\bibinfo {author} {\bibfnamefont {S.~G.}\ \bibnamefont
  {Schirmer}}, \bibinfo {author} {\bibfnamefont {H.}~\bibnamefont {Fu}},\ and\
  \bibinfo {author} {\bibfnamefont {A.~I.}\ \bibnamefont {Solomon}},\
  }\bibfield  {title} {\bibinfo {title} {Complete controllability of quantum
  systems},\ }\href {https://doi.org/10.1103/PhysRevA.63.063410} {\bibfield
  {journal} {\bibinfo  {journal} {Physical Review A}\ }\textbf {\bibinfo
  {volume} {63}},\ \bibinfo {pages} {063410} (\bibinfo {year}
  {2001})}\BibitemShut {NoStop}%
\bibitem [{\citenamefont {Deng}\ \emph {et~al.}(2017)\citenamefont {Deng},
  \citenamefont {Li},\ and\ \citenamefont {Das~Sarma}}]{DLD17NNent}%
  \BibitemOpen
  \bibfield  {author} {\bibinfo {author} {\bibfnamefont {D.-L.}\ \bibnamefont
  {Deng}}, \bibinfo {author} {\bibfnamefont {X.}~\bibnamefont {Li}},\ and\
  \bibinfo {author} {\bibfnamefont {S.}~\bibnamefont {Das~Sarma}},\ }\bibfield
  {title} {\bibinfo {title} {Quantum entanglement in neural network states},\
  }\href {https://doi.org/10.1103/PhysRevX.7.021021} {\bibfield  {journal}
  {\bibinfo  {journal} {Physical Review X}\ }\textbf {\bibinfo {volume} {7}},\
  \bibinfo {pages} {021021} (\bibinfo {year} {2017})}\BibitemShut {NoStop}%
\bibitem [{\citenamefont {Jia}\ \emph {et~al.}(2020)\citenamefont {Jia},
  \citenamefont {Wei}, \citenamefont {Wu}, \citenamefont {Guo},\ and\
  \citenamefont {Guo}}]{JWWGG20QNNent}%
  \BibitemOpen
  \bibfield  {author} {\bibinfo {author} {\bibfnamefont {Z.-A.}\ \bibnamefont
  {Jia}}, \bibinfo {author} {\bibfnamefont {L.}~\bibnamefont {Wei}}, \bibinfo
  {author} {\bibfnamefont {Y.-C.}\ \bibnamefont {Wu}}, \bibinfo {author}
  {\bibfnamefont {G.-C.}\ \bibnamefont {Guo}},\ and\ \bibinfo {author}
  {\bibfnamefont {G.-P.}\ \bibnamefont {Guo}},\ }\bibfield  {title} {\bibinfo
  {title} {Entanglement area law for shallow and deep quantum neural network
  states},\ }\href {https://doi.org/10.1088/1367-2630/ab8262} {\bibfield
  {journal} {\bibinfo  {journal} {New Journal of Physics}\ }\textbf {\bibinfo
  {volume} {22}},\ \bibinfo {pages} {053022} (\bibinfo {year}
  {2020})}\BibitemShut {NoStop}%
\bibitem [{\citenamefont {Larocca}\ \emph {et~al.}(2023)\citenamefont
  {Larocca}, \citenamefont {Ju}, \citenamefont {Garc{\'i}a-Mart{\'i}n},
  \citenamefont {Coles},\ and\ \citenamefont {Cerezo}}]{Larocca2023}%
  \BibitemOpen
  \bibfield  {author} {\bibinfo {author} {\bibfnamefont {M.}~\bibnamefont
  {Larocca}}, \bibinfo {author} {\bibfnamefont {N.}~\bibnamefont {Ju}},
  \bibinfo {author} {\bibfnamefont {D.}~\bibnamefont {Garc{\'i}a-Mart{\'i}n}},
  \bibinfo {author} {\bibfnamefont {P.~J.}\ \bibnamefont {Coles}},\ and\
  \bibinfo {author} {\bibfnamefont {M.}~\bibnamefont {Cerezo}},\ }\bibfield
  {title} {\bibinfo {title} {Theory of overparametrization in quantum neural
  networks},\ }\href {https://doi.org/10.1038/s43588-023-00467-6} {\bibfield
  {journal} {\bibinfo  {journal} {Nature Computational Science}\ }\textbf
  {\bibinfo {volume} {3}},\ \bibinfo {pages} {542} (\bibinfo {year}
  {2023})}\BibitemShut {NoStop}%
\bibitem [{\citenamefont {Larocca}\ \emph {et~al.}(2022)\citenamefont
  {Larocca}, \citenamefont {Czarnik}, \citenamefont {Sharma}, \citenamefont
  {Muraleedharan}, \citenamefont {Coles},\ and\ \citenamefont
  {Cerezo}}]{Larocca2022diagnosingbarren}%
  \BibitemOpen
  \bibfield  {author} {\bibinfo {author} {\bibfnamefont {M.}~\bibnamefont
  {Larocca}}, \bibinfo {author} {\bibfnamefont {P.}~\bibnamefont {Czarnik}},
  \bibinfo {author} {\bibfnamefont {K.}~\bibnamefont {Sharma}}, \bibinfo
  {author} {\bibfnamefont {G.}~\bibnamefont {Muraleedharan}}, \bibinfo {author}
  {\bibfnamefont {P.~J.}\ \bibnamefont {Coles}},\ and\ \bibinfo {author}
  {\bibfnamefont {M.}~\bibnamefont {Cerezo}},\ }\bibfield  {title} {\bibinfo
  {title} {Diagnosing barren plateaus with tools from quantum optimal
  control},\ }\href {https://doi.org/10.22331/q-2022-09-29-824} {\bibfield
  {journal} {\bibinfo  {journal} {{Quantum}}\ }\textbf {\bibinfo {volume}
  {6}},\ \bibinfo {pages} {824} (\bibinfo {year} {2022})}\BibitemShut {NoStop}%
\bibitem [{\citenamefont {McClean}\ \emph {et~al.}(2018)\citenamefont
  {McClean}, \citenamefont {Boixo}, \citenamefont {Smelyanskiy}, \citenamefont
  {Babbush},\ and\ \citenamefont {Neven}}]{MBSBN18BarrenP}%
  \BibitemOpen
  \bibfield  {author} {\bibinfo {author} {\bibfnamefont {J.~R.}\ \bibnamefont
  {McClean}}, \bibinfo {author} {\bibfnamefont {S.}~\bibnamefont {Boixo}},
  \bibinfo {author} {\bibfnamefont {V.~N.}\ \bibnamefont {Smelyanskiy}},
  \bibinfo {author} {\bibfnamefont {R.}~\bibnamefont {Babbush}},\ and\ \bibinfo
  {author} {\bibfnamefont {H.}~\bibnamefont {Neven}},\ }\bibfield  {title}
  {\bibinfo {title} {Barren plateaus in quantum neural network training
  landscapes},\ }\href {https://doi.org/10.1038/s41467-018-07090-4} {\bibfield
  {journal} {\bibinfo  {journal} {Nature Communications}\ }\textbf {\bibinfo
  {volume} {9}},\ \bibinfo {pages} {4812} (\bibinfo {year} {2018})}\BibitemShut
  {NoStop}%
\bibitem [{\citenamefont {Verstraete}\ \emph {et~al.}(2004)\citenamefont
  {Verstraete}, \citenamefont {Garc\'ia-Ripoll},\ and\ \citenamefont
  {Cirac}}]{VGC04MPDO}%
  \BibitemOpen
  \bibfield  {author} {\bibinfo {author} {\bibfnamefont {F.}~\bibnamefont
  {Verstraete}}, \bibinfo {author} {\bibfnamefont {J.~J.}\ \bibnamefont
  {Garc\'ia-Ripoll}},\ and\ \bibinfo {author} {\bibfnamefont {J.~I.}\
  \bibnamefont {Cirac}},\ }\bibfield  {title} {\bibinfo {title} {{Matrix
  product density operators: Simulation of finite-temperature and dissipative
  systems}},\ }\href {https://doi.org/10.1103/PhysRevLett.93.207204} {\bibfield
   {journal} {\bibinfo  {journal} {Physical Review Letters}\ }\textbf {\bibinfo
  {volume} {93}},\ \bibinfo {pages} {207204} (\bibinfo {year}
  {2004})}\BibitemShut {NoStop}%
\bibitem [{\citenamefont {Zwolak}\ and\ \citenamefont {Vidal}(2004)}]{ZV04MPO}%
  \BibitemOpen
  \bibfield  {author} {\bibinfo {author} {\bibfnamefont {M.}~\bibnamefont
  {Zwolak}}\ and\ \bibinfo {author} {\bibfnamefont {G.}~\bibnamefont {Vidal}},\
  }\bibfield  {title} {\bibinfo {title} {{Mixed-state dynamics in
  one-dimensional quantum lattice systems: A time-dependent superoperator
  renormalization algorithm}},\ }\href
  {https://doi.org/10.1103/PhysRevLett.93.207205} {\bibfield  {journal}
  {\bibinfo  {journal} {Physical Review Letters}\ }\textbf {\bibinfo {volume}
  {93}},\ \bibinfo {pages} {207205} (\bibinfo {year} {2004})}\BibitemShut
  {NoStop}%
\bibitem [{\citenamefont {Ba\~{n}uls}\ \emph {et~al.}(2009)\citenamefont
  {Ba\~{n}uls}, \citenamefont {Hastings}, \citenamefont {Verstraete},\ and\
  \citenamefont {Cirac}}]{BHVC09folding}%
  \BibitemOpen
  \bibfield  {author} {\bibinfo {author} {\bibfnamefont {M.-C.}\ \bibnamefont
  {Ba\~{n}uls}}, \bibinfo {author} {\bibfnamefont {M.~B.}\ \bibnamefont
  {Hastings}}, \bibinfo {author} {\bibfnamefont {F.}~\bibnamefont
  {Verstraete}},\ and\ \bibinfo {author} {\bibfnamefont {J.~I.}\ \bibnamefont
  {Cirac}},\ }\bibfield  {title} {\bibinfo {title} {{Matrix product states for
  dynamical simulation of infinite chains}},\ }\href
  {https://doi.org/10.1103/PhysRevLett.102.240603} {\bibfield  {journal}
  {\bibinfo  {journal} {Physical Review Letters}\ }\textbf {\bibinfo {volume}
  {102}},\ \bibinfo {pages} {240603} (\bibinfo {year} {2009})}\BibitemShut
  {NoStop}%
\bibitem [{\citenamefont {Ran}\ \emph {et~al.}(2012)\citenamefont {Ran},
  \citenamefont {Li}, \citenamefont {Xi}, \citenamefont {Zhang},\ and\
  \citenamefont {Su}}]{RLXZS12ODTNS}%
  \BibitemOpen
  \bibfield  {author} {\bibinfo {author} {\bibfnamefont {S.-J.}\ \bibnamefont
  {Ran}}, \bibinfo {author} {\bibfnamefont {W.}~\bibnamefont {Li}}, \bibinfo
  {author} {\bibfnamefont {B.}~\bibnamefont {Xi}}, \bibinfo {author}
  {\bibfnamefont {Z.}~\bibnamefont {Zhang}},\ and\ \bibinfo {author}
  {\bibfnamefont {G.}~\bibnamefont {Su}},\ }\bibfield  {title} {\bibinfo
  {title} {Optimized decimation of tensor networks with super-orthogonalization
  for two-dimensional quantum lattice models},\ }\href
  {https://doi.org/10.1103/PhysRevB.86.134429} {\bibfield  {journal} {\bibinfo
  {journal} {Physical Review B}\ }\textbf {\bibinfo {volume} {86}},\ \bibinfo
  {pages} {134429} (\bibinfo {year} {2012})}\BibitemShut {NoStop}%
\bibitem [{\citenamefont {Czarnik}\ \emph {et~al.}(2012)\citenamefont
  {Czarnik}, \citenamefont {Cincio},\ and\ \citenamefont
  {Dziarmaga}}]{CCD12FTPEPS}%
  \BibitemOpen
  \bibfield  {author} {\bibinfo {author} {\bibfnamefont {P.}~\bibnamefont
  {Czarnik}}, \bibinfo {author} {\bibfnamefont {L.}~\bibnamefont {Cincio}},\
  and\ \bibinfo {author} {\bibfnamefont {J.}~\bibnamefont {Dziarmaga}},\
  }\bibfield  {title} {\bibinfo {title} {Projected entangled pair states at
  finite temperature: Imaginary time evolution with ancillas},\ }\href
  {https://doi.org/10.1103/PhysRevB.86.245101} {\bibfield  {journal} {\bibinfo
  {journal} {Physical Review B}\ }\textbf {\bibinfo {volume} {86}},\ \bibinfo
  {pages} {245101} (\bibinfo {year} {2012})}\BibitemShut {NoStop}%
\bibitem [{\citenamefont {Hastings}\ and\ \citenamefont
  {Mahajan}(2015)}]{HM15folding}%
  \BibitemOpen
  \bibfield  {author} {\bibinfo {author} {\bibfnamefont {M.~B.}\ \bibnamefont
  {Hastings}}\ and\ \bibinfo {author} {\bibfnamefont {R.}~\bibnamefont
  {Mahajan}},\ }\bibfield  {title} {\bibinfo {title} {Connecting entanglement
  in time and space: Improving the folding algorithm},\ }\href
  {https://doi.org/10.1103/PhysRevA.91.032306} {\bibfield  {journal} {\bibinfo
  {journal} {Physical Review A}\ }\textbf {\bibinfo {volume} {91}},\ \bibinfo
  {pages} {032306} (\bibinfo {year} {2015})}\BibitemShut {NoStop}%
\bibitem [{\citenamefont {Kshetrimayum}\ \emph {et~al.}(2017)\citenamefont
  {Kshetrimayum}, \citenamefont {Weimer},\ and\ \citenamefont
  {Or{\'{u}}s}}]{KWO17MPSopen}%
  \BibitemOpen
  \bibfield  {author} {\bibinfo {author} {\bibfnamefont {A.}~\bibnamefont
  {Kshetrimayum}}, \bibinfo {author} {\bibfnamefont {H.}~\bibnamefont
  {Weimer}},\ and\ \bibinfo {author} {\bibfnamefont {R.}~\bibnamefont
  {Or{\'{u}}s}},\ }\bibfield  {title} {\bibinfo {title} {A simple tensor
  network algorithm for two-dimensional steady states},\ }\href
  {https://doi.org/10.1038/s41467-017-01511-6} {\bibfield  {journal} {\bibinfo
  {journal} {Nature Communications}\ }\textbf {\bibinfo {volume} {8}},\
  \bibinfo {pages} {1291} (\bibinfo {year} {2017})}\BibitemShut {NoStop}%
\bibitem [{\citenamefont {Huggins}\ \emph {et~al.}(2019)\citenamefont
  {Huggins}, \citenamefont {Patil}, \citenamefont {Mitchell}, \citenamefont
  {Whaley},\ and\ \citenamefont {Stoudenmire}}]{HPWS18TNQML}%
  \BibitemOpen
  \bibfield  {author} {\bibinfo {author} {\bibfnamefont {W.}~\bibnamefont
  {Huggins}}, \bibinfo {author} {\bibfnamefont {P.}~\bibnamefont {Patil}},
  \bibinfo {author} {\bibfnamefont {B.}~\bibnamefont {Mitchell}}, \bibinfo
  {author} {\bibfnamefont {K.~B.}\ \bibnamefont {Whaley}},\ and\ \bibinfo
  {author} {\bibfnamefont {E.~M.}\ \bibnamefont {Stoudenmire}},\ }\bibfield
  {title} {\bibinfo {title} {Towards quantum machine learning with tensor
  networks},\ }\href {https://doi.org/10.1088/2058-9565/aaea94} {\bibfield
  {journal} {\bibinfo  {journal} {Quantum Science and Technology}\ }\textbf
  {\bibinfo {volume} {4}},\ \bibinfo {pages} {024001} (\bibinfo {year}
  {2019})}\BibitemShut {NoStop}%
\bibitem [{Note12()}]{Note12}%
  \BibitemOpen
  \bibinfo {note} {A quantum gate is a specific operation on quantum state, and
  thus can be given by a quantum operator. When considering the quantum
  computation in practice, the quantum gates realizable on different quantum
  platforms (such as super-conducting circuits, ultra-cold atoms, and single
  photons) are different.}\BibitemShut {Stop}%
\bibitem [{\citenamefont {Shor}(1999)}]{S99shor}%
  \BibitemOpen
  \bibfield  {author} {\bibinfo {author} {\bibfnamefont {P.~W.}\ \bibnamefont
  {Shor}},\ }\bibfield  {title} {\bibinfo {title} {Polynomial-time algorithms
  for prime factorization and discrete logarithms on a quantum computer},\
  }\href {https://doi.org/10.1137/S0036144598347011} {\bibfield  {journal}
  {\bibinfo  {journal} {SIAM Review}\ }\textbf {\bibinfo {volume} {41}},\
  \bibinfo {pages} {303} (\bibinfo {year} {1999})}\BibitemShut {NoStop}%
\bibitem [{\citenamefont {Grover}(1997)}]{G97Grover}%
  \BibitemOpen
  \bibfield  {author} {\bibinfo {author} {\bibfnamefont {L.~K.}\ \bibnamefont
  {Grover}},\ }\bibfield  {title} {\bibinfo {title} {Quantum mechanics helps in
  searching for a needle in a haystack},\ }\href
  {https://doi.org/10.1103/PhysRevLett.79.325} {\bibfield  {journal} {\bibinfo
  {journal} {Physical Review Letters}\ }\textbf {\bibinfo {volume} {79}},\
  \bibinfo {pages} {325} (\bibinfo {year} {1997})}\BibitemShut {NoStop}%
\bibitem [{\citenamefont {Ran}(2020)}]{R20MPSencode}%
  \BibitemOpen
  \bibfield  {author} {\bibinfo {author} {\bibfnamefont {S.-J.}\ \bibnamefont
  {Ran}},\ }\bibfield  {title} {\bibinfo {title} {Encoding of matrix product
  states into quantum circuits of one- and two-qubit gates},\ }\href
  {https://doi.org/10.1103/PhysRevA.101.032310} {\bibfield  {journal} {\bibinfo
   {journal} {Physical Review A}\ }\textbf {\bibinfo {volume} {101}},\ \bibinfo
  {pages} {032310} (\bibinfo {year} {2020})}\BibitemShut {NoStop}%
\bibitem [{\citenamefont {Cerezo}\ \emph {et~al.}(2021)\citenamefont {Cerezo},
  \citenamefont {Arrasmith}, \citenamefont {Babbush}, \citenamefont {Benjamin},
  \citenamefont {Endo}, \citenamefont {Fujii}, \citenamefont {McClean},
  \citenamefont {Mitarai}, \citenamefont {Yuan}, \citenamefont {Cincio},\ and\
  \citenamefont {Coles}}]{CAB+21VQA}%
  \BibitemOpen
  \bibfield  {author} {\bibinfo {author} {\bibfnamefont {M.}~\bibnamefont
  {Cerezo}}, \bibinfo {author} {\bibfnamefont {A.}~\bibnamefont {Arrasmith}},
  \bibinfo {author} {\bibfnamefont {R.}~\bibnamefont {Babbush}}, \bibinfo
  {author} {\bibfnamefont {S.~C.}\ \bibnamefont {Benjamin}}, \bibinfo {author}
  {\bibfnamefont {S.}~\bibnamefont {Endo}}, \bibinfo {author} {\bibfnamefont
  {K.}~\bibnamefont {Fujii}}, \bibinfo {author} {\bibfnamefont {J.~R.}\
  \bibnamefont {McClean}}, \bibinfo {author} {\bibfnamefont {K.}~\bibnamefont
  {Mitarai}}, \bibinfo {author} {\bibfnamefont {X.}~\bibnamefont {Yuan}},
  \bibinfo {author} {\bibfnamefont {L.}~\bibnamefont {Cincio}},\ and\ \bibinfo
  {author} {\bibfnamefont {P.~J.}\ \bibnamefont {Coles}},\ }\bibfield  {title}
  {\bibinfo {title} {Variational quantum algorithms},\ }\href
  {https://doi.org/10.1038/s42254-021-00348-9} {\bibfield  {journal} {\bibinfo
  {journal} {Nature Reviews Physics}\ }\textbf {\bibinfo {volume} {3}},\
  \bibinfo {pages} {625} (\bibinfo {year} {2021})}\BibitemShut {NoStop}%
\bibitem [{\citenamefont {Liu}\ \emph {et~al.}(2020)\citenamefont {Liu},
  \citenamefont {Wang}, \citenamefont {Xue}, \citenamefont {Huang},
  \citenamefont {Fu}, \citenamefont {Qiang}, \citenamefont {Xu}, \citenamefont
  {Huang}, \citenamefont {Deng}, \citenamefont {Guo}, \citenamefont {Yang},\
  and\ \citenamefont {Wu}}]{LWX+20VQCtomo}%
  \BibitemOpen
  \bibfield  {author} {\bibinfo {author} {\bibfnamefont {Y.}~\bibnamefont
  {Liu}}, \bibinfo {author} {\bibfnamefont {D.}~\bibnamefont {Wang}}, \bibinfo
  {author} {\bibfnamefont {S.}~\bibnamefont {Xue}}, \bibinfo {author}
  {\bibfnamefont {A.}~\bibnamefont {Huang}}, \bibinfo {author} {\bibfnamefont
  {X.}~\bibnamefont {Fu}}, \bibinfo {author} {\bibfnamefont {X.}~\bibnamefont
  {Qiang}}, \bibinfo {author} {\bibfnamefont {P.}~\bibnamefont {Xu}}, \bibinfo
  {author} {\bibfnamefont {H.-L.}\ \bibnamefont {Huang}}, \bibinfo {author}
  {\bibfnamefont {M.}~\bibnamefont {Deng}}, \bibinfo {author} {\bibfnamefont
  {C.}~\bibnamefont {Guo}}, \bibinfo {author} {\bibfnamefont {X.}~\bibnamefont
  {Yang}},\ and\ \bibinfo {author} {\bibfnamefont {J.}~\bibnamefont {Wu}},\
  }\bibfield  {title} {\bibinfo {title} {Variational quantum circuits for
  quantum state tomography},\ }\href
  {https://doi.org/10.1103/PhysRevA.101.052316} {\bibfield  {journal} {\bibinfo
   {journal} {Physical Review A}\ }\textbf {\bibinfo {volume} {101}},\ \bibinfo
  {pages} {052316} (\bibinfo {year} {2020})}\BibitemShut {NoStop}%
\bibitem [{\citenamefont {Zhou}\ \emph {et~al.}(2021)\citenamefont {Zhou},
  \citenamefont {Hong},\ and\ \citenamefont {Ran}}]{ZHR21ADQC}%
  \BibitemOpen
  \bibfield  {author} {\bibinfo {author} {\bibfnamefont {P.-F.}\ \bibnamefont
  {Zhou}}, \bibinfo {author} {\bibfnamefont {R.}~\bibnamefont {Hong}},\ and\
  \bibinfo {author} {\bibfnamefont {S.-J.}\ \bibnamefont {Ran}},\ }\bibfield
  {title} {\bibinfo {title} {Automatically differentiable quantum circuit for
  many-qubit state preparation},\ }\href
  {https://doi.org/10.1103/PhysRevA.104.042601} {\bibfield  {journal} {\bibinfo
   {journal} {Physical Review A}\ }\textbf {\bibinfo {volume} {104}},\ \bibinfo
  {pages} {042601} (\bibinfo {year} {2021})}\BibitemShut {NoStop}%
\bibitem [{\citenamefont {Peruzzo}\ \emph {et~al.}(2014)\citenamefont
  {Peruzzo}, \citenamefont {McClean}, \citenamefont {Shadbolt}, \citenamefont
  {Yung}, \citenamefont {Zhou}, \citenamefont {Love}, \citenamefont
  {Aspuru-Guzik},\ and\ \citenamefont {O'Brien}}]{PMSY+14VQE}%
  \BibitemOpen
  \bibfield  {author} {\bibinfo {author} {\bibfnamefont {A.}~\bibnamefont
  {Peruzzo}}, \bibinfo {author} {\bibfnamefont {J.}~\bibnamefont {McClean}},
  \bibinfo {author} {\bibfnamefont {P.}~\bibnamefont {Shadbolt}}, \bibinfo
  {author} {\bibfnamefont {M.-H.}\ \bibnamefont {Yung}}, \bibinfo {author}
  {\bibfnamefont {X.-Q.}\ \bibnamefont {Zhou}}, \bibinfo {author}
  {\bibfnamefont {P.~J.}\ \bibnamefont {Love}}, \bibinfo {author}
  {\bibfnamefont {A.}~\bibnamefont {Aspuru-Guzik}},\ and\ \bibinfo {author}
  {\bibfnamefont {J.~L.}\ \bibnamefont {O'Brien}},\ }\bibfield  {title}
  {\bibinfo {title} {A variational eigenvalue solver on a photonic quantum
  processor},\ }\href {https://doi.org/10.1038/ncomms5213} {\bibfield
  {journal} {\bibinfo  {journal} {Nature Communications}\ }\textbf {\bibinfo
  {volume} {5}},\ \bibinfo {pages} {4213} (\bibinfo {year} {2014})}\BibitemShut
  {NoStop}%
\bibitem [{\citenamefont {Tilly}\ \emph {et~al.}(2022)\citenamefont {Tilly},
  \citenamefont {Chen}, \citenamefont {Cao}, \citenamefont {Picozzi},
  \citenamefont {Setia}, \citenamefont {Li}, \citenamefont {Grant},
  \citenamefont {Wossnig}, \citenamefont {Rungger}, \citenamefont {Booth},\
  and\ \citenamefont {Tennyson}}]{TCCP+22VQErev}%
  \BibitemOpen
  \bibfield  {author} {\bibinfo {author} {\bibfnamefont {J.}~\bibnamefont
  {Tilly}}, \bibinfo {author} {\bibfnamefont {H.}~\bibnamefont {Chen}},
  \bibinfo {author} {\bibfnamefont {S.}~\bibnamefont {Cao}}, \bibinfo {author}
  {\bibfnamefont {D.}~\bibnamefont {Picozzi}}, \bibinfo {author} {\bibfnamefont
  {K.}~\bibnamefont {Setia}}, \bibinfo {author} {\bibfnamefont
  {Y.}~\bibnamefont {Li}}, \bibinfo {author} {\bibfnamefont {E.}~\bibnamefont
  {Grant}}, \bibinfo {author} {\bibfnamefont {L.}~\bibnamefont {Wossnig}},
  \bibinfo {author} {\bibfnamefont {I.}~\bibnamefont {Rungger}}, \bibinfo
  {author} {\bibfnamefont {G.~H.}\ \bibnamefont {Booth}},\ and\ \bibinfo
  {author} {\bibfnamefont {J.}~\bibnamefont {Tennyson}},\ }\bibfield  {title}
  {\bibinfo {title} {The variational quantum eigensolver: A review of methods
  and best practices},\ }\href
  {https://doi.org/https://doi.org/10.1016/j.physrep.2022.08.003} {\bibfield
  {journal} {\bibinfo  {journal} {Physics Reports}\ }\textbf {\bibinfo {volume}
  {986}},\ \bibinfo {pages} {1} (\bibinfo {year} {2022})}\BibitemShut {NoStop}%
\bibitem [{\citenamefont {McClean}\ \emph {et~al.}(2016)\citenamefont
  {McClean}, \citenamefont {Romero}, \citenamefont {Babbush},\ and\
  \citenamefont {Aspuru-Guzik}}]{MRBA16VQA}%
  \BibitemOpen
  \bibfield  {author} {\bibinfo {author} {\bibfnamefont {J.~R.}\ \bibnamefont
  {McClean}}, \bibinfo {author} {\bibfnamefont {J.}~\bibnamefont {Romero}},
  \bibinfo {author} {\bibfnamefont {R.}~\bibnamefont {Babbush}},\ and\ \bibinfo
  {author} {\bibfnamefont {A.}~\bibnamefont {Aspuru-Guzik}},\ }\bibfield
  {title} {\bibinfo {title} {The theory of variational hybrid quantum-classical
  algorithms},\ }\href {https://doi.org/10.1088/1367-2630/18/2/023023}
  {\bibfield  {journal} {\bibinfo  {journal} {New Journal of Physics}\ }\textbf
  {\bibinfo {volume} {18}},\ \bibinfo {pages} {023023} (\bibinfo {year}
  {2016})}\BibitemShut {NoStop}%
\bibitem [{\citenamefont {Markov}\ and\ \citenamefont
  {Shi}(2008)}]{MS08TNQcomp}%
  \BibitemOpen
  \bibfield  {author} {\bibinfo {author} {\bibfnamefont {I.~L.}\ \bibnamefont
  {Markov}}\ and\ \bibinfo {author} {\bibfnamefont {Y.}~\bibnamefont {Shi}},\
  }\bibfield  {title} {\bibinfo {title} {Simulating quantum computation by
  contracting tensor networks},\ }\href {https://doi.org/10.1137/050644756}
  {\bibfield  {journal} {\bibinfo  {journal} {SIAM Journal on Computing}\
  }\textbf {\bibinfo {volume} {38}},\ \bibinfo {pages} {963} (\bibinfo {year}
  {2008})}\BibitemShut {NoStop}%
\bibitem [{\citenamefont {Huang}\ \emph
  {et~al.}(2021{\natexlab{a}})\citenamefont {Huang}, \citenamefont {Zhang},
  \citenamefont {Newman}, \citenamefont {Ni}, \citenamefont {Ding},
  \citenamefont {Cai}, \citenamefont {Gao}, \citenamefont {Wang}, \citenamefont
  {Wu}, \citenamefont {Zhang} \emph {et~al.}}]{HZNN+21TNQC}%
  \BibitemOpen
  \bibfield  {author} {\bibinfo {author} {\bibfnamefont {C.}~\bibnamefont
  {Huang}}, \bibinfo {author} {\bibfnamefont {F.}~\bibnamefont {Zhang}},
  \bibinfo {author} {\bibfnamefont {M.}~\bibnamefont {Newman}}, \bibinfo
  {author} {\bibfnamefont {X.}~\bibnamefont {Ni}}, \bibinfo {author}
  {\bibfnamefont {D.}~\bibnamefont {Ding}}, \bibinfo {author} {\bibfnamefont
  {J.}~\bibnamefont {Cai}}, \bibinfo {author} {\bibfnamefont {X.}~\bibnamefont
  {Gao}}, \bibinfo {author} {\bibfnamefont {T.}~\bibnamefont {Wang}}, \bibinfo
  {author} {\bibfnamefont {F.}~\bibnamefont {Wu}}, \bibinfo {author}
  {\bibfnamefont {G.}~\bibnamefont {Zhang}}, \emph {et~al.},\ }\bibfield
  {title} {\bibinfo {title} {Efficient parallelization of tensor network
  contraction for simulating quantum computation},\ }\href
  {https://doi.org/10.1038/s43588-021-00119-7} {\bibfield  {journal} {\bibinfo
  {journal} {Nature Computational Science}\ }\textbf {\bibinfo {volume} {1}},\
  \bibinfo {pages} {578} (\bibinfo {year} {2021}{\natexlab{a}})}\BibitemShut
  {NoStop}%
\bibitem [{\citenamefont {Haghshenas}\ \emph {et~al.}(2022)\citenamefont
  {Haghshenas}, \citenamefont {Gray}, \citenamefont {Potter},\ and\
  \citenamefont {Chan}}]{HGPC22TNQC}%
  \BibitemOpen
  \bibfield  {author} {\bibinfo {author} {\bibfnamefont {R.}~\bibnamefont
  {Haghshenas}}, \bibinfo {author} {\bibfnamefont {J.}~\bibnamefont {Gray}},
  \bibinfo {author} {\bibfnamefont {A.~C.}\ \bibnamefont {Potter}},\ and\
  \bibinfo {author} {\bibfnamefont {G.~K.-L.}\ \bibnamefont {Chan}},\
  }\bibfield  {title} {\bibinfo {title} {Variational power of quantum circuit
  tensor networks},\ }\href {https://doi.org/10.1103/PhysRevX.12.011047}
  {\bibfield  {journal} {\bibinfo  {journal} {Physical Review X}\ }\textbf
  {\bibinfo {volume} {12}},\ \bibinfo {pages} {011047} (\bibinfo {year}
  {2022})}\BibitemShut {NoStop}%
\bibitem [{\citenamefont {Guo}\ \emph {et~al.}(2021)\citenamefont {Guo},
  \citenamefont {Zhao},\ and\ \citenamefont {Huang}}]{GZH21TNQC}%
  \BibitemOpen
  \bibfield  {author} {\bibinfo {author} {\bibfnamefont {C.}~\bibnamefont
  {Guo}}, \bibinfo {author} {\bibfnamefont {Y.}~\bibnamefont {Zhao}},\ and\
  \bibinfo {author} {\bibfnamefont {H.-L.}\ \bibnamefont {Huang}},\ }\bibfield
  {title} {\bibinfo {title} {Verifying random quantum circuits with arbitrary
  geometry using tensor network states algorithm},\ }\href
  {https://doi.org/10.1103/PhysRevLett.126.070502} {\bibfield  {journal}
  {\bibinfo  {journal} {Physical Review Letters}\ }\textbf {\bibinfo {volume}
  {126}},\ \bibinfo {pages} {070502} (\bibinfo {year} {2021})}\BibitemShut
  {NoStop}%
\bibitem [{\citenamefont {Vincent}\ \emph {et~al.}(2022)\citenamefont
  {Vincent}, \citenamefont {O'Riordan}, \citenamefont {Andrenkov},
  \citenamefont {Brown}, \citenamefont {Killoran}, \citenamefont {Qi},\ and\
  \citenamefont {Dhand}}]{VOA+22TNQC}%
  \BibitemOpen
  \bibfield  {author} {\bibinfo {author} {\bibfnamefont {T.}~\bibnamefont
  {Vincent}}, \bibinfo {author} {\bibfnamefont {L.~J.}\ \bibnamefont
  {O'Riordan}}, \bibinfo {author} {\bibfnamefont {M.}~\bibnamefont
  {Andrenkov}}, \bibinfo {author} {\bibfnamefont {J.}~\bibnamefont {Brown}},
  \bibinfo {author} {\bibfnamefont {N.}~\bibnamefont {Killoran}}, \bibinfo
  {author} {\bibfnamefont {H.}~\bibnamefont {Qi}},\ and\ \bibinfo {author}
  {\bibfnamefont {I.}~\bibnamefont {Dhand}},\ }\bibfield  {title} {\bibinfo
  {title} {Jet: {F}ast quantum circuit simulations with parallel task-based
  tensor-network contraction},\ }\href
  {https://doi.org/10.22331/q-2022-05-09-709} {\bibfield  {journal} {\bibinfo
  {journal} {{Quantum}}\ }\textbf {\bibinfo {volume} {6}},\ \bibinfo {pages}
  {709} (\bibinfo {year} {2022})}\BibitemShut {NoStop}%
\bibitem [{\citenamefont {Pan}\ and\ \citenamefont {Zhang}(2022)}]{PZ22TNQC}%
  \BibitemOpen
  \bibfield  {author} {\bibinfo {author} {\bibfnamefont {F.}~\bibnamefont
  {Pan}}\ and\ \bibinfo {author} {\bibfnamefont {P.}~\bibnamefont {Zhang}},\
  }\bibfield  {title} {\bibinfo {title} {Simulation of quantum circuits using
  the big-batch tensor network method},\ }\href
  {https://doi.org/10.1103/PhysRevLett.128.030501} {\bibfield  {journal}
  {\bibinfo  {journal} {Physical Review Letters}\ }\textbf {\bibinfo {volume}
  {128}},\ \bibinfo {pages} {030501} (\bibinfo {year} {2022})}\BibitemShut
  {NoStop}%
\bibitem [{\citenamefont {Lykov}\ \emph {et~al.}(2022)\citenamefont {Lykov},
  \citenamefont {Schutski}, \citenamefont {Galda}, \citenamefont {Vinokur},\
  and\ \citenamefont {Alexeev}}]{LSGVA22TNQC}%
  \BibitemOpen
  \bibfield  {author} {\bibinfo {author} {\bibfnamefont {D.}~\bibnamefont
  {Lykov}}, \bibinfo {author} {\bibfnamefont {R.}~\bibnamefont {Schutski}},
  \bibinfo {author} {\bibfnamefont {A.}~\bibnamefont {Galda}}, \bibinfo
  {author} {\bibfnamefont {V.}~\bibnamefont {Vinokur}},\ and\ \bibinfo {author}
  {\bibfnamefont {Y.}~\bibnamefont {Alexeev}},\ }\bibfield  {title} {\bibinfo
  {title} {Tensor network quantum simulator with step-dependent
  parallelization},\ }in\ \href@noop {} {\emph {\bibinfo {booktitle} {2022 IEEE
  International Conference on Quantum Computing and Engineering (QCE)}}}\
  (\bibinfo {year} {2022})\ pp.\ \bibinfo {pages} {582--593}\BibitemShut
  {NoStop}%
\bibitem [{\citenamefont {Preskill}(2018)}]{P18NISQ}%
  \BibitemOpen
  \bibfield  {author} {\bibinfo {author} {\bibfnamefont {J.}~\bibnamefont
  {Preskill}},\ }\bibfield  {title} {\bibinfo {title} {Quantum computing in the
  {NISQ} era and beyond},\ }\href {https://doi.org/10.22331/q-2018-08-06-79}
  {\bibfield  {journal} {\bibinfo  {journal} {{Quantum}}\ }\textbf {\bibinfo
  {volume} {2}},\ \bibinfo {pages} {79} (\bibinfo {year} {2018})}\BibitemShut
  {NoStop}%
\bibitem [{\citenamefont {Arute}\ \emph {et~al.}(2019)\citenamefont {Arute},
  \citenamefont {Arya}, \citenamefont {Babbush}, \citenamefont {Bacon},
  \citenamefont {Bardin}, \citenamefont {Barends}, \citenamefont {Biswas},
  \citenamefont {Boixo}, \citenamefont {Brandao}, \citenamefont {Buell},
  \citenamefont {Burkett}, \citenamefont {Chen}, \citenamefont {Chen},
  \citenamefont {Chiaro}, \citenamefont {Collins}, \citenamefont {Courtney},
  \citenamefont {Dunsworth}, \citenamefont {Farhi}, \citenamefont {Foxen},
  \citenamefont {Fowler}, \citenamefont {Gidney}, \citenamefont {Giustina},
  \citenamefont {Graff}, \citenamefont {Guerin}, \citenamefont {Habegger},
  \citenamefont {Harrigan}, \citenamefont {Hartmann}, \citenamefont {Ho},
  \citenamefont {Hoffmann}, \citenamefont {Huang}, \citenamefont {Humble},
  \citenamefont {Isakov}, \citenamefont {Jeffrey}, \citenamefont {Jiang},
  \citenamefont {Kafri}, \citenamefont {Kechedzhi}, \citenamefont {Kelly},
  \citenamefont {Klimov}, \citenamefont {Knysh}, \citenamefont {Korotkov},
  \citenamefont {Kostritsa}, \citenamefont {Landhuis}, \citenamefont
  {Lindmark}, \citenamefont {Lucero}, \citenamefont {Lyakh}, \citenamefont
  {Mandr{\`a}}, \citenamefont {McClean}, \citenamefont {McEwen}, \citenamefont
  {Megrant}, \citenamefont {Mi}, \citenamefont {Michielsen}, \citenamefont
  {Mohseni}, \citenamefont {Mutus}, \citenamefont {Naaman}, \citenamefont
  {Neeley}, \citenamefont {Neill}, \citenamefont {Niu}, \citenamefont {Ostby},
  \citenamefont {Petukhov}, \citenamefont {Platt}, \citenamefont {Quintana},
  \citenamefont {Rieffel}, \citenamefont {Roushan}, \citenamefont {Rubin},
  \citenamefont {Sank}, \citenamefont {Satzinger}, \citenamefont {Smelyanskiy},
  \citenamefont {Sung}, \citenamefont {Trevithick}, \citenamefont
  {Vainsencher}, \citenamefont {Villalonga}, \citenamefont {White},
  \citenamefont {Yao}, \citenamefont {Yeh}, \citenamefont {Zalcman},
  \citenamefont {Neven},\ and\ \citenamefont {Martinis}}]{AABB+2019GoogleQ}%
  \BibitemOpen
  \bibfield  {author} {\bibinfo {author} {\bibfnamefont {F.}~\bibnamefont
  {Arute}}, \bibinfo {author} {\bibfnamefont {K.}~\bibnamefont {Arya}},
  \bibinfo {author} {\bibfnamefont {R.}~\bibnamefont {Babbush}}, \bibinfo
  {author} {\bibfnamefont {D.}~\bibnamefont {Bacon}}, \bibinfo {author}
  {\bibfnamefont {J.~C.}\ \bibnamefont {Bardin}}, \bibinfo {author}
  {\bibfnamefont {R.}~\bibnamefont {Barends}}, \bibinfo {author} {\bibfnamefont
  {R.}~\bibnamefont {Biswas}}, \bibinfo {author} {\bibfnamefont
  {S.}~\bibnamefont {Boixo}}, \bibinfo {author} {\bibfnamefont {F.~G. S.~L.}\
  \bibnamefont {Brandao}}, \bibinfo {author} {\bibfnamefont {D.~A.}\
  \bibnamefont {Buell}}, \bibinfo {author} {\bibfnamefont {B.}~\bibnamefont
  {Burkett}}, \bibinfo {author} {\bibfnamefont {Y.}~\bibnamefont {Chen}},
  \bibinfo {author} {\bibfnamefont {Z.}~\bibnamefont {Chen}}, \bibinfo {author}
  {\bibfnamefont {B.}~\bibnamefont {Chiaro}}, \bibinfo {author} {\bibfnamefont
  {R.}~\bibnamefont {Collins}}, \bibinfo {author} {\bibfnamefont
  {W.}~\bibnamefont {Courtney}}, \bibinfo {author} {\bibfnamefont
  {A.}~\bibnamefont {Dunsworth}}, \bibinfo {author} {\bibfnamefont
  {E.}~\bibnamefont {Farhi}}, \bibinfo {author} {\bibfnamefont
  {B.}~\bibnamefont {Foxen}}, \bibinfo {author} {\bibfnamefont
  {A.}~\bibnamefont {Fowler}}, \bibinfo {author} {\bibfnamefont
  {C.}~\bibnamefont {Gidney}}, \bibinfo {author} {\bibfnamefont
  {M.}~\bibnamefont {Giustina}}, \bibinfo {author} {\bibfnamefont
  {R.}~\bibnamefont {Graff}}, \bibinfo {author} {\bibfnamefont
  {K.}~\bibnamefont {Guerin}}, \bibinfo {author} {\bibfnamefont
  {S.}~\bibnamefont {Habegger}}, \bibinfo {author} {\bibfnamefont {M.~P.}\
  \bibnamefont {Harrigan}}, \bibinfo {author} {\bibfnamefont {M.~J.}\
  \bibnamefont {Hartmann}}, \bibinfo {author} {\bibfnamefont {A.}~\bibnamefont
  {Ho}}, \bibinfo {author} {\bibfnamefont {M.}~\bibnamefont {Hoffmann}},
  \bibinfo {author} {\bibfnamefont {T.}~\bibnamefont {Huang}}, \bibinfo
  {author} {\bibfnamefont {T.~S.}\ \bibnamefont {Humble}}, \bibinfo {author}
  {\bibfnamefont {S.~V.}\ \bibnamefont {Isakov}}, \bibinfo {author}
  {\bibfnamefont {E.}~\bibnamefont {Jeffrey}}, \bibinfo {author} {\bibfnamefont
  {Z.}~\bibnamefont {Jiang}}, \bibinfo {author} {\bibfnamefont
  {D.}~\bibnamefont {Kafri}}, \bibinfo {author} {\bibfnamefont
  {K.}~\bibnamefont {Kechedzhi}}, \bibinfo {author} {\bibfnamefont
  {J.}~\bibnamefont {Kelly}}, \bibinfo {author} {\bibfnamefont {P.~V.}\
  \bibnamefont {Klimov}}, \bibinfo {author} {\bibfnamefont {S.}~\bibnamefont
  {Knysh}}, \bibinfo {author} {\bibfnamefont {A.}~\bibnamefont {Korotkov}},
  \bibinfo {author} {\bibfnamefont {F.}~\bibnamefont {Kostritsa}}, \bibinfo
  {author} {\bibfnamefont {D.}~\bibnamefont {Landhuis}}, \bibinfo {author}
  {\bibfnamefont {M.}~\bibnamefont {Lindmark}}, \bibinfo {author}
  {\bibfnamefont {E.}~\bibnamefont {Lucero}}, \bibinfo {author} {\bibfnamefont
  {D.}~\bibnamefont {Lyakh}}, \bibinfo {author} {\bibfnamefont
  {S.}~\bibnamefont {Mandr{\`a}}}, \bibinfo {author} {\bibfnamefont {J.~R.}\
  \bibnamefont {McClean}}, \bibinfo {author} {\bibfnamefont {M.}~\bibnamefont
  {McEwen}}, \bibinfo {author} {\bibfnamefont {A.}~\bibnamefont {Megrant}},
  \bibinfo {author} {\bibfnamefont {X.}~\bibnamefont {Mi}}, \bibinfo {author}
  {\bibfnamefont {K.}~\bibnamefont {Michielsen}}, \bibinfo {author}
  {\bibfnamefont {M.}~\bibnamefont {Mohseni}}, \bibinfo {author} {\bibfnamefont
  {J.}~\bibnamefont {Mutus}}, \bibinfo {author} {\bibfnamefont
  {O.}~\bibnamefont {Naaman}}, \bibinfo {author} {\bibfnamefont
  {M.}~\bibnamefont {Neeley}}, \bibinfo {author} {\bibfnamefont
  {C.}~\bibnamefont {Neill}}, \bibinfo {author} {\bibfnamefont {M.~Y.}\
  \bibnamefont {Niu}}, \bibinfo {author} {\bibfnamefont {E.}~\bibnamefont
  {Ostby}}, \bibinfo {author} {\bibfnamefont {A.}~\bibnamefont {Petukhov}},
  \bibinfo {author} {\bibfnamefont {J.~C.}\ \bibnamefont {Platt}}, \bibinfo
  {author} {\bibfnamefont {C.}~\bibnamefont {Quintana}}, \bibinfo {author}
  {\bibfnamefont {E.~G.}\ \bibnamefont {Rieffel}}, \bibinfo {author}
  {\bibfnamefont {P.}~\bibnamefont {Roushan}}, \bibinfo {author} {\bibfnamefont
  {N.~C.}\ \bibnamefont {Rubin}}, \bibinfo {author} {\bibfnamefont
  {D.}~\bibnamefont {Sank}}, \bibinfo {author} {\bibfnamefont {K.~J.}\
  \bibnamefont {Satzinger}}, \bibinfo {author} {\bibfnamefont {V.}~\bibnamefont
  {Smelyanskiy}}, \bibinfo {author} {\bibfnamefont {K.~J.}\ \bibnamefont
  {Sung}}, \bibinfo {author} {\bibfnamefont {M.~D.}\ \bibnamefont
  {Trevithick}}, \bibinfo {author} {\bibfnamefont {A.}~\bibnamefont
  {Vainsencher}}, \bibinfo {author} {\bibfnamefont {B.}~\bibnamefont
  {Villalonga}}, \bibinfo {author} {\bibfnamefont {T.}~\bibnamefont {White}},
  \bibinfo {author} {\bibfnamefont {Z.~J.}\ \bibnamefont {Yao}}, \bibinfo
  {author} {\bibfnamefont {P.}~\bibnamefont {Yeh}}, \bibinfo {author}
  {\bibfnamefont {A.}~\bibnamefont {Zalcman}}, \bibinfo {author} {\bibfnamefont
  {H.}~\bibnamefont {Neven}},\ and\ \bibinfo {author} {\bibfnamefont {J.~M.}\
  \bibnamefont {Martinis}},\ }\bibfield  {title} {\bibinfo {title} {Quantum
  supremacy using a programmable superconducting processor},\ }\href
  {https://doi.org/10.1038/s41586-019-1666-5} {\bibfield  {journal} {\bibinfo
  {journal} {Nature}\ }\textbf {\bibinfo {volume} {574}},\ \bibinfo {pages}
  {505} (\bibinfo {year} {2019})}\BibitemShut {NoStop}%
\bibitem [{\citenamefont {Guo}\ \emph {et~al.}(2019)\citenamefont {Guo},
  \citenamefont {Liu}, \citenamefont {Xiong}, \citenamefont {Xue},
  \citenamefont {Fu}, \citenamefont {Huang}, \citenamefont {Qiang},
  \citenamefont {Xu}, \citenamefont {Liu}, \citenamefont {Zheng}, \citenamefont
  {Huang}, \citenamefont {Deng}, \citenamefont {Poletti}, \citenamefont {Bao},\
  and\ \citenamefont {Wu}}]{GLXX+19PEPSQC}%
  \BibitemOpen
  \bibfield  {author} {\bibinfo {author} {\bibfnamefont {C.}~\bibnamefont
  {Guo}}, \bibinfo {author} {\bibfnamefont {Y.}~\bibnamefont {Liu}}, \bibinfo
  {author} {\bibfnamefont {M.}~\bibnamefont {Xiong}}, \bibinfo {author}
  {\bibfnamefont {S.}~\bibnamefont {Xue}}, \bibinfo {author} {\bibfnamefont
  {X.}~\bibnamefont {Fu}}, \bibinfo {author} {\bibfnamefont {A.}~\bibnamefont
  {Huang}}, \bibinfo {author} {\bibfnamefont {X.}~\bibnamefont {Qiang}},
  \bibinfo {author} {\bibfnamefont {P.}~\bibnamefont {Xu}}, \bibinfo {author}
  {\bibfnamefont {J.}~\bibnamefont {Liu}}, \bibinfo {author} {\bibfnamefont
  {S.}~\bibnamefont {Zheng}}, \bibinfo {author} {\bibfnamefont {H.-L.}\
  \bibnamefont {Huang}}, \bibinfo {author} {\bibfnamefont {M.}~\bibnamefont
  {Deng}}, \bibinfo {author} {\bibfnamefont {D.}~\bibnamefont {Poletti}},
  \bibinfo {author} {\bibfnamefont {W.-S.}\ \bibnamefont {Bao}},\ and\ \bibinfo
  {author} {\bibfnamefont {J.}~\bibnamefont {Wu}},\ }\bibfield  {title}
  {\bibinfo {title} {General-purpose quantum circuit simulator with projected
  entangled-pair states and the quantum supremacy frontier},\ }\href
  {https://doi.org/10.1103/PhysRevLett.123.190501} {\bibfield  {journal}
  {\bibinfo  {journal} {Physical Review Letters}\ }\textbf {\bibinfo {volume}
  {123}},\ \bibinfo {pages} {190501} (\bibinfo {year} {2019})}\BibitemShut
  {NoStop}%
\bibitem [{\citenamefont {Liu}\ \emph {et~al.}(2021{\natexlab{b}})\citenamefont
  {Liu}, \citenamefont {Liu}, \citenamefont {Li}, \citenamefont {Fu},
  \citenamefont {Yang}, \citenamefont {Song}, \citenamefont {Zhao},
  \citenamefont {Wang}, \citenamefont {Peng}, \citenamefont {Chen},
  \citenamefont {Guo}, \citenamefont {Huang}, \citenamefont {Wu},\ and\
  \citenamefont {Chen}}]{LLLF+21TNQC}%
  \BibitemOpen
  \bibfield  {author} {\bibinfo {author} {\bibfnamefont {Y.~A.}\ \bibnamefont
  {Liu}}, \bibinfo {author} {\bibfnamefont {X.~L.}\ \bibnamefont {Liu}},
  \bibinfo {author} {\bibfnamefont {F.~N.}\ \bibnamefont {Li}}, \bibinfo
  {author} {\bibfnamefont {H.}~\bibnamefont {Fu}}, \bibinfo {author}
  {\bibfnamefont {Y.}~\bibnamefont {Yang}}, \bibinfo {author} {\bibfnamefont
  {J.}~\bibnamefont {Song}}, \bibinfo {author} {\bibfnamefont {P.}~\bibnamefont
  {Zhao}}, \bibinfo {author} {\bibfnamefont {Z.}~\bibnamefont {Wang}}, \bibinfo
  {author} {\bibfnamefont {D.}~\bibnamefont {Peng}}, \bibinfo {author}
  {\bibfnamefont {H.}~\bibnamefont {Chen}}, \bibinfo {author} {\bibfnamefont
  {C.}~\bibnamefont {Guo}}, \bibinfo {author} {\bibfnamefont {H.}~\bibnamefont
  {Huang}}, \bibinfo {author} {\bibfnamefont {W.}~\bibnamefont {Wu}},\ and\
  \bibinfo {author} {\bibfnamefont {D.}~\bibnamefont {Chen}},\ }\bibfield
  {title} {\bibinfo {title} {Closing the ``quantum supremacy'' gap: Achieving
  real-time simulation of a random quantum circuit using a new sunway
  supercomputer},\ }in\ \href@noop {} {\emph {\bibinfo {booktitle} {Proceedings
  of the International Conference for High Performance Computing, Networking,
  Storage and Analysis}}},\ \bibinfo {series and number} {SC '21}\ (\bibinfo
  {publisher} {Association for Computing Machinery},\ \bibinfo {address} {New
  York, NY, USA},\ \bibinfo {year} {2021})\BibitemShut {NoStop}%
\bibitem [{\citenamefont {Gray}\ and\ \citenamefont
  {Kourtis}(2021)}]{GK21QCTNC}%
  \BibitemOpen
  \bibfield  {author} {\bibinfo {author} {\bibfnamefont {J.}~\bibnamefont
  {Gray}}\ and\ \bibinfo {author} {\bibfnamefont {S.}~\bibnamefont {Kourtis}},\
  }\bibfield  {title} {\bibinfo {title} {Hyper-optimized tensor network
  contraction},\ }\href {https://doi.org/10.22331/q-2021-03-15-410} {\bibfield
  {journal} {\bibinfo  {journal} {{Quantum}}\ }\textbf {\bibinfo {volume}
  {5}},\ \bibinfo {pages} {410} (\bibinfo {year} {2021})}\BibitemShut {NoStop}%
\bibitem [{\citenamefont {Pan}\ \emph {et~al.}(2022)\citenamefont {Pan},
  \citenamefont {Chen},\ and\ \citenamefont {Zhang}}]{PCZ22TNQC}%
  \BibitemOpen
  \bibfield  {author} {\bibinfo {author} {\bibfnamefont {F.}~\bibnamefont
  {Pan}}, \bibinfo {author} {\bibfnamefont {K.}~\bibnamefont {Chen}},\ and\
  \bibinfo {author} {\bibfnamefont {P.}~\bibnamefont {Zhang}},\ }\bibfield
  {title} {\bibinfo {title} {Solving the sampling problem of the sycamore
  quantum circuits},\ }\href {https://doi.org/10.1103/PhysRevLett.129.090502}
  {\bibfield  {journal} {\bibinfo  {journal} {Physical Review Letters}\
  }\textbf {\bibinfo {volume} {129}},\ \bibinfo {pages} {090502} (\bibinfo
  {year} {2022})}\BibitemShut {NoStop}%
\bibitem [{\citenamefont {Benedetti}\ \emph {et~al.}(2019)\citenamefont
  {Benedetti}, \citenamefont {Lloyd}, \citenamefont {Sack},\ and\ \citenamefont
  {Fiorentini}}]{BLSF19VQCML}%
  \BibitemOpen
  \bibfield  {author} {\bibinfo {author} {\bibfnamefont {M.}~\bibnamefont
  {Benedetti}}, \bibinfo {author} {\bibfnamefont {E.}~\bibnamefont {Lloyd}},
  \bibinfo {author} {\bibfnamefont {S.}~\bibnamefont {Sack}},\ and\ \bibinfo
  {author} {\bibfnamefont {M.}~\bibnamefont {Fiorentini}},\ }\bibfield  {title}
  {\bibinfo {title} {Parameterized quantum circuits as machine learning
  models},\ }\href {https://doi.org/10.1088/2058-9565/ab4eb5} {\bibfield
  {journal} {\bibinfo  {journal} {Quantum Science and Technology}\ }\textbf
  {\bibinfo {volume} {4}},\ \bibinfo {pages} {043001} (\bibinfo {year}
  {2019})}\BibitemShut {NoStop}%
\bibitem [{\citenamefont {Lin}\ \emph {et~al.}(2021)\citenamefont {Lin},
  \citenamefont {Dilip}, \citenamefont {Green}, \citenamefont {Smith},\ and\
  \citenamefont {Pollmann}}]{LDGSP21TNQC}%
  \BibitemOpen
  \bibfield  {author} {\bibinfo {author} {\bibfnamefont {S.-H.}\ \bibnamefont
  {Lin}}, \bibinfo {author} {\bibfnamefont {R.}~\bibnamefont {Dilip}}, \bibinfo
  {author} {\bibfnamefont {A.~G.}\ \bibnamefont {Green}}, \bibinfo {author}
  {\bibfnamefont {A.}~\bibnamefont {Smith}},\ and\ \bibinfo {author}
  {\bibfnamefont {F.}~\bibnamefont {Pollmann}},\ }\bibfield  {title} {\bibinfo
  {title} {Real- and imaginary-time evolution with compressed quantum
  circuits},\ }\href {https://doi.org/10.1103/PRXQuantum.2.010342} {\bibfield
  {journal} {\bibinfo  {journal} {PRX Quantum}\ }\textbf {\bibinfo {volume}
  {2}},\ \bibinfo {pages} {010342} (\bibinfo {year} {2021})}\BibitemShut
  {NoStop}%
\bibitem [{\citenamefont {Shirakawa}\ \emph {et~al.}(2021)\citenamefont
  {Shirakawa}, \citenamefont {Ueda},\ and\ \citenamefont
  {Yunoki}}]{shirakawa2021automatic}%
  \BibitemOpen
  \bibfield  {author} {\bibinfo {author} {\bibfnamefont {T.}~\bibnamefont
  {Shirakawa}}, \bibinfo {author} {\bibfnamefont {H.}~\bibnamefont {Ueda}},\
  and\ \bibinfo {author} {\bibfnamefont {S.}~\bibnamefont {Yunoki}},\
  }\href@noop {} {\bibinfo {title} {Automatic quantum circuit encoding of a
  given arbitrary quantum state}} (\bibinfo {year} {2021}),\ \Eprint
  {https://arxiv.org/abs/2112.14524} {arXiv:2112.14524 [quant-ph]} \BibitemShut
  {NoStop}%
\bibitem [{\citenamefont {Dov}\ \emph {et~al.}(2022)\citenamefont {Dov},
  \citenamefont {Shnaiderov}, \citenamefont {Makmal},\ and\ \citenamefont
  {Torre}}]{dov2022approximate}%
  \BibitemOpen
  \bibfield  {author} {\bibinfo {author} {\bibfnamefont {M.~B.}\ \bibnamefont
  {Dov}}, \bibinfo {author} {\bibfnamefont {D.}~\bibnamefont {Shnaiderov}},
  \bibinfo {author} {\bibfnamefont {A.}~\bibnamefont {Makmal}},\ and\ \bibinfo
  {author} {\bibfnamefont {E.~G.~D.}\ \bibnamefont {Torre}},\ }\href@noop {}
  {\bibinfo {title} {Approximate encoding of quantum states using shallow
  circuits}} (\bibinfo {year} {2022}),\ \Eprint
  {https://arxiv.org/abs/2207.00028} {arXiv:2207.00028 [quant-ph]} \BibitemShut
  {NoStop}%
\bibitem [{\citenamefont {Rudolph}\ \emph {et~al.}(2022)\citenamefont
  {Rudolph}, \citenamefont {Chen}, \citenamefont {Miller}, \citenamefont
  {Acharya},\ and\ \citenamefont {Perdomo-Ortiz}}]{rudolph2022decomposition}%
  \BibitemOpen
  \bibfield  {author} {\bibinfo {author} {\bibfnamefont {M.~S.}\ \bibnamefont
  {Rudolph}}, \bibinfo {author} {\bibfnamefont {J.}~\bibnamefont {Chen}},
  \bibinfo {author} {\bibfnamefont {J.}~\bibnamefont {Miller}}, \bibinfo
  {author} {\bibfnamefont {A.}~\bibnamefont {Acharya}},\ and\ \bibinfo {author}
  {\bibfnamefont {A.}~\bibnamefont {Perdomo-Ortiz}},\ }\href@noop {} {\bibinfo
  {title} {Decomposition of matrix product states into shallow quantum
  circuits}} (\bibinfo {year} {2022}),\ \Eprint
  {https://arxiv.org/abs/2209.00595} {arXiv:2209.00595 [quant-ph]} \BibitemShut
  {NoStop}%
\bibitem [{\citenamefont {Chong}\ \emph {et~al.}(2017)\citenamefont {Chong},
  \citenamefont {Franklin},\ and\ \citenamefont {Martonosi}}]{CFM17Qcompile}%
  \BibitemOpen
  \bibfield  {author} {\bibinfo {author} {\bibfnamefont {F.~T.}\ \bibnamefont
  {Chong}}, \bibinfo {author} {\bibfnamefont {D.}~\bibnamefont {Franklin}},\
  and\ \bibinfo {author} {\bibfnamefont {M.}~\bibnamefont {Martonosi}},\
  }\bibfield  {title} {\bibinfo {title} {Programming languages and compiler
  design for realistic quantum hardware},\ }\href
  {https://doi.org/10.1038/nature23459} {\bibfield  {journal} {\bibinfo
  {journal} {Nature}\ }\textbf {\bibinfo {volume} {549}},\ \bibinfo {pages}
  {180} (\bibinfo {year} {2017})}\BibitemShut {NoStop}%
\bibitem [{\citenamefont {Chen}\ \emph {et~al.}(2020)\citenamefont {Chen},
  \citenamefont {Huang}, \citenamefont {Hsing},\ and\ \citenamefont
  {Kao}}]{chen2020hybrid}%
  \BibitemOpen
  \bibfield  {author} {\bibinfo {author} {\bibfnamefont {S.~Y.-C.}\
  \bibnamefont {Chen}}, \bibinfo {author} {\bibfnamefont {C.-M.}\ \bibnamefont
  {Huang}}, \bibinfo {author} {\bibfnamefont {C.-W.}\ \bibnamefont {Hsing}},\
  and\ \bibinfo {author} {\bibfnamefont {Y.-J.}\ \bibnamefont {Kao}},\
  }\href@noop {} {\bibinfo {title} {Hybrid quantum-classical classifier based
  on tensor network and variational quantum circuit}} (\bibinfo {year}
  {2020}),\ \Eprint {https://arxiv.org/abs/2011.14651} {arXiv:2011.14651
  [quant-ph]} \BibitemShut {NoStop}%
\bibitem [{\citenamefont {Huang}\ \emph
  {et~al.}(2021{\natexlab{b}})\citenamefont {Huang}, \citenamefont {Tan},\ and\
  \citenamefont {Xu}}]{HUANG202189}%
  \BibitemOpen
  \bibfield  {author} {\bibinfo {author} {\bibfnamefont {R.}~\bibnamefont
  {Huang}}, \bibinfo {author} {\bibfnamefont {X.}~\bibnamefont {Tan}},\ and\
  \bibinfo {author} {\bibfnamefont {Q.}~\bibnamefont {Xu}},\ }\bibfield
  {title} {\bibinfo {title} {Variational quantum tensor networks classifiers},\
  }\href {https://doi.org/https://doi.org/10.1016/j.neucom.2021.04.074}
  {\bibfield  {journal} {\bibinfo  {journal} {Neurocomputing}\ }\textbf
  {\bibinfo {volume} {452}},\ \bibinfo {pages} {89} (\bibinfo {year}
  {2021}{\natexlab{b}})}\BibitemShut {NoStop}%
\bibitem [{\citenamefont {Kardashin}\ \emph {et~al.}(2021)\citenamefont
  {Kardashin}, \citenamefont {Uvarov},\ and\ \citenamefont
  {Biamonte}}]{10.3389/fphy.2020.586374}%
  \BibitemOpen
  \bibfield  {author} {\bibinfo {author} {\bibfnamefont {A.}~\bibnamefont
  {Kardashin}}, \bibinfo {author} {\bibfnamefont {A.}~\bibnamefont {Uvarov}},\
  and\ \bibinfo {author} {\bibfnamefont {J.}~\bibnamefont {Biamonte}},\
  }\bibfield  {title} {\bibinfo {title} {Quantum machine learning tensor
  network states},\ }\bibfield  {journal} {\bibinfo  {journal} {Frontiers in
  Physics}\ }\textbf {\bibinfo {volume} {8}},\ \href
  {https://doi.org/10.3389/fphy.2020.586374} {10.3389/fphy.2020.586374}
  (\bibinfo {year} {2021})\BibitemShut {NoStop}%
\bibitem [{\citenamefont {Araz}\ and\ \citenamefont
  {Spannowsky}(2022)}]{PhysRevA.106.062423}%
  \BibitemOpen
  \bibfield  {author} {\bibinfo {author} {\bibfnamefont {J.~Y.}\ \bibnamefont
  {Araz}}\ and\ \bibinfo {author} {\bibfnamefont {M.}~\bibnamefont
  {Spannowsky}},\ }\bibfield  {title} {\bibinfo {title} {Classical versus
  quantum: Comparing tensor-network-based quantum circuits on large hadron
  collider data},\ }\href {https://doi.org/10.1103/PhysRevA.106.062423}
  {\bibfield  {journal} {\bibinfo  {journal} {Physical Review A}\ }\textbf
  {\bibinfo {volume} {106}},\ \bibinfo {pages} {062423} (\bibinfo {year}
  {2022})}\BibitemShut {NoStop}%
\bibitem [{\citenamefont {Wall}\ \emph {et~al.}(2021)\citenamefont {Wall},
  \citenamefont {Abernathy},\ and\ \citenamefont {Quiroz}}]{WAQ21GTNexp}%
  \BibitemOpen
  \bibfield  {author} {\bibinfo {author} {\bibfnamefont {M.~L.}\ \bibnamefont
  {Wall}}, \bibinfo {author} {\bibfnamefont {M.~R.}\ \bibnamefont
  {Abernathy}},\ and\ \bibinfo {author} {\bibfnamefont {G.}~\bibnamefont
  {Quiroz}},\ }\bibfield  {title} {\bibinfo {title} {Generative machine
  learning with tensor networks: Benchmarks on near-term quantum computers},\
  }\href {https://doi.org/10.1103/PhysRevResearch.3.023010} {\bibfield
  {journal} {\bibinfo  {journal} {Physical Review Research}\ }\textbf {\bibinfo
  {volume} {3}},\ \bibinfo {pages} {023010} (\bibinfo {year}
  {2021})}\BibitemShut {NoStop}%
\bibitem [{\citenamefont {Lazzarin}\ \emph {et~al.}(2022)\citenamefont
  {Lazzarin}, \citenamefont {Galli},\ and\ \citenamefont
  {Prati}}]{LAZZARIN2022128056}%
  \BibitemOpen
  \bibfield  {author} {\bibinfo {author} {\bibfnamefont {M.}~\bibnamefont
  {Lazzarin}}, \bibinfo {author} {\bibfnamefont {D.~E.}\ \bibnamefont
  {Galli}},\ and\ \bibinfo {author} {\bibfnamefont {E.}~\bibnamefont {Prati}},\
  }\bibfield  {title} {\bibinfo {title} {Multi-class quantum classifiers with
  tensor network circuits for quantum phase recognition},\ }\href
  {https://doi.org/https://doi.org/10.1016/j.physleta.2022.128056} {\bibfield
  {journal} {\bibinfo  {journal} {Physics Letters A}\ }\textbf {\bibinfo
  {volume} {434}},\ \bibinfo {pages} {128056} (\bibinfo {year}
  {2022})}\BibitemShut {NoStop}%
\bibitem [{Note13()}]{Note13}%
  \BibitemOpen
  \bibinfo {note} {Quantum computing is expected to possess higher parallelism
  over classical computing. One operation on an entangled quantum state can
  manage to process all the exponentially-many coefficients, even this
  operation might be just on one qubit or a local part of the state. In
  comparison, such parallelism cannot be gained if the state has no
  entanglement (i.e., a product state). This is one reason for regarding
  entanglement as the source for the superior power of quantum
  computing.}\BibitemShut {Stop}%
\bibitem [{\citenamefont {Nishino}(1995)}]{N95TMRG2Dclassic}%
  \BibitemOpen
  \bibfield  {author} {\bibinfo {author} {\bibfnamefont {T.}~\bibnamefont
  {Nishino}},\ }\bibfield  {title} {\bibinfo {title} {{Density matrix
  renormalization group method for 2D classical models}},\ }\href
  {https://doi.org/10.1143/jpsj.64.3598} {\bibfield  {journal} {\bibinfo
  {journal} {Journal of the Physical Society of Japan}\ }\textbf {\bibinfo
  {volume} {64}},\ \bibinfo {pages} {3598} (\bibinfo {year}
  {1995})}\BibitemShut {NoStop}%
\bibitem [{\citenamefont {Levin}\ and\ \citenamefont {Nave}(2007)}]{LN07TRG}%
  \BibitemOpen
  \bibfield  {author} {\bibinfo {author} {\bibfnamefont {M.}~\bibnamefont
  {Levin}}\ and\ \bibinfo {author} {\bibfnamefont {C.~P.}\ \bibnamefont
  {Nave}},\ }\bibfield  {title} {\bibinfo {title} {Tensor renormalization group
  approach to two-dimensional classical lattice models},\ }\href
  {https://doi.org/10.1103/PhysRevLett.99.120601} {\bibfield  {journal}
  {\bibinfo  {journal} {Physical Review Letters}\ }\textbf {\bibinfo {volume}
  {99}},\ \bibinfo {pages} {120601} (\bibinfo {year} {2007})}\BibitemShut
  {NoStop}%
\bibitem [{\citenamefont {Evenbly}\ and\ \citenamefont
  {Vidal}(2015)}]{EV15TNR}%
  \BibitemOpen
  \bibfield  {author} {\bibinfo {author} {\bibfnamefont {G.}~\bibnamefont
  {Evenbly}}\ and\ \bibinfo {author} {\bibfnamefont {G.}~\bibnamefont
  {Vidal}},\ }\bibfield  {title} {\bibinfo {title} {Tensor network
  renormalization},\ }\href {https://doi.org/10.1103/PhysRevLett.115.180405}
  {\bibfield  {journal} {\bibinfo  {journal} {Physical Review Letters}\
  }\textbf {\bibinfo {volume} {115}},\ \bibinfo {pages} {180405} (\bibinfo
  {year} {2015})}\BibitemShut {NoStop}%
\bibitem [{\citenamefont {Chen}\ \emph
  {et~al.}(2018{\natexlab{a}})\citenamefont {Chen}, \citenamefont {Cheng},
  \citenamefont {Xie}, \citenamefont {Wang},\ and\ \citenamefont
  {Xiang}}]{CCXWX18MPSbolzmann}%
  \BibitemOpen
  \bibfield  {author} {\bibinfo {author} {\bibfnamefont {J.}~\bibnamefont
  {Chen}}, \bibinfo {author} {\bibfnamefont {S.}~\bibnamefont {Cheng}},
  \bibinfo {author} {\bibfnamefont {H.}~\bibnamefont {Xie}}, \bibinfo {author}
  {\bibfnamefont {L.}~\bibnamefont {Wang}},\ and\ \bibinfo {author}
  {\bibfnamefont {T.}~\bibnamefont {Xiang}},\ }\bibfield  {title} {\bibinfo
  {title} {Equivalence of restricted boltzmann machines and tensor network
  states},\ }\href {https://doi.org/10.1103/PhysRevB.97.085104} {\bibfield
  {journal} {\bibinfo  {journal} {Physical Review B}\ }\textbf {\bibinfo
  {volume} {97}},\ \bibinfo {pages} {085104} (\bibinfo {year}
  {2018}{\natexlab{a}})}\BibitemShut {NoStop}%
\bibitem [{\citenamefont {Cheng}\ \emph
  {et~al.}(2018{\natexlab{b}})\citenamefont {Cheng}, \citenamefont {Chen},\
  and\ \citenamefont {Wang}}]{cheng2018information}%
  \BibitemOpen
  \bibfield  {author} {\bibinfo {author} {\bibfnamefont {S.}~\bibnamefont
  {Cheng}}, \bibinfo {author} {\bibfnamefont {J.}~\bibnamefont {Chen}},\ and\
  \bibinfo {author} {\bibfnamefont {L.}~\bibnamefont {Wang}},\ }\bibfield
  {title} {\bibinfo {title} {Information perspective to probabilistic modeling:
  Boltzmann machines versus born machines},\ }\href
  {https://doi.org/10.3390/e20080583} {\bibfield  {journal} {\bibinfo
  {journal} {Entropy}\ }\textbf {\bibinfo {volume} {20}},\ \bibinfo {pages}
  {583} (\bibinfo {year} {2018}{\natexlab{b}})}\BibitemShut {NoStop}%
\bibitem [{\citenamefont {Zheng}\ \emph {et~al.}(2019)\citenamefont {Zheng},
  \citenamefont {He}, \citenamefont {Regnault},\ and\ \citenamefont
  {Bernevig}}]{ZHRB19BolMPS}%
  \BibitemOpen
  \bibfield  {author} {\bibinfo {author} {\bibfnamefont {Y.}~\bibnamefont
  {Zheng}}, \bibinfo {author} {\bibfnamefont {H.}~\bibnamefont {He}}, \bibinfo
  {author} {\bibfnamefont {N.}~\bibnamefont {Regnault}},\ and\ \bibinfo
  {author} {\bibfnamefont {B.~A.}\ \bibnamefont {Bernevig}},\ }\bibfield
  {title} {\bibinfo {title} {Restricted boltzmann machines and matrix product
  states of one-dimensional translationally invariant stabilizer codes},\
  }\href {https://doi.org/10.1103/PhysRevB.99.155129} {\bibfield  {journal}
  {\bibinfo  {journal} {Physical Review B}\ }\textbf {\bibinfo {volume} {99}},\
  \bibinfo {pages} {155129} (\bibinfo {year} {2019})}\BibitemShut {NoStop}%
\bibitem [{\citenamefont {Li}\ \emph {et~al.}(2021)\citenamefont {Li},
  \citenamefont {Pan}, \citenamefont {Zhou},\ and\ \citenamefont
  {Zhang}}]{LPZZ21TNBM}%
  \BibitemOpen
  \bibfield  {author} {\bibinfo {author} {\bibfnamefont {S.}~\bibnamefont
  {Li}}, \bibinfo {author} {\bibfnamefont {F.}~\bibnamefont {Pan}}, \bibinfo
  {author} {\bibfnamefont {P.}~\bibnamefont {Zhou}},\ and\ \bibinfo {author}
  {\bibfnamefont {P.}~\bibnamefont {Zhang}},\ }\bibfield  {title} {\bibinfo
  {title} {Boltzmann machines as two-dimensional tensor networks},\ }\href
  {https://doi.org/10.1103/PhysRevB.104.075154} {\bibfield  {journal} {\bibinfo
   {journal} {Physical Review B}\ }\textbf {\bibinfo {volume} {104}},\ \bibinfo
  {pages} {075154} (\bibinfo {year} {2021})}\BibitemShut {NoStop}%
\bibitem [{\citenamefont {Glasser}\ \emph {et~al.}(2018)\citenamefont
  {Glasser}, \citenamefont {Pancotti}, \citenamefont {August}, \citenamefont
  {Rodriguez},\ and\ \citenamefont {Cirac}}]{GPARC18MLstringbond}%
  \BibitemOpen
  \bibfield  {author} {\bibinfo {author} {\bibfnamefont {I.}~\bibnamefont
  {Glasser}}, \bibinfo {author} {\bibfnamefont {N.}~\bibnamefont {Pancotti}},
  \bibinfo {author} {\bibfnamefont {M.}~\bibnamefont {August}}, \bibinfo
  {author} {\bibfnamefont {I.~D.}\ \bibnamefont {Rodriguez}},\ and\ \bibinfo
  {author} {\bibfnamefont {J.~I.}\ \bibnamefont {Cirac}},\ }\bibfield  {title}
  {\bibinfo {title} {Neural-network quantum states, string-bond states, and
  chiral topological states},\ }\href
  {https://doi.org/10.1103/PhysRevX.8.011006} {\bibfield  {journal} {\bibinfo
  {journal} {Physical Review X}\ }\textbf {\bibinfo {volume} {8}},\ \bibinfo
  {pages} {011006} (\bibinfo {year} {2018})}\BibitemShut {NoStop}%
\bibitem [{\citenamefont {Medina}\ \emph {et~al.}(2021)\citenamefont {Medina},
  \citenamefont {Vasseur},\ and\ \citenamefont {Serbyn}}]{MVS21QSRBM}%
  \BibitemOpen
  \bibfield  {author} {\bibinfo {author} {\bibfnamefont {R.}~\bibnamefont
  {Medina}}, \bibinfo {author} {\bibfnamefont {R.}~\bibnamefont {Vasseur}},\
  and\ \bibinfo {author} {\bibfnamefont {M.}~\bibnamefont {Serbyn}},\
  }\bibfield  {title} {\bibinfo {title} {Entanglement transitions from
  restricted boltzmann machines},\ }\href
  {https://doi.org/10.1103/PhysRevB.104.104205} {\bibfield  {journal} {\bibinfo
   {journal} {Physical Review B}\ }\textbf {\bibinfo {volume} {104}},\ \bibinfo
  {pages} {104205} (\bibinfo {year} {2021})}\BibitemShut {NoStop}%
\bibitem [{\citenamefont {Wang}\ \emph {et~al.}(2023)\citenamefont {Wang},
  \citenamefont {Pan}, \citenamefont {Xu}, \citenamefont {Yang}, \citenamefont
  {Li},\ and\ \citenamefont {Cichocki}}]{wang2023tensor}%
  \BibitemOpen
  \bibfield  {author} {\bibinfo {author} {\bibfnamefont {M.}~\bibnamefont
  {Wang}}, \bibinfo {author} {\bibfnamefont {Y.}~\bibnamefont {Pan}}, \bibinfo
  {author} {\bibfnamefont {Z.}~\bibnamefont {Xu}}, \bibinfo {author}
  {\bibfnamefont {X.}~\bibnamefont {Yang}}, \bibinfo {author} {\bibfnamefont
  {G.}~\bibnamefont {Li}},\ and\ \bibinfo {author} {\bibfnamefont
  {A.}~\bibnamefont {Cichocki}},\ }\href@noop {} {\bibinfo {title} {Tensor
  networks meet neural networks: A survey and future perspectives}} (\bibinfo
  {year} {2023}),\ \Eprint {https://arxiv.org/abs/2302.09019} {arXiv:2302.09019
  [cs.LG]} \BibitemShut {NoStop}%
\bibitem [{\citenamefont {Zhao}\ \emph {et~al.}(2016)\citenamefont {Zhao},
  \citenamefont {Zhou}, \citenamefont {Xie}, \citenamefont {Zhang},\ and\
  \citenamefont {Cichocki}}]{ZZXZ+16tensorring}%
  \BibitemOpen
  \bibfield  {author} {\bibinfo {author} {\bibfnamefont {Q.}~\bibnamefont
  {Zhao}}, \bibinfo {author} {\bibfnamefont {G.}~\bibnamefont {Zhou}}, \bibinfo
  {author} {\bibfnamefont {S.}~\bibnamefont {Xie}}, \bibinfo {author}
  {\bibfnamefont {L.}~\bibnamefont {Zhang}},\ and\ \bibinfo {author}
  {\bibfnamefont {A.}~\bibnamefont {Cichocki}},\ }\href@noop {} {\bibinfo
  {title} {Tensor ring decomposition}} (\bibinfo {year} {2016}),\ \Eprint
  {https://arxiv.org/abs/1606.05535} {arXiv:1606.05535 [cs.NA]} \BibitemShut
  {NoStop}%
\bibitem [{\citenamefont {Chen}\ \emph {et~al.}(2019)\citenamefont {Chen},
  \citenamefont {Batselier}, \citenamefont {Ko},\ and\ \citenamefont
  {Wong}}]{CBKW19TTSVM}%
  \BibitemOpen
  \bibfield  {author} {\bibinfo {author} {\bibfnamefont {C.}~\bibnamefont
  {Chen}}, \bibinfo {author} {\bibfnamefont {K.}~\bibnamefont {Batselier}},
  \bibinfo {author} {\bibfnamefont {C.-Y.}\ \bibnamefont {Ko}},\ and\ \bibinfo
  {author} {\bibfnamefont {N.}~\bibnamefont {Wong}},\ }\bibfield  {title}
  {\bibinfo {title} {A support tensor train machine},\ }in\ \href@noop {}
  {\emph {\bibinfo {booktitle} {2019 International Joint Conference on Neural
  Networks (IJCNN)}}}\ (\bibinfo {year} {2019})\ pp.\ \bibinfo {pages}
  {1--8}\BibitemShut {NoStop}%
\bibitem [{\citenamefont {Chen}\ \emph {et~al.}(2022)\citenamefont {Chen},
  \citenamefont {Batselier}, \citenamefont {Yu},\ and\ \citenamefont
  {Wong}}]{CBYW22TTSVM}%
  \BibitemOpen
  \bibfield  {author} {\bibinfo {author} {\bibfnamefont {C.}~\bibnamefont
  {Chen}}, \bibinfo {author} {\bibfnamefont {K.}~\bibnamefont {Batselier}},
  \bibinfo {author} {\bibfnamefont {W.}~\bibnamefont {Yu}},\ and\ \bibinfo
  {author} {\bibfnamefont {N.}~\bibnamefont {Wong}},\ }\bibfield  {title}
  {\bibinfo {title} {Kernelized support tensor train machines},\ }\href
  {https://doi.org/https://doi.org/10.1016/j.patcog.2021.108337} {\bibfield
  {journal} {\bibinfo  {journal} {Pattern Recognition}\ }\textbf {\bibinfo
  {volume} {122}},\ \bibinfo {pages} {108337} (\bibinfo {year}
  {2022})}\BibitemShut {NoStop}%
\bibitem [{\citenamefont {Qiu}\ \emph {et~al.}(2022)\citenamefont {Qiu},
  \citenamefont {Zhou}, \citenamefont {Huang}, \citenamefont {Zhao},\ and\
  \citenamefont {Xie}}]{QZHZX22TTPCA}%
  \BibitemOpen
  \bibfield  {author} {\bibinfo {author} {\bibfnamefont {Y.}~\bibnamefont
  {Qiu}}, \bibinfo {author} {\bibfnamefont {G.}~\bibnamefont {Zhou}}, \bibinfo
  {author} {\bibfnamefont {Z.}~\bibnamefont {Huang}}, \bibinfo {author}
  {\bibfnamefont {Q.}~\bibnamefont {Zhao}},\ and\ \bibinfo {author}
  {\bibfnamefont {S.}~\bibnamefont {Xie}},\ }\bibfield  {title} {\bibinfo
  {title} {Efficient tensor robust pca under hybrid model of tucker and tensor
  train},\ }\href {https://doi.org/10.1109/LSP.2022.3143721} {\bibfield
  {journal} {\bibinfo  {journal} {IEEE Signal Processing Letters}\ }\textbf
  {\bibinfo {volume} {29}},\ \bibinfo {pages} {627} (\bibinfo {year}
  {2022})}\BibitemShut {NoStop}%
\bibitem [{\citenamefont {Yang}\ \emph {et~al.}(2017)\citenamefont {Yang},
  \citenamefont {Krompass},\ and\ \citenamefont {Tresp}}]{YKT17TTRNN}%
  \BibitemOpen
  \bibfield  {author} {\bibinfo {author} {\bibfnamefont {Y.}~\bibnamefont
  {Yang}}, \bibinfo {author} {\bibfnamefont {D.}~\bibnamefont {Krompass}},\
  and\ \bibinfo {author} {\bibfnamefont {V.}~\bibnamefont {Tresp}},\ }\bibfield
   {title} {\bibinfo {title} {Tensor-train recurrent neural networks for video
  classification},\ }in\ \href@noop {} {\emph {\bibinfo {booktitle}
  {Proceedings of the 34th International Conference on Machine Learning}}},\
  \bibinfo {series} {Proceedings of Machine Learning Research}, Vol.~\bibinfo
  {volume} {70},\ \bibinfo {editor} {edited by\ \bibinfo {editor}
  {\bibfnamefont {D.}~\bibnamefont {Precup}}\ and\ \bibinfo {editor}
  {\bibfnamefont {Y.~W.}\ \bibnamefont {Teh}}}\ (\bibinfo  {publisher}
  {Proceedings of Machine Learning Research},\ \bibinfo {year} {2017})\ pp.\
  \bibinfo {pages} {3891--3900}\BibitemShut {NoStop}%
\bibitem [{\citenamefont {Chen}\ \emph
  {et~al.}(2018{\natexlab{b}})\citenamefont {Chen}, \citenamefont {Batselier},
  \citenamefont {Suykens},\ and\ \citenamefont {Wong}}]{CBSW18TTML}%
  \BibitemOpen
  \bibfield  {author} {\bibinfo {author} {\bibfnamefont {Z.}~\bibnamefont
  {Chen}}, \bibinfo {author} {\bibfnamefont {K.}~\bibnamefont {Batselier}},
  \bibinfo {author} {\bibfnamefont {J.~A.~K.}\ \bibnamefont {Suykens}},\ and\
  \bibinfo {author} {\bibfnamefont {N.}~\bibnamefont {Wong}},\ }\bibfield
  {title} {\bibinfo {title} {Parallelized tensor train learning of polynomial
  classifiers},\ }\href {https://doi.org/10.1109/TNNLS.2017.2771264} {\bibfield
   {journal} {\bibinfo  {journal} {IEEE Transactions on Neural Networks and
  Learning Systems}\ }\textbf {\bibinfo {volume} {29}},\ \bibinfo {pages}
  {4621} (\bibinfo {year} {2018}{\natexlab{b}})}\BibitemShut {NoStop}%
\bibitem [{\citenamefont {Wang}\ \emph {et~al.}(2018)\citenamefont {Wang},
  \citenamefont {Sun}, \citenamefont {Eriksson}, \citenamefont {Wang},\ and\
  \citenamefont {Aggarwal}}]{WSEWA18TRNN}%
  \BibitemOpen
  \bibfield  {author} {\bibinfo {author} {\bibfnamefont {W.}~\bibnamefont
  {Wang}}, \bibinfo {author} {\bibfnamefont {Y.}~\bibnamefont {Sun}}, \bibinfo
  {author} {\bibfnamefont {B.}~\bibnamefont {Eriksson}}, \bibinfo {author}
  {\bibfnamefont {W.}~\bibnamefont {Wang}},\ and\ \bibinfo {author}
  {\bibfnamefont {V.}~\bibnamefont {Aggarwal}},\ }\bibfield  {title} {\bibinfo
  {title} {Wide compression: tensor ring nets},\ }in\ \href@noop {} {\emph
  {\bibinfo {booktitle} {Proceedings of the IEEE Conference on Computer Vision
  and Pattern Recognition}}}\ (\bibinfo {year} {2018})\BibitemShut {NoStop}%
\bibitem [{\citenamefont {Su}\ \emph {et~al.}(2020)\citenamefont {Su},
  \citenamefont {Byeon}, \citenamefont {Kossaifi}, \citenamefont {Huang},
  \citenamefont {Kautz},\ and\ \citenamefont {Anandkumar}}]{SBKH20TTLSTM}%
  \BibitemOpen
  \bibfield  {author} {\bibinfo {author} {\bibfnamefont {J.}~\bibnamefont
  {Su}}, \bibinfo {author} {\bibfnamefont {W.}~\bibnamefont {Byeon}}, \bibinfo
  {author} {\bibfnamefont {J.}~\bibnamefont {Kossaifi}}, \bibinfo {author}
  {\bibfnamefont {F.}~\bibnamefont {Huang}}, \bibinfo {author} {\bibfnamefont
  {J.}~\bibnamefont {Kautz}},\ and\ \bibinfo {author} {\bibfnamefont
  {A.}~\bibnamefont {Anandkumar}},\ }\bibfield  {title} {\bibinfo {title}
  {Convolutional tensor-train lstm for spatio-temporal learning},\ }in\
  \href@noop {} {\emph {\bibinfo {booktitle} {Advances in Neural Information
  Processing Systems}}},\ Vol.~\bibinfo {volume} {33},\ \bibinfo {editor}
  {edited by\ \bibinfo {editor} {\bibfnamefont {H.}~\bibnamefont {Larochelle}},
  \bibinfo {editor} {\bibfnamefont {M.}~\bibnamefont {Ranzato}}, \bibinfo
  {editor} {\bibfnamefont {R.}~\bibnamefont {Hadsell}}, \bibinfo {editor}
  {\bibfnamefont {M.}~\bibnamefont {Balcan}},\ and\ \bibinfo {editor}
  {\bibfnamefont {H.}~\bibnamefont {Lin}}}\ (\bibinfo  {publisher} {Curran
  Associates, Inc.},\ \bibinfo {year} {2020})\ pp.\ \bibinfo {pages}
  {13714--13726}\BibitemShut {NoStop}%
\bibitem [{\citenamefont {Meng}\ \emph {et~al.}(2023)\citenamefont {Meng},
  \citenamefont {Zhang}, \citenamefont {Zhang}, \citenamefont {Gao},\ and\
  \citenamefont {Ran}}]{MZZGR23ResMPS}%
  \BibitemOpen
  \bibfield  {author} {\bibinfo {author} {\bibfnamefont {Y.-M.}\ \bibnamefont
  {Meng}}, \bibinfo {author} {\bibfnamefont {J.}~\bibnamefont {Zhang}},
  \bibinfo {author} {\bibfnamefont {P.}~\bibnamefont {Zhang}}, \bibinfo
  {author} {\bibfnamefont {C.}~\bibnamefont {Gao}},\ and\ \bibinfo {author}
  {\bibfnamefont {S.-J.}\ \bibnamefont {Ran}},\ }\bibfield  {title} {\bibinfo
  {title} {{Residual matrix product state for machine learning}},\ }\href
  {https://doi.org/10.21468/SciPostPhys.14.6.142} {\bibfield  {journal}
  {\bibinfo  {journal} {SciPost Physics}\ }\textbf {\bibinfo {volume} {14}},\
  \bibinfo {pages} {142} (\bibinfo {year} {2023})}\BibitemShut {NoStop}%
\bibitem [{\citenamefont {Wu}\ \emph {et~al.}(2023)\citenamefont {Wu},
  \citenamefont {Rossi}, \citenamefont {Vicentini},\ and\ \citenamefont
  {Carleo}}]{wu2023tensor}%
  \BibitemOpen
  \bibfield  {author} {\bibinfo {author} {\bibfnamefont {D.}~\bibnamefont
  {Wu}}, \bibinfo {author} {\bibfnamefont {R.}~\bibnamefont {Rossi}}, \bibinfo
  {author} {\bibfnamefont {F.}~\bibnamefont {Vicentini}},\ and\ \bibinfo
  {author} {\bibfnamefont {G.}~\bibnamefont {Carleo}},\ }\href@noop {}
  {\bibinfo {title} {From tensor network quantum states to tensorial recurrent
  neural networks}} (\bibinfo {year} {2023}),\ \Eprint
  {https://arxiv.org/abs/2206.12363} {arXiv:2206.12363 [quant-ph]} \BibitemShut
  {NoStop}%
\bibitem [{\citenamefont {Novikov}\ \emph {et~al.}(2015)\citenamefont
  {Novikov}, \citenamefont {Podoprikhin}, \citenamefont {Osokin},\ and\
  \citenamefont {Vetrov}}]{NPOV15TTNN}%
  \BibitemOpen
  \bibfield  {author} {\bibinfo {author} {\bibfnamefont {A.}~\bibnamefont
  {Novikov}}, \bibinfo {author} {\bibfnamefont {D.}~\bibnamefont
  {Podoprikhin}}, \bibinfo {author} {\bibfnamefont {A.}~\bibnamefont
  {Osokin}},\ and\ \bibinfo {author} {\bibfnamefont {D.~P.}\ \bibnamefont
  {Vetrov}},\ }\bibfield  {title} {\bibinfo {title} {Tensorizing neural
  networks},\ }in\ \href@noop {} {\emph {\bibinfo {booktitle} {Advances in
  Neural Information Processing Systems}}},\ Vol.~\bibinfo {volume} {28},\
  \bibinfo {editor} {edited by\ \bibinfo {editor} {\bibfnamefont
  {C.}~\bibnamefont {Cortes}}, \bibinfo {editor} {\bibfnamefont
  {N.}~\bibnamefont {Lawrence}}, \bibinfo {editor} {\bibfnamefont
  {D.}~\bibnamefont {Lee}}, \bibinfo {editor} {\bibfnamefont {M.}~\bibnamefont
  {Sugiyama}},\ and\ \bibinfo {editor} {\bibfnamefont {R.}~\bibnamefont
  {Garnett}}}\ (\bibinfo  {publisher} {Curran Associates, Inc.},\ \bibinfo
  {year} {2015})\BibitemShut {NoStop}%
\bibitem [{\citenamefont {Hayashi}\ \emph {et~al.}(2019)\citenamefont
  {Hayashi}, \citenamefont {Yamaguchi}, \citenamefont {Sugawara},\ and\
  \citenamefont {Maeda}}]{HYSM19TNCNN}%
  \BibitemOpen
  \bibfield  {author} {\bibinfo {author} {\bibfnamefont {K.}~\bibnamefont
  {Hayashi}}, \bibinfo {author} {\bibfnamefont {T.}~\bibnamefont {Yamaguchi}},
  \bibinfo {author} {\bibfnamefont {Y.}~\bibnamefont {Sugawara}},\ and\
  \bibinfo {author} {\bibfnamefont {S.-i.}\ \bibnamefont {Maeda}},\ }\bibfield
  {title} {\bibinfo {title} {Exploring unexplored tensor network decompositions
  for convolutional neural networks},\ }in\ \href@noop {} {\emph {\bibinfo
  {booktitle} {Advances in Neural Information Processing Systems}}},\
  Vol.~\bibinfo {volume} {32},\ \bibinfo {editor} {edited by\ \bibinfo {editor}
  {\bibfnamefont {H.}~\bibnamefont {Wallach}}, \bibinfo {editor} {\bibfnamefont
  {H.}~\bibnamefont {Larochelle}}, \bibinfo {editor} {\bibfnamefont
  {A.}~\bibnamefont {Beygelzimer}}, \bibinfo {editor} {\bibfnamefont
  {F.}~\bibnamefont {d\textquotesingle Alch\'{e}-Buc}}, \bibinfo {editor}
  {\bibfnamefont {E.}~\bibnamefont {Fox}},\ and\ \bibinfo {editor}
  {\bibfnamefont {R.}~\bibnamefont {Garnett}}}\ (\bibinfo  {publisher} {Curran
  Associates, Inc.},\ \bibinfo {year} {2019})\BibitemShut {NoStop}%
\bibitem [{\citenamefont {Hawkins}\ and\ \citenamefont
  {Zhang}(2021)}]{HZ21BTNN}%
  \BibitemOpen
  \bibfield  {author} {\bibinfo {author} {\bibfnamefont {C.}~\bibnamefont
  {Hawkins}}\ and\ \bibinfo {author} {\bibfnamefont {Z.}~\bibnamefont
  {Zhang}},\ }\bibfield  {title} {\bibinfo {title} {Bayesian tensorized neural
  networks with automatic rank selection},\ }\href
  {https://doi.org/https://doi.org/10.1016/j.neucom.2021.04.117} {\bibfield
  {journal} {\bibinfo  {journal} {Neurocomputing}\ }\textbf {\bibinfo {volume}
  {453}},\ \bibinfo {pages} {172} (\bibinfo {year} {2021})}\BibitemShut
  {NoStop}%
\bibitem [{\citenamefont {Liu}\ and\ \citenamefont {Ng}(2022)}]{LK22TuckerNN}%
  \BibitemOpen
  \bibfield  {author} {\bibinfo {author} {\bibfnamefont {Y.}~\bibnamefont
  {Liu}}\ and\ \bibinfo {author} {\bibfnamefont {M.~K.}\ \bibnamefont {Ng}},\
  }\bibfield  {title} {\bibinfo {title} {Deep neural network compression by
  tucker decomposition with nonlinear response},\ }\href
  {https://doi.org/https://doi.org/10.1016/j.knosys.2022.108171} {\bibfield
  {journal} {\bibinfo  {journal} {Knowledge-Based Systems}\ }\textbf {\bibinfo
  {volume} {241}},\ \bibinfo {pages} {108171} (\bibinfo {year}
  {2022})}\BibitemShut {NoStop}%
\bibitem [{\citenamefont {Brown}\ \emph {et~al.}(2020)\citenamefont {Brown},
  \citenamefont {Mann}, \citenamefont {Ryder}, \citenamefont {Subbiah},
  \citenamefont {Kaplan}, \citenamefont {Dhariwal}, \citenamefont
  {Neelakantan}, \citenamefont {Shyam}, \citenamefont {Sastry}, \citenamefont
  {Askell}, \citenamefont {Agarwal}, \citenamefont {Herbert-Voss},
  \citenamefont {Krueger}, \citenamefont {Henighan}, \citenamefont {Child},
  \citenamefont {Ramesh}, \citenamefont {Ziegler}, \citenamefont {Wu},
  \citenamefont {Winter}, \citenamefont {Hesse}, \citenamefont {Chen},
  \citenamefont {Sigler}, \citenamefont {Litwin}, \citenamefont {Gray},
  \citenamefont {Chess}, \citenamefont {Clark}, \citenamefont {Berner},
  \citenamefont {McCandlish}, \citenamefont {Radford}, \citenamefont
  {Sutskever},\ and\ \citenamefont {Amodei}}]{BMR+20GPT}%
  \BibitemOpen
  \bibfield  {author} {\bibinfo {author} {\bibfnamefont {T.}~\bibnamefont
  {Brown}}, \bibinfo {author} {\bibfnamefont {B.}~\bibnamefont {Mann}},
  \bibinfo {author} {\bibfnamefont {N.}~\bibnamefont {Ryder}}, \bibinfo
  {author} {\bibfnamefont {M.}~\bibnamefont {Subbiah}}, \bibinfo {author}
  {\bibfnamefont {J.~D.}\ \bibnamefont {Kaplan}}, \bibinfo {author}
  {\bibfnamefont {P.}~\bibnamefont {Dhariwal}}, \bibinfo {author}
  {\bibfnamefont {A.}~\bibnamefont {Neelakantan}}, \bibinfo {author}
  {\bibfnamefont {P.}~\bibnamefont {Shyam}}, \bibinfo {author} {\bibfnamefont
  {G.}~\bibnamefont {Sastry}}, \bibinfo {author} {\bibfnamefont
  {A.}~\bibnamefont {Askell}}, \bibinfo {author} {\bibfnamefont
  {S.}~\bibnamefont {Agarwal}}, \bibinfo {author} {\bibfnamefont
  {A.}~\bibnamefont {Herbert-Voss}}, \bibinfo {author} {\bibfnamefont
  {G.}~\bibnamefont {Krueger}}, \bibinfo {author} {\bibfnamefont
  {T.}~\bibnamefont {Henighan}}, \bibinfo {author} {\bibfnamefont
  {R.}~\bibnamefont {Child}}, \bibinfo {author} {\bibfnamefont
  {A.}~\bibnamefont {Ramesh}}, \bibinfo {author} {\bibfnamefont
  {D.}~\bibnamefont {Ziegler}}, \bibinfo {author} {\bibfnamefont
  {J.}~\bibnamefont {Wu}}, \bibinfo {author} {\bibfnamefont {C.}~\bibnamefont
  {Winter}}, \bibinfo {author} {\bibfnamefont {C.}~\bibnamefont {Hesse}},
  \bibinfo {author} {\bibfnamefont {M.}~\bibnamefont {Chen}}, \bibinfo {author}
  {\bibfnamefont {E.}~\bibnamefont {Sigler}}, \bibinfo {author} {\bibfnamefont
  {M.}~\bibnamefont {Litwin}}, \bibinfo {author} {\bibfnamefont
  {S.}~\bibnamefont {Gray}}, \bibinfo {author} {\bibfnamefont {B.}~\bibnamefont
  {Chess}}, \bibinfo {author} {\bibfnamefont {J.}~\bibnamefont {Clark}},
  \bibinfo {author} {\bibfnamefont {C.}~\bibnamefont {Berner}}, \bibinfo
  {author} {\bibfnamefont {S.}~\bibnamefont {McCandlish}}, \bibinfo {author}
  {\bibfnamefont {A.}~\bibnamefont {Radford}}, \bibinfo {author} {\bibfnamefont
  {I.}~\bibnamefont {Sutskever}},\ and\ \bibinfo {author} {\bibfnamefont
  {D.}~\bibnamefont {Amodei}},\ }\bibfield  {title} {\bibinfo {title} {Language
  models are few-shot learners},\ }in\ \href@noop {} {\emph {\bibinfo
  {booktitle} {Advances in Neural Information Processing Systems}}},\
  Vol.~\bibinfo {volume} {33},\ \bibinfo {editor} {edited by\ \bibinfo {editor}
  {\bibfnamefont {H.}~\bibnamefont {Larochelle}}, \bibinfo {editor}
  {\bibfnamefont {M.}~\bibnamefont {Ranzato}}, \bibinfo {editor} {\bibfnamefont
  {R.}~\bibnamefont {Hadsell}}, \bibinfo {editor} {\bibfnamefont
  {M.}~\bibnamefont {Balcan}},\ and\ \bibinfo {editor} {\bibfnamefont
  {H.}~\bibnamefont {Lin}}}\ (\bibinfo  {publisher} {Curran Associates, Inc.},\
  \bibinfo {year} {2020})\ pp.\ \bibinfo {pages} {1877--1901}\BibitemShut
  {NoStop}%
\end{thebibliography}
%

\end{document}